\documentclass{tlp}
\usepackage{times} 
\usepackage{amsfonts} 
\usepackage{amssymb} 
\usepackage{latexsym} 
\usepackage{url} 
\usepackage{listings} 
\usepackage[dvips]{graphicx}
\usepackage{color}
\DeclareSymbolFontAlphabet{\mathbb}{AMSb}

\DeclareMathSymbol{\FORALL}   {\mathord}{symbols}{"38}
\DeclareMathSymbol{\EXISTS}   {\mathord}{symbols}{"39}
\DeclareMathSymbol{\SUCHTHAT} {\mathbin}{symbols}{"01}
\def\Forall#1#2{{\FORALL#1} \SUCHTHAT #2}
\def\Exists#1#2{{\EXISTS#1} \SUCHTHAT #2}
\newcommand{\set}[1]{\ensuremath{\{#1\}}}
\newcommand{\hbase}[1]{{\ensuremath{{\cal B}_{#1}}}}
\newcommand{\hbases}[1]{{\ensuremath{{\cal B}_{#1}\cup\neg{\cal B}_{#1}}}}
\newcommand{\setmin}[2]{\ensuremath{#1\!\setminus\!#2}}
\newcommand{\aset}[2]{\mathit{aset}({#1},\linebreak[0]{#2})}
\newcommand{\expansion}[2]{\mathit{\mu}(#1,\linebreak[2]{#2})}
\newcommand{\witness}[1]{\ensuremath{\omega({#1})}}
\newcommand{\down}[1]{\ensuremath{\mathit{down}(#1)}}
\newcommand{\pair}[2]{\langle#1,\linebreak[2]#2\rangle}
\newtheorem{principle}{Principle}
\newenvironment{principlerecap}[1]{\vspace{1ex}\par{\bf Principle #1}\itshape} 
		{\vspace{1ex}\par}
\newtheorem{example}{Example}

\newtheorem{definition}{Definition}
 
\newtheorem{theorem}{Theorem}
\newtheorem{lemma}{Lemma}
\newtheorem{corollary}{Corollary}
\newenvironment{theoremrecap}[1]
	{\intheoremtrue\normalfont\rmfamily
	\trivlist\pagebreak[3]\item[\hskip \labelsep{\normalfont\itshape Theorem\ #1}]%
	\item[]%
	}
	{\endtrivlist\intheoremfalse}
\newenvironment{lemmarecap}[1]
	{\intheoremtrue\normalfont\rmfamily
	\trivlist\pagebreak[3]\item[\hskip \labelsep{\normalfont\itshape Lemma\ #1}]%
	\item[]%
	}
	{\endtrivlist\intheoremfalse}

\newenvironment{program}{\[\begin{array}{r@{\:\gets\:}l}}{\end{array}\]}
\newenvironment{program2c}{\[\begin{array}{r@{\:\gets\:}lr@{\:\gets\:}l}}{\end{array}\]}
\newenvironment{program3c}{\[\begin{array}{r@{\:\gets\:}lr@{\:\gets\:}lr@{\:\gets\:}l}}{\end{array}\]}
\newcommand{\srule}[2]{\ensuremath{#1 & #2}}        
\newcommand{\ssrule}[2]{\ensuremath{\mathit{#1} & \gets & \mathit{#2}}}        
\newcommand{\crrule}[2]{\ensuremath{\mathit{#1} & \gets_{cr} & \mathit{#2}}}        
\newcommand{\acrrule}[2]{\ensuremath{\mathit{#1} \gets_{cr} \mathit{#2}}}        

\newcommand{\prule}[2]{\ensuremath{\mathit{#1}\gets \mathit{#2}}}
\newcommand{\NAF}{\ensuremath{\textit{not}}}
\newcommand{\Naf}[1]{\ensuremath{\textit{not}\:#1}}
\newcommand{\naf}[1]{\ensuremath{\textit{not}\:#1}}
\newcommand{\anot}[1]{\ensuremath{\textit{not}_{#1}}}
\newcommand{\lit}[1]{\ensuremath{\hat{#1}}}
\newcommand{\olp}[2]{\ensuremath{\langle#1,#2\rangle}}
\newcommand{\lrule}[3]{\mbox{\ensuremath{#1\!:#2\gets #3}}}

\newcommand{\rleq}{\ensuremath{\sqsubseteq}}
\newcommand{\rlt}{\ensuremath{\sqsubset}}

\newcommand{\HEAD}[1]{\ensuremath{H_{#1}}}		
\newcommand{\BODY}[1]{\ensuremath{B_{#1}}}		

\submitted{November 27, 2003}
\revised{June 6, 2004}
\accepted{July 19, 2004}

\begin{document}

\bibliographystyle{acmtrans}

\title{Preferred Answer Sets for Ordered Logic Programs}

\author[Davy Van Nieuwenborgh and Dirk Vermeir]{DAVY VAN NIEUWENBORGH\thanks{Supported by the FWO} and
	DIRK VERMEIR\thanks{This work was partially funded by the 
	Information Society Technologies programme of the European Commission, 
	Future and Emerging Technologies under the IST-2001-37004 WASP project}\\
	Vrije Universiteit Brussel\\
	Dept. of Computer Science \\
	Pleinlaan 2, B-1050 Brussel\\
	Belgium\\
	\email{dvnieuwe@vub.ac.be}\\
	\email{dvermeir@vub.ac.be}}

\maketitle

\begin{abstract}
We extend answer set semantics to deal with inconsistent programs
(containing classical negation), by finding a ``best'' answer set.
Within the context of inconsistent programs, it is natural to
have a partial order on rules, representing a preference for
satisfying certain rules, possibly at the cost of violating less
important ones.
We show that such a rule order
induces a natural order on extended answer sets, the minimal
elements of which we call preferred answer sets.
We characterize the expressiveness of the resulting semantics and show
that it can simulate negation as failure, disjunction and some
other formalisms such as logic programs with ordered disjunction.
The approach is shown to be useful in several application areas,
e.g. repairing database, where minimal repairs correspond to 
preferred answer sets.
\par
To appear in Theory and Practice of Logic Programming (TPLP).
\end{abstract}

\begin{keywords}
nonmonotonic reasoning, knowledge representation, answer set programming, preference
\end{keywords}

\section{Introduction}
\par
The intuition behind the stable model semantics~\cite{gelfond88}, and, more generally,
behind answer set semantics~\cite{gelfond91} for (extended) logic programs is both
intuitive and elegant. Given a program $P$ and a candidate answer set
$M$, one computes a reduct program $P_M$ of a simpler type for
which a semantics $P^\star_M$ is known.
The reduct $P_M$ is obtained from $P$ by taking into
account the consequences of accepting the proposed truth values of the
literals in $M$.
The candidate set $M$ is then
an answer set just when $P_M^\star = M$, i.e. $M$ is
``self-supporting''.
\par
In this paper, we apply this reduction technique to deal with
inconsistent programs, e.g. programs with (only) classical negation (denoted
as $\neg$) where the immediate consequence operator would yield inconsistent
interpretations. For example, computing the least fixpoint of the
program $\set{\prule{a}{},\: \prule{b},\: \prule{\neg a}{b}}$, where
negative literals $\neg a$ are considered as fresh atoms, yields
the inconsistent \set{a, b, \neg a}. To prevent this, we will allow
for a rule to be defeated by an opposing rule w.r.t. an
interpretation. In the example, \set{a,b} will be accepted because
the rule \prule{\neg a}{b} is defeated by the rule \prule{a}{}.
The definition of answer set remains the same (see, e.g., 
\cite{lif2000}), but the reduct
is restricted to rules that are not defeated. We show that the
\emph{extended answer set semantics} thus obtained can be simulated by an extended logic
program $E(P)$ that is trivially constructed from the original program $P$.
\par
The above technique can be generalized to \emph{ordered programs} where
a partial order, representing preference or specificity, is defined on
the rules of a program.
E.g. one may prefer certain ``constraint'' rules to be satisfied,
possibly at the expense of defeating less important ``optional'' or
``default'' rules. 
We show that such a preference structure on the rules induces
a natural partial order on the reducts of the program, and hence on
its candidate (extended) answer sets.
Minimal elements in this induced partial order are called 
\emph{preferred answer sets}.
\par
Intuitively, an answer set $M_1$ is preferred over an answer set $M_2$
if any rule $r_2$ that is satisfied by $M_2$ but not by $M_1$,
is ``countered'' by a more preferred (than $r_2$) rule $r_1$ that
is satisfied by $M_1$ but not by $M_2$. In other words, with preferred
answer sets, one tries to maximize rule satisfaction, taking into
account the relative ``priority'' of  the rules.
The approach has some immediate applications in e.g. diagnostic
systems, as illustrated in the example below.
\begin{example}\label{ex:light}
Consider the problem of diagnosing a simple system where 
the light fails to come on. The normal operation of the system is
described using the rules $r_1$, $r_2$ and $r_3$. 
\[
\begin{array}{lrll}
r_1 & \ssrule{\mathit{light}}{\mathit{power}, \mathit{bulb}} \\
r_2 & \ssrule{\mathit{power}}{} \\
r_3 & \ssrule{\mathit{bulb}}{} \\
\end{array}
\]
which, by themselves, yield $\mathit{light}$. 
The fault model
is given by the following rules ($r_4$ and $r_5$), which indicate that the power
may fail and the bulb may be broken.
\[
\begin{array}{lrll}
r_4 & \ssrule{\neg \mathit{power}}{} \\
r_5 & \ssrule{\neg \mathit{bulb}}{} \\
\end{array}
\]
Finally, the observation that something is wrong is encoded by the
constraint-like rule $r_6$, i.e. a rule that can only be satisfied when
there is no light.
\[
\begin{array}{lrll}
r_6 & \ssrule{\neg\mathit{light}}{\mathit{light}} \\
\end{array}
\]
Obviously, the program \set{r_1, r_2, r_3 , r_4 , r_5 , r_6} is
inconsistent. On the other hand, it is natural to structure the rules
in a preference hierarchy where $r_1$ (the ``law'' governing the
system) and $r_6$ (the observation) are most preferred. Slightly
less preferred (than $r_1$ and $r_6$) are the assumptions $r_2$ and $r_3$
representing normal system operation. Finally, the fault rules $r_4$
and $r_5$ are least preferred, indicating that, if the program is
inconsistent, such rules 
will be the first to be considered for defeat.
\par
Without the observation $r_6$, the program is still inconsistent and,
following the preference relation on the rules, 
$M_1 = \set{\mathit{bulb}, \mathit{power}, \mathit{light}}$ will
be a preferred answer set satisfying all but the least preferred 
($r_4$ and $r_5$) rules. If we take into account the observation $r_6$, 
$M_1$, which does not satisfy $r_6$, will turn out to be less preferred
than either of
$M_2 = \set{\mathit{\neg bulb}, \mathit{power}}$ 
or
$M_3 = \set{\mathit{bulb}, \mathit{\neg power}}$.
E.g. $M_2$ is preferred over $M_1$ because, unlike $M_1$, $M_2$ satisfies
$r_6$ which counters the non-satisfaction of $r_3$ by $M_2$.
It can be verified that both $M_2$ and $M_3$ are preferred answer 
sets, with each corresponding to a minimal explanation of the observation.
\end{example}
\par
Besides diagnostic systems, ordered logic programs may be useful
in other application areas.
E.g. we show that the minimal repairs of a database $D$~\cite{arenas2000} w.r.t. a set of constraints $C$
correspond with the preferred answer sets of an ordered program
where the constraints $C$ are preferred over
$D$, which is itself preferred over $\neg D$, the latter representing
the negation of the facts in $D$.
\par
Although simple, ordered programs turn out to be rather
expressive under the preferred answer set semantics.
E.g., it is possible to simulate both negation as failure
and disjunction in classical non-ordered programs.
\par
Negation as failure has a long history, starting from the Clark
completion~\cite{clark77}, over stable model semantics~\cite{gelfond88} and well-founded
semantics~\cite{gelder88}, to answer set programming~\cite{gelfond91,lif2000}. It
is well-known that adding negation as failure to programs results
in a more expressive formalism. However, in the context of
disjunctive logic programming~\cite{przymusinski91,leone97}, \cite{inoue98} demonstrated that
adding negation as failure positively in a program, i.e. in the head
of the rules, yields no extra computational power to the formalism.
One of the more interesting features of negation as failure in the head 
is that answers no longer have to be minimal w.r.t. subset inclusion
(e.g. the program \set{\prule{a\lor\naf{a}}{}} has both $\set{a}$ and
$\emptyset$ as answer sets).
Indeed, such minimality turns out to be too demanding to express
certain problems, e.g. in the areas of abductive logic
programming~\cite{kakas92,inoue96} or logic programming with ordered
disjunction~\cite{brewka2002a,brewka2002b}.
\par
In light of the above, it is natural to consider \emph{extended
ordered programs} where
negation as failure is allowed in both the head and the body of a clause.
Just as for disjunctive logic programs, adding negation as 
failure positively results in a formalism where answer sets 
are not anymore guaranteed to be subset minimal.
Nevertheless, we will present a construction that translates
an extended ordered program into a semantically equivalent ordered program without
negation as failure, thus demonstrating that negation as failure does
not increase the expressiveness of ordered programs.
\par
Although extended ordered programs do not improve on ordered programs
w.r.t. computational power, they can be used
profitably to express certain problems in a more natural way.
They also support the simulation of certain extensions of answer set
programming, which we
illustrate by two intuitive transformations that translate
more complex concepts, i.e. ordered disjunction
\cite{brewka2002a,brewka2002b} and
consistency-restoring rules \cite{balduccini2003a,balduccini2003b},
into equivalent extended ordered programs. This demonstrates that
ordered programs can be used successfully as an implementation vehicle
for such high level extensions of answer set programming, where
the translation is processed by an ordered 
logic program solver such as \textsc{olps} (Section~\ref{compute}).
\par
The remainder of this paper is organized as follows.
After some preliminary notions and notations,
Section~\ref{slp} presents
an extension of the usual answer set semantics to cover also inconsistent
simple programs (without disjunction or negation as failure).
\par
In
Section~\ref{olp-basic},we introduce ordered programs where rules are partially 
ordered according to preference. It is shown that the rule-order induces 
a partial order on extended answer sets. The minimal elements in the latter 
order are called preferred answer sets.
We characterize the expressiveness of the resulting semantics and show
that it can simulate negation as failure as well as disjunction.
Section~\ref{compute} proposes an algorithm to compute such preferred answer
sets and shows that the complexity is the same as for disjunctive logic programming.
In Section~\ref{olp-naf}, we show that adding negation as failure to ordered
programs does not yield any extra expressive power.
\par
The relation of
preferred answer set semantics with similar formalisms from
the literature is discussed in Section~\ref{relationships}.
We consider Brewka's preferred answer sets~\cite{brewka99} in 
Section~\ref{b-pref}, together with D- and W-preferred answer sets
\cite{delgrande2000,wang2000}. It turns out that these semantics are
not related to our framework as they yield, in general, different
preferred answer sets, that are sometimes less intuitive than
the ones resulting from the semantics in Section \ref{olp-basic}.
Section~\ref{lpod} shows that logic
programs with ordered disjunction~\cite{brewka2002a} have a natural
simulation using ordered programs with preferred answer sets.
In Section~\ref{crrules} we elaborate on the simulation of 
consistency-restoring rules \cite{balduccini2003a} using the preferred answer set semantics.
In Section~\ref{doldlp}, we compare
our semantics with $\mathcal{DOL}$~\cite{bucca98},
$\mathcal{DLP^<}$~\cite{bucca99}
and ordered logic~\cite{laenens90b}.
Section~\ref{db-repair} illustrates another application of the preferred
answer set semantics: the minimal repairs of a database $D$
w.r.t. a set of constraints $C$ can be obtained as the preferred answer
sets of an ordered program $P(C,D)$.
\par
In Section~\ref{conclusions} we conclude and give
some directions for further research.
\par
To increase readability, several proofs and lemmas have
been moved to the appendix.
%
%
\section{Extended Answer Sets for Simple Programs}\label{slp}
\subsection{Preliminaries and Notation}\label{prelim}
\par
We use the following basic definitions and notation.
\subsection*{Literals}
\par
A \textit{literal} is an \textit{atom} $a$ or a negated atom $\neg a$.
An \textit{extended literal} is a literal or of the form $\naf{l}$ where $l$
is a literal. The latter form is called a \textit{naf-literal} and denotes negation as failure:
\naf{l} is interpreted as ``$l$ is not true''.
We use $\lit{l}$ to denote the ordinary literal underlying
an extended literal, i.e. $\lit{l} = a$ if $l=\naf{a}$
while $\lit{a} = a$ if $a$ is an ordinary literal. Both notations
are extended to sets so 
$\lit{X} = \set{\lit{e} \mid e\in X}$, with $X$ a set of
extended literals, while
$\naf{Y} = \set{\naf{l} \mid l\in Y}$ for any set of (ordinary)
literals $Y$.
\par
For a set of (ordinary)
literals $X$ we use $\neg X$ to denote \set{\neg p \mid p\in X}
where $\neg(\neg a)\equiv a$. 
Also, $X^+$ denotes the positive part of $X$, i.e.
$X^+ = \set{a \in X \mid a \mbox{ is an atom}}$.
The \textit{Herbrand base} of $X$, denoted
\hbase{X}, contains all atoms appearing in $X$, i.e.
$\hbase{X} = (X\cup \neg X)^+$.
A set $I$ of literals is \textit{consistent}
if $I\cap\neg I=\emptyset$.
\par
For a set of extended literals $X$, we use $X^-$ to denote
the literals underlying elements of $X$ that are not ordinary literals, i.e.
$X^- = \set{l \mid \naf{l}\in X}$. We say that $X$ is
consistent iff the set of ordinary literals
$\neg{X^{-}}\cup(\setmin{X}\naf{X^{-}})$ is consistent.
\subsection*{Programs}
\par
An \textit{extended disjunctive logic program} (EDLP,
see e.g. \cite{lif2000}) 
is a countable set of rules of the
form \prule{\alpha}{\beta} where $\alpha\cup\beta$ is a finite
set of extended literals. 
In a \textit{disjunctive logic program} (DLP), the
head $\alpha$ of each rule \prule{\alpha}{\beta} must contain
only ordinary literals.
\par
If always $|\alpha|\le 1$, i.e. $\alpha$ is a singleton or empty, 
we drop the ``disjunctive'' qualification.
If, for all rules,
all literals in $\alpha\cup\lit{\beta}$ are atoms,
the program is called \textit{seminegative} and if, furthermore, each
rule satisfies
$\beta^- =\emptyset$, the program is said to be \textit{positive}. 
\subsection*{Answer sets}
\par
The \textit{Herbrand base}
\hbase{P} of and EDLP $P$ contains all atoms appearing in $P$. An
\textit{interpretation} $I$ of $P$ is any consistent subset of 
$\hbase{P}\cup\neg\hbase{P}$ (for seminegative programs, we can restrict to a set of
atoms). 
An interpretation $I$ is \textit{total}
if $\hbase{P}\subseteq I\cup\neg I$.
\par
An extended literal $l$ is true w.r.t. an interpretation $I$,
denoted $I\models l$ if $l\in I$ in case $l$ is ordinary, or
$I\not\models a$ if $l = \naf{a}$ for some ordinary literal $a$.
As usual, $I\models X$ for some set of (extended) literals
$l$ iff $\Forall{l\in X}{I\models l}$.
\par
A rule $r = \prule{\alpha}{\beta}$ is \textit{satisfied} by $I$,
denoted $I\models r$,
if $I\models l$ for some $l\in\alpha$ and $\alpha\neq\emptyset$,
whenever $I\models\beta$, i.e.
if $r$ is \textit{applicable} ($I\models\beta$), then
it must be \textit{applied} ($\Exists{l\in\alpha}{I\models \beta\cup\set{l}}$).
As a consequence, a \textit{constraint}, i.e. a rule with empty head ($\alpha=\emptyset$),
can only be satisfied if it is not applicable ($I\not\models\beta$).
\par
For a DLP $P$ without negation as failure ($\beta^{-} =
\emptyset$), an \textit{answer set} is a minimal (w.r.t. set
inclusion) interpretation $I$ that is
\textit{closed} under the rules of $P$ (i.e. $\Forall{r\in P}{I\models r}$).
\par
For an EDLP $P$ containing negation
as failure and an interpretation $I$, the Gelfond-Lifschitz
transformation~\cite{gelfond88} yields the
\textit{GL-reduct} program $P^I$ that consists of those rules
\prule{(\setmin{\alpha}{\naf{\alpha^{-}}})}{(\setmin{\beta}{\naf{\beta^{-}}})} 
where $\prule{\alpha}{\beta}$
is in $P$,
$I\models \naf{\beta^{-}}$ and
$I\models \alpha^{-}$.
\par
Thus, $P^I$ is obtained from $P$ by 
(a)~removing all true naf-literals \naf{a}, $a\not\in I$, from the bodies
of rules in $P$,
(b)~removing all false naf-literals \naf{a}, $a\in I$ from the heads of
rules in $P$, and
(c)~keeping in $P^I$ only the transformed rules that are free from
negation as failure.
An interpretation $I$ is then an \textit{answer set} 
of $P$ iff $I$ is an answer set of the reduct $P^I$.
\subsection*{Reducts}
\par
In this paper, we use the term ``reduct'' of a program,
w.r.t. an interpretation, 
to denote the set of rules that are satisfied w.r.t.
the interpretation.
\begin{definition}\label{def:reduct}
Let $P$ be an EDLP program.
The \textbf{reduct} $P_I \subseteq P$ of $P$ w.r.t. an
interpretation $I$ 
contains just the rules satisfied by $I$, i.e.
$P_I = \set{r\in P \mid I\models r}$. 
\end{definition}
\par
Naturally, $P_M = P$ for any answer set $M$ of $P$.
\subsection{Simple Programs and Extended Answer Sets}\label{slp_eas}
\par
In this section, we consider simple logic programs which are logic
programs with only classical negation and no disjunction in the head
of a rule.
\begin{definition}\label{def:simple-program}
A \textbf{simple logic program} (SLP) is a countable set $P$ of \textbf{rules}
of the form \prule{\alpha}{\beta} where $\alpha\cup\beta$ is a finite set of
literals\footnote{
  As usual, we assume that programs have already been grounded.
  } and $|\alpha|\le 1$, i.e. $\alpha$ is a singleton or empty.
\par
A rule $r = \prule{a}{\beta}$ is \textbf{defeated} w.r.t. an
interpretation $I$ iff there exists
an applied (w.r.t. $I$) \textbf{competing rule} $\prule{\neg a}{\beta'}$
in P; such a rule is said to \textbf{defeat} $r$.
\end{definition}
\par
We will often confuse a singleton set with its sole element,
writing rules as \prule{a}{\beta} or \prule{}{\beta}.
Thus, a rule $r = \prule{a}{\beta}$ cannot be left unsatisfied unless
one accepts the opposite conclusion $\neg a$ which is motivated
by a competing applied rule \prule{\neg a}{\beta'} that 
\textit{defeats} $r$. Obviously, it follows that a constraint can 
never be defeated.
\begin{example}\label{ex0}
Consider the SLP $P$ containing the following rules.
\begin{program2c}
\srule{\neg a}{} & \srule{\neg b}{} \\
\srule{a}{\neg b} & \srule{b}{\neg a}
\end{program2c}
For the interpretation $I = \set{ \neg a, b}$ we have that
$I$ satisfies all rules in $P$ but one: 
\prule{\neg a}{} and \prule{b}{\neg a} are applied while
\prule{a}{\neg b} is not applicable. The unsatisfied
rule \prule{\neg b}{} is defeated by \prule{b}{\neg a}.
\end{example}
\par
For a set of rules $R$, we use $R^\star$ to denote the unique
minimal~\cite{emdkow76} model of the positive logic program consisting
of the rules in $R$ where (a) negative literals $\neg a$
are considered as fresh atoms and (b) constraint rules \prule{}{\beta}
are replaced by rules of the form \prule{\perp}{\beta}.
Besides the normal notion of inconsistency, a set of literals 
containing $\perp$ is also considered inconsistent.
Clearly, the $^\star$ operator is monotonic.
\par
For the program of Example~\ref{ex0}, we have that
$P^\star = \set{\neg a, \neg b, a, b}$ is inconsistent.
The following definition allows us to not apply certain rules,
when computing a consistent interpretation for programs such
as $P$.
\begin{definition}\label{def:slp-aset}
\par
An interpretation $I$ of a SLP $P$
is \textbf{founded} iff $P_I^\star = I$.
A founded interpretation $I$ is an \textbf{extended answer set}
of $P$ if all rules
in $P$ are satisfied or defeated.
\end{definition}
\par
The following is a straightforward consequence of the above definition and the
fact that a simple (reduct) program has at most one answer set.
\begin{theorem}\label{slp-aset-characterization}
An interpretation $M$ is an extended answer set of a SLP $P$
iff $M$ is the unique answer set (Section~\ref{prelim}) of $P_M$
and every rule in \setmin{P}{P_M} is defeated w.r.t $M$.
\end{theorem}
\par
Thus, the extended answer set semantics deals with inconsistency in a
simple yet intuitive way: when faced with contradictory applicable
rules, just select one for application and ignore (defeat) the other.
In the absence of extra information (e.g. regarding a preference
for satisfying certain rules at the expense of others), this 
seems a reasonable strategy for extracting a consistent semantics
from inconsistent programs.
\par
Using the above definition, it is easy to verify that the program
$P$ from Example~\ref{ex0} has three extended answer sets, namely
$M_1 = \set{\neg a,b}$, $M_2 = \set{a, \neg b}$ and
$M_3 = \set{\neg a, \neg b}$.
Note that $P_{M_1} = \setmin{P}{\set{\prule{\neg b}{}}}$ while
$P_{M_2} = \setmin{P}{\set{\prule{\neg a}{}}}$, and
$P_{M_3} = \setmin{P}{\set{\prule{a}{\neg b},\prule{b}{\neg a}}}$, i.e.
\prule{\neg b}{} is defeated w.r.t. $M_1$,
\prule{\neg a}{} is defeated w.r.t. $M_2$ and
both \prule{a}{\neg b} and \prule{b}{\neg a} are defeated
w.r.t. $M_3$.
\par
The definition of extended answer set is rather similar to the
definition of answer sets for (non-disjunctive) programs
without negation as failure:
the only non-technical difference
being that, for extended answer sets,
a rule may be left unsatisfied if it is defeated by
a competing (i.e.  a rule with opposite head) rule. 
This is confirmed by the following theorem.
\begin{theorem}\label{th-slp-aset}
Let $P$ be a SLP and let $M$
be an answer set of $P$.
Then, $M$ is the unique extended answer set of $P$.
\end{theorem}
\begin{proof}
By definition, $M$ is a minimal consistent interpretation that
satisfies all rules in $P$.
The latter implies that $P_M = P$. Because $M$ is minimal, it
follows that $M = P_M^\star$, making $M$ founded. Obviously,
$M$ must be unique.
\end{proof}
\par
While allowing for $P_M$, with $M$ an extended answer set, to be
a strict subset of $P$, Definition~\ref{def:slp-aset}
still maximizes the set of satisfied rules w.r.t. an extended answer set.
\begin{theorem}\label{slp-max-reduct}
Let $P$ be a SLP and let $M$ be an extended answer set
for $P$. Then, $P_M$ is maximal w.r.t. $\subseteq$ among the
reducts of founded interpretations of $P$.
\end{theorem}
\begin{proof}
Assume that, on the contrary, $P_M$ is not maximal, i.e. there exists a
founded interpretation $N$ such that $P_M \subset P_N$.
From the monotonicity of the $\star$-operator, it follows that $M\subseteq N$.
As $M$ is an extended answer set, all constraints are included in
$P_M$, thus, by $P_M\subset P_N$, also in $P_N$.
So, $\setmin{P_N}{P_M}$ does not contain any constraint.
Let $r = (\prule{a}{\beta})\in\setmin{P_N}{P_M}$.
Since $r$ is not satisfied w.r.t. $M$, it must be the case that
$\beta\subseteq M$ while $a\not\in M$.
Because $M$ is an extended answer set, $r$ must have been defeated by
an applied rule $r' = (\prule{\neg a}{\beta'})\in P_M\subset P_N$
and, consequently, $\neg a\in M\subseteq N$. On the other hand,
$\beta\subseteq N$ and thus, since $r\in P_N$, $r$ must be
applied w.r.t. $N$, yielding that $a\in N$. This makes $N$
inconsistent, a contradiction.
\end{proof}
\par
The reverse of Theorem~\ref{slp-max-reduct} does not hold in general,
as can be seen from the following example.
\begin{example}\label{ex4}
Consider the program $P$ containing the following rules.
\begin{program}
\srule{\neg a}{} \\
\srule{b}{} \\
\srule{\neg b}{\neg a}
\end{program}
The interpretation $N = \set{b}$ is founded with
$P_N = \set{ \prule{b}{},\: \prule{\neg b}{\neg a} }$ which is
obviously maximal since $P^\star$ is inconsistent.
Still, $N$ is not an extended answer set because
\prule{\neg a}{} is not defeated.
\end{example}
\par
However, when considering simple programs without constraints,
for total interpretations, founded interpretations with
maximal reducts are extended answer sets.
\begin{theorem}\label{slp-total-max-reduct-is-aset}
Let $P$ be a SLP without constraints and let $M$ be a total founded 
interpretation such that $P_M$ is maximal
among the reducts of founded interpretations of $P$.
Then, $M$ is an extended answer set.
\end{theorem}
\begin{proof}
It suffices to show that each unsatisfied rule is defeated w.r.t. $M$.
Assume that, on the contrary,
$r = (\prule{a}{\beta})\in\setmin{P}{P_M}$ is
not defeated, i.e. $a\not\in M$ while $\beta\subseteq M$
and there is no applied competitor 
\prule{\neg a}{\beta'}. But then also $\neg a\not\in M$,
contradicting the fact that $M$ is total.
\end{proof}
\par
The need for programs to be constraint free in the previous
theorem is demonstrated by the following example.
\begin{example}\label{ex3}
Consider the program $P$ containing the following rules.
\begin{program3c}
\srule{a}{} & \srule{\neg a}{} & \srule{}{a}
\end{program3c}
The total interpretation $N = \set{a}$ is founded with
$P_N = \set{ \prule{a}{} }$ which is
obviously maximal. However, $N$ is not an extended answer 
set as the constraint \prule{}{a} is neither satisfied
nor defeated w.r.t. $N$.
\end{example}
\par
The computation of extended answer sets reduces to the computation
of answer sets for seminegative non-disjunctive logic programs, using
the following transformation, which is
similar to the one used in \cite{kowalski90} for logic
programs with exceptions.
\par
\begin{definition}\label{slp-to-elp}
Let $P$ be a SLP. The \textbf{extended version} $E(P)$ of $P$ is the
(non-disjunctive) logic program obtained from
$P$ by replacing each rule \prule{a}{\beta} by its extended version
\prule{a}{\beta,\:\naf{\neg a}}.
\end{definition}
\par
Note that the above definition captures our intuition about defeat:
one can ignore an applicable rule $\prule{a}{\beta}$ if
it is defeated by evidence for the contrary $\neg a$, thus
making \naf{\neg a} false and the rule
\prule{a}{\beta, \naf{\neg a}} not applicable. 
\begin{theorem}\label{slp-vs-elp}
Let $P$ be a SLP. The extended answer sets of $P$ coincide with the
answer sets of $E(P)$.
\end{theorem}
\par
When considering programs without constraints, the extended answer set semantics is universal.

\begin{theorem}
Each simple logic program without constraints has extended answer sets.
\end{theorem}
\begin{proof}
Let $P$ be a simple logic program without constraints.
Define $\delta_P : 2^\hbase{P} \rightarrow 2^\hbase{P}$
by
\[
\delta_P(I) = 
\set{ a\not\in I \mid \neg a \not\in I \land 
  \Exists{(\prule{a}{\beta})\in P}{\beta\subseteq I}}
\]
Then, clearly, any sequence $I_0 = \emptyset, I_1 , \ldots$ where,
for $i\geq 0$, $I_{i+1} = I_i \cup \set{a}$ for some
$a\in\delta_P(I_i)$ if
$\delta_P(I_i)\neq\emptyset$, and $I_{i+1} = I_i$ otherwise,
is monotonically increasing and thus reaches a fixpoint $I^\star$
which is easily verified to be an extended answer set.
\par
Note that a similar result is well-known for normal default
logic\cite{reiter80}.
\end{proof}
%
%
%
\section{Ordered Programs and Preferred Answer Sets}\label{olp}
\subsection{Definitions and Basic Results}\label{olp-basic}
\par
When constructing extended answer sets
for simple logic programs, one can defeat any rule for which there
is an applied competing rule. In many cases, however,
there is a clear preference among rules in the sense that
one would rather defeat less preferred rules in order to
keep the more preferred ones satisfied. 
\par
As an example, reconsider the program $P$ from Example~\ref{ex0}
and assume that we prefer not to defeat the rules with positive
conclusion (\set{\prule{a}{\neg b},\:\prule{b}{\neg a}}).
Semantically, this should result in the rejection of
$M_3 = \set{\neg a, \neg b}$ in favor of either
$M_1 = \set{\neg a, b}$ or $M_2 = \set{a, \neg b}$ because
the latter two sets are consistent with our preferences.
\par
In ordered programs, such preferences are represented by a partial
order on the rules of the program.
\begin{definition}\label{def:olp}
An \textbf{ordered logic program} (OLP) is a pair $\olp{R}{<}$ 
where $R$ is a a simple program and 
$<$ is a well-founded strict\footnote{
  A strict partial order $<$ on a set $X$ is a binary relation on $X$ that
  is antisymmetric, anti-reflexive and transitive.
  The relation $<$ is well-founded if every nonempty subset of $X$ has
  a $<$-minimal element.
  } 
partial order on the rules in $R$\footnote{
  Strictly speaking, we should allow $R$ to be a multiset or,
  equivalently, have labeled rules, so that
  the same rule can appear in several positions in the order.
  For the sake of simplicity of notation, we will ignore this issue in
  the present paper: all results also hold for the general multiset
  case.
  }.
\end{definition}
\par
Intuitively, $r_1 < r_2$ indicates that $r_1$ is more preferred than
$r_2$.
In the examples we will often represent the order implicitly using
the format
\[
\begin{array}{rll}
& \ldots & \\
\hline
& R_2 & \\
\hline
& R_1 & \\
\hline
& R_0 & 
\end{array}
\]
where each $R_i$, $i\geq 0$, represents a set of rules,
indicating that all rules below a line are more preferred than any of
the rules above the line, i.e.
$\Forall{i\geq 0}{
 \Forall{r_i \in R_i , r_{i+1}\in R_{i+1}}{r_i < r_{i+1}}}$ or
$\Forall{i\geq 0}{R_i < R_{i+1}}$ for short.
\begin{example}\label{ex1}
Consider the OLP $P = \olp{R}{<}$ where $<$ is as shown below.
\begin{program}
\srule{f}{b} \\
\hline
\srule{\neg f}{p} \\
\hline
\srule{b}{p} \\
\srule{p}{} \\
\end{program}
The program uses the preference order to indicate that the rule
\prule{f}{b} (``birds fly'')  should be considered 
a ``default'', i.e. the rule
\prule{\neg f}{p} (``penguins don't fly'') is more preferred.
The lowest rules, i.e.
\prule{b}{p} (``penguins are birds'') and \prule{p}{} (``the bird
under consideration is a penguin''),
are the ``strongest'' (minimal): an extended answer set for
$P$ that respects the
preference order should satisfy these minimal rules, if at all possible.
\par
For the interpretations $I_1 = \set{p, b, f}$
and $I_2 = \set{p, b, \neg f}$,
the reducts are 
$R_{I_1} = \set{
\prule{f}{b},\:
\prule{b}{p},\:
\prule{p}{}
}$ 
and
$R_{I_2} = \set{
\prule{\neg f}{p},\:
\prule{b}{p},\:
\prule{p}{}
}$, respectively.
Both $I_1$ and $I_2$ are extended answer sets of $P$:
for $I_1$, the unsatisfied rule
\prule{\neg f}{p} is defeated by \prule{f}{b} while
for $I_2$, the reverse holds: \prule{f}{b} is defeated by 
\prule{\neg f}{p}.
\par
Intuitively, if we take the preference order $<$ into account,
$I_2$ is to be preferred over $I_1$ because
$I_2$ defeats less preferred rules than does $I_1$. Specifically,
$I_2$ compensates for defeating \prule{f}{b} by
satisfying the stronger \prule{\neg f}{p} which is itself defeated w.r.t.
$I_1$.
\end{example}
\par
The following definition formalizes the above intuition by
defining a preference relation between reducts.
\begin{definition}\label{def:reduct-order}
Let $P = \olp{R}{<}$ be an OLP.
For subsets $R_1$ and $R_2$ of $R$
we define \mbox{$R_1 \rleq R_2$} iff
$\Forall{r_2\in \setmin{R_2}{R_1}}{
  \Exists{r_1\in \setmin{R_1}{R_2}}{r_1 < r_2}
  }$.
We write $R_1 \rlt R_2$ just when $R_1 \rleq R_2$ and not $R_2 \rleq R_1$.
\end{definition}
\par
Intuitively, a reduct $R_1$ is preferred over a reduct $R_2$ if every
rule $r_2$ which is in $R_2$ but not in $R_1$ is ``countered''
by a stronger rule $r_1 < r_2$ from $R_1$ which is not in $R_2$.
\par
According to the above definition, we obtain that, indeed,
$R_{I_2} \rlt R_{I_1}$, for the program $P$ from
Example~\ref{ex1}.
\par
Note that, unlike other approaches, e.g. \cite{laenens92}, we do not
require that the stronger rule 
$r_1 \in \setmin{R_1}{R_2}$ that counters a weaker rule
$r_1 < r_2\in\setmin{R_2}{R_1}$, is applied and neither does
$r_1$ need to be a competitor of $r_2$. Thus, unlike the other 
approaches, we do not consider rule application as somehow 
"stronger" than satisfaction. This is illustrated in the following 
example.
\begin{example}\label{ex6}
Consider $P = \olp{R}{<}$, were $R$ is shown below and the
interpretations $M_1 = \set{\mathit{study},\mathit{pass}}$,
$M_2 = \set{\neg\mathit{study},\mathit{pass}}$,
$M_3 = \set{\neg\mathit{study},\neg\mathit{pass}}$,
and
$M_4 = \set{\mathit{study},\neg\mathit{pass}}$.
The program indicates a preference for not studying,
a strong desire to pass\footnote{
Note that, while the rule \prule{\mathit{pass}}{\neg\mathit{pass}}
can only be satisfied by an interpretation containing
\textit{pass}, it does not provide a justification
for \textit{pass}. Thus such rules act like constraints. 
However, note that, depending on where such a rule occurs, 
it may, unlike traditional constraints, be defeated.
} and an equally strong (and uncomfortable)
suspicion that not studying leads to failure.
\[
\begin{array}{lll}
r_4 &  : & \prule{\mathit{pass}}{\mathit{study}} \\
r_3 &  : & \prule{\mathit{study}}{} \\
\hline
r_2 &  : & \prule{\neg\mathit{study}}{} \\
\hline
r_1 &  : & \prule{\neg\mathit{pass}}{\neg\mathit{study}} \\
r_0 &  : & \prule{\mathit{pass}}{\neg\mathit{pass}} \\
\end{array}
\]
It is easily verified that $R_{M_1}\rlt R_{M_2}$, 
$R_{M_1}\rlt R_{M_3}$,
$R_{M_1}\rlt R_{M_4}$ (vacuously) and $R_{M_3}\rlt R_{M_4}$.
Here, e.g. 
$R_{M_1} = \set{r_0, r_1, r_3, r_4} \rlt R_{M_2} = \set{r_0 , r_2, r_4}$
because $r_2\in\setmin{R_{M_2}}{R_{M_1}}$ 
is countered by
$r_1\in\setmin{R_{M_1}}{R_{M_2}}$ which is neither applied nor
a competitor of $r_2$.
\end{example}
\par
The following theorem implies that the relation $\rleq$ 
is a partial order on reducts.
\begin{theorem}\label{thm1}
Let $<$ be a well-founded strict partial order
on a set $X$. The binary relation $\rleq$ on $2^X$  defined by
\mbox{$X_1 \rleq X_2$} iff
$\Forall{x_2\in \setmin{X_2}{X_1}}{
  \Exists{x_1\in \setmin{X_1}{X_2}}{x_1 < x_2}}$
is a partial order.
\end{theorem}
\par
Theorem~\ref{thm1} can be used to define a partial order
on extended answer sets of $R$,
where \olp{R}{<} is an ordered logic program.
\begin{definition}\label{def:aset-order}
Let $P = \olp{R}{<}$ be an OLP.
For $M_1 ,M_2$ extended answer sets of $R$, we define
$M_1 \rleq M_2$ iff $R_{M_1}\rleq R_{M_2}$.
As usual, $M_1 \rlt M_2$ iff $M_1 \rleq M_2$ and not $M_2 \rleq M_1$.
\end{definition}
\par
Preferred answer sets for ordered programs correspond to
minimal (according to $\rleq$) extended answer sets.
\begin{definition}\label{def:preferred-aset}
Let $P = \olp{R}{<}$ be an OLP.
An \textbf{answer set} for $P$ is any extended answer set of $R$.
An answer set for $P$ is called \textbf{preferred} if it
is minimal w.r.t. $\rleq$.
An answer set
is called \textbf{proper} if it satisfies all minimal (according to $<$) rules
in $R$.
\end{definition}
\par
Proper answer sets respect the strongest (minimal) rules of the
program.
\begin{lemma}\label{lemma-proper}
Let $M$ be a proper answer set of an OLP $P$. Then
any more preferred answer set $N\rlt M$ is also proper.
\end{lemma}
\begin{proof}
Assume that, on the contrary, $N\rlt M$ for some answer set $N$ which
is not proper. It follows that there is some minimal rule
$r\in\setmin{P_M}{P_N}$ which cannot be countered by $N$,
contradicting that $N\rlt M$.
\end{proof}
\par
The following theorem confirms that taking the minimal (according to $\rleq$)
elements among the proper answer sets is equivalent to selecting the proper
elements among the preferred answer sets.
\begin{theorem}
Let $P$ be an OLP. The set of minimal proper answer sets
of $P$ coincides with the set of proper preferred answer sets of $P$.
\end{theorem}
\begin{proof}
Let $M$ be a  minimal proper answer set and suppose that, on the
contrary, $M$ is not a proper preferred answer set. Since $M$ is
proper, this would imply that $M$ is not preferred, i.e.
$N\rlt M$ for some answer set $N$. From Lemma~\ref{lemma-proper}, we
obtain that $N$ must also be proper, contradicting that $M$ is a
minimal proper answer set.
\par
To show the reverse, let $M$ be a proper preferred answer set of $P$.
If $M$ were not a minimal proper answer set, there would exist
a proper answer set $N\rlt M$, contradicting that $M$ is preferred.
\end{proof}

\par
The program from Example~\ref{ex1} has a single preferred answer set
\set{p,b,\neg f} which is also proper. In Example~\ref{ex6}, 
$M_1 = \set{\mathit{pass}, \mathit{study}}$
is the only proper preferred answer set.
\par
While all the previous examples have a linear ordering, the semantics
also yields intuitively correct solutions in case of non-linear
orderings, as witnessed by the following example.
\par
\begin{example}\label{examplebalduccini}
Consider a problem taken from \cite{balduccini2003b}.
We need to take full-body exercise. Full-body exercise is achieved 
either by combining swimming and ball playing, or by combining weight
lifting and running. We prefer running to swimming and ball playing 
to weight lifting, but we do not like to do more than necessary
to achieve our full-body exercise. This last condition implies that we
cannot have a solution containing our two most preferred sports as in
that case we also need a third sport to have a full-body exercise.
The ordered program $P$ corresponding to this problem is shown below using a 
straightforward extension of the graphical representation 
defined before.
\par
\begin{center}
\begin{picture}(0,0)%
\includegraphics{example7.pstex}%
\end{picture}%
\setlength{\unitlength}{3552sp}%
\begingroup\makeatletter\ifx\SetFigFont\undefined%
\gdef\SetFigFont#1#2#3#4#5{%
  \reset@font\fontsize{#1}{#2pt}%
  \fontfamily{#3}\fontseries{#4}\fontshape{#5}%
  \selectfont}%
\fi\endgroup%
\begin{picture}(3792,2729)(438,-2037)
\put(615,562){\makebox(0,0)[lb]{\smash{{\SetFigFont{10}{12.0}{\rmdefault}{\mddefault}{\updefault}{\color[rgb]{0,0,0}\prule{lift\_weights}{}}%
}}}}
\put(798,260){\makebox(0,0)[lb]{\smash{{\SetFigFont{10}{12.0}{\rmdefault}{\mddefault}{\updefault}{\color[rgb]{0,0,0}\prule{play\_ball}{}}%
}}}}
\put(1099,-40){\makebox(0,0)[lb]{\smash{{\SetFigFont{10}{12.0}{\rmdefault}{\mddefault}{\updefault}{\color[rgb]{0,0,0}\prule{\neg full\_body\_exercise}{}}%
}}}}
\put(2630,262){\makebox(0,0)[lb]{\smash{{\SetFigFont{10}{12.0}{\rmdefault}{\mddefault}{\updefault}{\color[rgb]{0,0,0}\prule{swim}{}}%
}}}}
\put(2764,562){\makebox(0,0)[lb]{\smash{{\SetFigFont{10}{12.0}{\rmdefault}{\mddefault}{\updefault}{\color[rgb]{0,0,0}\prule{run}{}}%
}}}}
\put(2494,-936){\makebox(0,0)[lb]{\smash{{\SetFigFont{10}{12.0}{\rmdefault}{\mddefault}{\updefault}{\color[rgb]{0,0,0}\prule{\neg swim}{}}%
}}}}
\put(2644,-492){\makebox(0,0)[lb]{\smash{{\SetFigFont{10}{12.0}{\rmdefault}{\mddefault}{\updefault}{\color[rgb]{0,0,0}\prule{\neg run}{}}%
}}}}
\put(841,-485){\makebox(0,0)[lb]{\smash{{\SetFigFont{10}{12.0}{\rmdefault}{\mddefault}{\updefault}{\color[rgb]{0,0,0}\prule{\neg play\_ball}{}}%
}}}}
\put(660,-935){\makebox(0,0)[lb]{\smash{{\SetFigFont{10}{12.0}{\rmdefault}{\mddefault}{\updefault}{\color[rgb]{0,0,0}\prule{\neg lift\_weights}{}}%
}}}}
\put(745,-1383){\makebox(0,0)[lb]{\smash{{\SetFigFont{10}{12.0}{\rmdefault}{\mddefault}{\updefault}{\color[rgb]{0,0,0}\prule{full\_body\_exercise}{lift\_weights,run}}%
}}}}
\put(739,-1684){\makebox(0,0)[lb]{\smash{{\SetFigFont{10}{12.0}{\rmdefault}{\mddefault}{\updefault}{\color[rgb]{0,0,0}\prule{full\_body\_exercise}{play\_ball,swim}}%
}}}}
\put(729,-1989){\makebox(0,0)[lb]{\smash{{\SetFigFont{10}{12.0}{\rmdefault}{\mddefault}{\updefault}{\color[rgb]{0,0,0}\prule{full\_body\_exercise}{\neg full\_body\_exercise}}%
}}}}
\end{picture}%

\end{center}
\par
The rules in the least preferred component indicate a reluctance to
do any sport; they will be used only to satisfy more preferred rules.
On the other hand, the
rules in the most preferred component contain the conditions for 
a full body exercise, together with a constraint-like rule that
demands such an exercise. The rules in the middle components
represent our preferences for certain sports. Note that,
in order to minimize the sports we need to do,
preferences are expressed on the negated facts.
Consequently, e.g. a preference for running over swimming is
encoded as a preference for not swimming over not running.
\par
Consider the following extended answer sets:
\[
\begin{array}{rcl}
M_1 & = & \set{\mathit{full\_body\_exercise,lift\_weights,run,\neg swim,\neg play\_ball}}\enspace ,\\
M_2 & = & \set{\mathit{full\_body\_exercise,swim,play\_ball,\neg lift\_weights,\neg run}}\enspace ,\\
M_3 & = & \set{\mathit{full\_body\_exercise,lift\_weights,run,swim,\neg play\_ball}}\enspace .\\
\end{array}
\]
Clearly, all extended answer sets satisfy the three most specific rules.
Comparing the reducts $P_{M_1}$ and $P_{M_3}$ yields that
$\setmin{P_{M_3}}{P_{M_1}}=\set{\prule{swim}{}}$
and
$\setmin{P_{M_1}}{P_{M_3}}=\set{\prule{\neg swim}{}}$. 
From $\prule{\neg swim}{} < \prule{swim}{}$, it then
follows that $M_1$ is preferred over $M_3$, fitting our desire that we do not like to do more
than necessary.
\par
As for $M_1$ and $M_2$, it appears that
the rule $\prule{\neg play\_ball}{}\in\setmin{P_{M_1}}{P_{M_2}}$
is countered by the rule $\prule{\neg lift\_weights}{}\in\setmin{P_{M_2}}{P_{M_1}}$.
However, there is no rule in in $\setmin{P_{M_2}}{P_{M_1}}$ to counter 
$\prule{\neg swim}{}\in\setmin{P_{M_1}}{P_{M_2}}$, and thus
$M_2\not\rleq M_1$. 
On the other hand, $M_1$ cannot counter
$\prule{\neg lift\_weights}{}\in\setmin{P_{M_2}}{P_{M_1}}$, and thus
$M_1\not\rleq M_2$, making $M_1$ and $M_2$ incomparable.
It can be verified that both $M_1$ and $M_2$ are minimal w.r.t.
$\rleq$, making them preferred extended answer sets.
\end{example}
\par
\begin{example}\label{ex5}
Consider the ordered program \olp{P}{<} where $P$
is as in Example~\ref{ex0} and $<$ is as shown below.
\begin{program}
\srule{\neg a}{} \\
\srule{\neg b}{} \\
\hline
\srule{a}{\neg b} \\
\srule{b}{\neg a} \\
\end{program}
The reducts of the extended answer sets of $P$ are
$P_{M_1} = \setmin{P}{\set{\prule{\neg b}{}}}$,
$P_{M_2} = \setmin{P}{\set{\prule{\neg a}{}}}$, and
$P_{M_3} = \setmin{P}{\set{\prule{a}{\neg b},\prule{b}{\neg a}}}$
which are ordered by
$P_{M_1}\rlt P_{M_3}$ and
$P_{M_2}\rlt P_{M_3}$. Thus \olp{P}{<} has two
(proper) preferred answer sets: 
$M_1 = \set{\neg a,b}$ and $M_2 = \set{a, \neg b}$.
\end{example}
\par
Note that, in the above example, the preferred answer sets
correspond to the stable models (answer sets) of the logic program
\set{\prule{a}{\naf{b}}, \prule{b}{\naf{a}}}, i.e. the
stronger rules of \olp{P}{<} where negation as failure (\NAF)
replaces classical negation ($\neg$).
In fact, the ordering of $P$,
which makes the rules \prule{\neg a}{}
and \prule{\neg b}{} less preferred, causes 
$\neg$ to behave as negation as failure, under the
preferred answer set semantics.
\par
In general, we can easily simulate negation as failure
using classical negation and a trivial ordering.
\begin{theorem}\label{naf-olp}
Let $P$ be an (non-disjunctive) seminegative logic program
The ordered version of $P$, denoted $N(P)$ is defined by
$N(P) = \olp{P' \cup P_\neg}{<}$ with
$P_\neg = \{\prule{\neg a}{} \mid a\in\hbase{P}\}$
and $P'$ is obtained from $P$ by replacing each negated literal
$\naf{p}$ by $\neg p$. The order is defined by
$P' < P_\neg$, i.e. $\Forall{r\in P',r'\in P_\neg}{r<r'}$ (note that $P'\cap
P_\neg=\emptyset$). Then $M$ is a stable model of $P$ iff
$M\cup\neg(\setmin{\hbase{P}}{M})$ is a proper preferred answer set of
$N(P)$.
\end{theorem}
\par
Note that 2-level programs as above can also be used to support an
extension of simple programs with ``strict'' rules. Such a program has
the form  \olp{P_s \cup P_d}{<} where $P_s$ contains strict rules that
may not be defeated and $P_d$ contains ``default'' rules. The order
$<$ is defined by $P_s < P_d$, i.e. $r_s < r_d$ for all $r_s \in P_s$,
$r_d \in P_d$. The proper preferred answer sets then provide an
intuitive semantics for such programs.
\par
Interestingly, preference can also simulate disjunction.
\par
\begin{definition}
Let $P$ be a positive disjunctive logic program.
The ordered version of
$P$, denoted $D(P)$, is defined by 
$D(P) = \langle P_+ \cup P_- \cup P_p, < \rangle$ where
$P_+ = \set{\prule{a}{}\;\mid a\in\hbase{P}}$,
$P_- = \set{\prule{\neg a}{}\;\mid a\in\hbase{P}}$,
$P_p = \set{\prule{a}{\beta\cup\neg(\setmin{\alpha}{\set{a}})}\mid
  (\prule{\alpha}{\beta})\in P \land a\in\alpha }$, and
$P_p < P_- < P_+$.
\end{definition}
\par
Intuitively, the rules from $P_+\cup P_-$ guess a total interpretation
$I$ of $P$ while the rules in $P_p$ ensure that $I^+$ is a model of
$P$. Minimality is assured by the fact that negations are preferred.
\par
\begin{example}\label{ex7}
Consider the disjunctive program
$P = \set{\prule{a\lor b},\; \prule{a}{b},\;\prule{b}{a}}$.
This program illustrates that the shifted version\footnote{
 The shifted version of a disjunctive program is a seminegative
 program where each disjunctive rule \prule{\alpha}{\beta} is
 replaced by the set of rules containing
 \prule{a}{\beta\cup\naf{(\setmin{\alpha}{\set{a}})}} for
 each $a\in\alpha$.
 }
of a disjunctive program need not have the same models, see
e.g.~\cite{dix96,mdv98a}.
The program $D(P)$ is represented below.
\begin{program2c}
\srule{a}{} & \srule{b}{} \\
\hline
\srule{\neg a}{} & \srule{\neg b}{} \\
\hline
\srule{b}{\neg a} & \srule{a}{\neg b} \\
\srule{a}{b} & \srule{b}{a} \\
\end{program2c}
$D(P)$ has a single proper preferred answer set \set{a,b} which is
also the unique minimal model of $P$, while the shifted version
yields no models at all. Note that both
\prule{\neg a}{} and \prule{\neg b}{} are defeated because
minimization is overridden by satisfaction of more preferred
non-disjunctive rules.
\end{example}
\begin{theorem}\label{disj-olp}
Let $P$ be a positive disjunctive logic program. $M$ is a minimal model
of $P$ iff $M'=M\cup\neg(\hbase{P}\setminus M)$ is a
proper preferred answer set of $D(P)$.
\end{theorem}
\par
In view of Theorem~\ref{naf-olp} and Theorem~\ref{disj-olp}, it
is natural to
try to simulate programs that combine negation as failure
and disjunction.
\par
\begin{definition}
Let $P$ be a seminegative disjunctive logic program.
The ordered version of
$P$, denoted $D_n(P)$, is defined by 
$D_n(P) = \langle P_c \cup P_- \cup P_p, < \rangle$ where
$P_c = \set{\prule{a}{\beta'}\;\mid (\prule{\alpha}{\beta})\in P \land
  a\in\alpha}$,
$P_- = \set{\prule{\neg a}{}\;\mid a\in\hbase{P}}$,
$P_p = \{\prule{a}{\beta'\cup\neg(\setmin{\alpha}{\set{a}})}\mid
  (\prule{\alpha}{\beta})\in P \land a\in\alpha \}$, and
$P_p < P_- < P_c$. Here, $\beta'$ is obtained from $\beta$ by
replacing all occurrences of $\naf{a}\in\beta$ by
$\neg a\in\beta'$.
\end{definition}
\par
Intuitively, the rules in $P_c$ apply disjunctive rules by choosing
a literal from the head of the original rule; rules in $P_p$
ensure that any proper answer set is a model and the preference 
$P_- < P_c$ supports minimization.
\begin{theorem}\label{dlp-olp}
Let $P$ be a seminegative disjunctive logic program. If $M$
is an answer set of $P$ then $M\cup\neg{(\setmin{\hbase{P}}{M})}$
is a proper preferred answer set of $D_n(P)$.
\end{theorem}
\begin{proof}
The theorem immediately follows from Theorem~\ref{dnp-pm}
and Proposition~3.3 in \cite{sakama94}.
\end{proof}
\par
Unfortunately, $D_n(P)$ may have too many proper preferred answer sets,
as illustrated by the following example.
\begin{example}\label{ex8}
Consider the seminegative disjunctive program
$P = \{\prule{a\lor b}{}, \mbox{\prule{b}{a}},\linebreak[0]
\prule{a}{\naf{a}}\}$.
This program does not have an answer set. Indeed, any answer set
$M$ would need to contain $a$ and thus, by the rule \prule{b}{a}, also
$b$, thus $M=\set{a,b}$. But the reduct $P^M = 
\set{ \prule{a\lor b}{},\; \prule{b}{a} }$ has only one minimal
answer set $\set{b}\neq M$.
\par
However, \set{a,b} is the unique minimal preferred answer set of
$D_n(P)$ which is shown below.
\begin{program2c}
\srule{a}{} & \srule{b}{} \\
\hline
\srule{\neg a}{} & \srule{\neg b}{} \\
\hline
\srule{b}{a} & \srule{a}{\neg a} \\
\srule{a}{\neg b} & \srule{b}{\neg a} \\
\end{program2c}
\end{example}
\par
In fact, the preferred answer sets semantics of $D_n(P)$ corresponds
to the possible model semantics of \cite{sakama94}.

\begin{definition}
For a seminegative disjunctive logic program $P$, we define a \textbf{split program}
as the (non-disjunctive) seminegative program obtained from $P$ by
replacing each rule $\prule{\alpha}{\beta}\in P$ with 
$\prule{a}{\beta}$ for every $a\in S$, where $S$ is some non-empty
subset of $\alpha$.
Now, a \textbf{possible model} of $P$ is any answer set of any split
program of $P$.
\end{definition}

\begin{theorem}\label{dnp-pm}
Let $P$ be a seminegative disjunctive logic program. An interpretation $M$ is a
proper preferred answer set of $D_n(P)$ iff $M^+$ is a minimal
possible model of $P$.
\end{theorem}
\par
In the next subsection, we'll see that, nevertheless, the expressiveness
of the preferred answer set semantics of OLP is similar to that
of seminegative disjunctive programs.

\subsection{Computing Preferred Answer Sets}\label{compute}
\par
In this subsection, we only consider finite programs (corresponding
to datalog-like rules).
\begin{definition}
Let \olp{P}{<} be a partially ordered set (where $<$ is strict).
The \textbf{downward closure} of a subset $X\subseteq P$ is
defined by $\down{X} = \set{ u \in P\mid \Exists{x\in X}{u < x} }$.
A set $X\subseteq P$ is \textbf{downward closed} iff
$\down{X}\subseteq X$.
\end{definition}
\par
A \textit{specification} for an ordered program poses restrictions on
the sets of rules that should be satisfied, respectively defeated,
by a conforming extended answer set.
\begin{definition}\label{def-specification}
Let \olp{P}{<} be an ordered program. A \textbf{specification}
for $P$ is pair $\pair{R_i}{R_o}$
of disjoint subsets of $P$ such that $R_i \cup R_o$ is
downward closed.
\par
A specification $\pair{R'_i}{R'_o}$ \textbf{extends} another
specification $\pair{R_i}{R_o}$,
denoted
$\pair{R_i}{R_o} \preceq \pair{R'_i}{R'_o}$,
iff $R_i \subseteq R'_i$ and $R_o\subseteq R'_o$.
\par
A set of rules $R\subseteq P$
\textbf{satisfies} a specification $\pair{R_i}{R_o}$,
denoted 
$R\models \pair{R_i}{R_o}$, iff
$R^\star$ is an extended answer set for $P$,
$R_i \subseteq R$,
$R_o \cap R = \emptyset$ and, moreover,
$\Forall{r\in R_o}{R^* \not\models r}$.
\end{definition}
\par
Obviously, if $R$ satisfies $\pair{R_i}{R_o}$ then
$R$ satisfies any weaker specification
$\pair{R'_i}{R'_o} \preceq \pair{R_i}{R_o}$.
\par
In the remainder of this section, we will use
the term ``extended answer set'' for both the interpretation $I$
and the corresponding set of rules $\set{r\in P\mid I\models r}$
that it satisfies.
\par
To force a conforming extended answer set
to satisfy at least one out of a collection of sets of rules,
we define a constraint.
Such constraints will be used to ensure that conforming extended answer
sets are not smaller (w.r.t. $\rlt$) than others.
\par
\begin{definition}\label{def-constraint}
Let \olp{P}{<} be an ordered program.
A \textbf{constraint} is a set of sets of rules $C\subseteq 2^P$.
A specification $\pair{R_i}{R_o}$
is consistent with a constraint iff
$\Exists{c\in C}{R_o \cap c = \emptyset}$.
A rule set $R$ satisfies a constraint $C$,
denoted $R\models C$, iff
$\Exists{c\in C}{c\subseteq R}$.
\par
The \textbf{expansion} of a specification and
a constraint, denoted
$\expansion{\pair{R_i}{R_o}}{C}$,
is defined by
\[
\expansion{\pair{R_i}{R_o}}{C} =
 \set{ R\subseteq P \mid R\models \pair{R_i}{R_o} \land R\models C  }\enspace .
\]
We use $\min\expansion{\pair{R_i}{R_o}}{C}$ to denote
the minimal elements of $\expansion{\pair{R_i}{R_o}}{C}$
w.r.t. the \rlt-order.
\end{definition}
\par
By definition, $\min\expansion{\pair{R_i}{R_o}}{C}$
contains minimal (according to $\rlt$) answer sets
that satisfy both the specification $\pair{R_i}{R_o}$
and the constraint $C$.
\begin{definition}\label{witness}
Let \olp{P}{<} be an ordered program and $R\subseteq P$ 
and $T\subseteq P$ be sets of rules. A
\textbf{witness} of $R$ against $T$ is any rule
$r\in\setmin{R}{T}$ such that
$\Forall{t\in\setmin{T}{R}}{t\not< r}$.
\par
We use
\witness{T} to denote the set
\set{ \set{r}\cup (\down{\set{r}}\cap T) \mid r\in P\setminus T}.
\end{definition}
\par
In Lemma~\ref{lemma-witness} from the Appendix, it is shown that
$T\not\rlt R$ iff $R$ has a witness against $T$,
which is itself equivalent to
$\Exists{X\in\witness{T}}{X\subseteq R}$.
\par
The basic algorithm to compute preferred answer sets
is shown in Figure~\ref{basic-algo}.
Intuitively, $\aset{\pair{R_i}{R_o}}{C}$ returns the
minimal elements among the extended answer sets from $P$
that satisfy both the specification
$\pair{R_i}{R_o}$ and the constraint $C$.
\par
This is achieved as follows:
\begin{itemize}
\item If $\pair{R_i}{R_o}$ is inconsistent with
  $C$ then, obviously, there are no elements
  satisfying both.
\item If $R_i\cup R_o = P$, $R_i$
  should be returned, if it is an extended answer set.
\item Otherwise, we first compute the set $M$ of minimal extended answer sets
  containing a minimal rule $r$ from $\setmin{P}{(R_i\cup R_o)}$, using the call
  $\aset{\pair{R_i\cup\set{r}}{R_o}}{C}$.
\item Next, we compute the minimal extended answer sets
  not containing $r$. Such answer sets must contain a witness
  against each $m\in M$. This is ensured by appropriately extending the
  constraint $C$ to $C'$. The missing solutions are then computed
  using $\aset{\pair{R_i}{R_o\cup\set{r}}}{C'}$.
\end{itemize}
\par
Formally, we will have that
$\aset{\pair{R_i}{R_o}}{C} = 
\min\expansion{\pair{R_i}{R_o}}{C}$ from which,
since the preferred answer sets of $P$ obviously
correspond to
$\min\expansion{\pair{\emptyset}{\emptyset}}{\set{\emptyset}}$,
it follows that
$\aset{\pair{\emptyset}{\emptyset}}{\set{\emptyset}}$ returns
exactly the preferred answer sets of $P$.

\begin{figure}[htb]
\begin{center}
\begin{minipage}{10cm}
\begin{lstlisting}{}
set<RuleSet>
aset($\pair{R_i}{R_o}$, C) {
// precondition: $C\neq\emptyset$

if ( $\pair{R_i}{R_o}$ and $C$ are inconsistent )
  return $\emptyset$

if ($(R_i \cup R_o) = P$ )
  if ( $R_i^\star$ is an extended answer set of $P$)
    return $\set{R_i}$
  else
    return $\emptyset$

choose $r$ minimal in $\setmin{P}{(R_i\cup R_o)}$

// compute preferred answer sets containing $r$
$M$ = $\aset{\pair{R_i\cup\set{r}}{R_o}}{C}$;

// add constraints that guarantee that each element in
// $\expansion{\pair{R_i}{R_o\cup\set{r}}}{C'}$ has a witness against any $m\in M$.
$C' = C$
for each $m \in M$
  $C'$  = $\set{ c\cup x \mid c\in C' ~\land~ x\in\witness{m} ~\land~ r\not\in x}$
if ($C'=\emptyset$) // $\Exists{m\in M}{\Forall{x\in\witness{m}}{r\in x}}$
  return $M$
// compute preferred answer sets not containing $r$
return $M \cup \aset{\pair{R_i}{R_o \cup\set{r}}}{C'}$;
}
\end{lstlisting}
\end{minipage}
\end{center}
\caption{Basic Algorithm\label{basic-algo}}
\end{figure}
\begin{theorem}\label{poas-eq-aset}
Let \olp{P}{<} be an ordered program,
$\pair{R_i}{R_o}$ be a specification and $C$ a constraint.
Then $\aset{\pair{R_i}{R_o}}{C} = 
\min\expansion{\pair{R_i}{R_o}}{C}$.
\end{theorem}
\par
Since a preferred answer set is minimal w.r.t.
the extended answer sets that satisfy the ``empty''
specification and constraint, we obtain the
following corollary.
\begin{corollary}\label{poas-eq-aset-empty}
Let \olp{P}{<} be an ordered program.
The preferred answer sets of $P$ are computed
by $\aset{\pair{\emptyset}{\emptyset}}{\set{\emptyset}}$.
\end{corollary}
\par
Clearly, the algorithm of Figure~\ref{basic-algo} can be further optimized, e.g. by
proactively computing certain conditions, such as the
consistency of $R_i$, etc.
\par
A first implementation of an ordered logic program solver
(\textsc{olps}) is available under the \textsc{gpl} at \url{http://tinf2.vub.ac.be/olp/}.
After grounding, 
\textsc{olps} computes (a selection of) the proper preferred answer sets of a
finite ordered program which is described using a sequence of
module definitions and order assertions. 
A module is specified using a module name followed by
a set of rules, enclosed in braces while
an order assertion is of the form
$m_0 < m_1 < \ldots < m_n$, $n>0$, where each $m_i$, $0\leq i\leq n$
is a module name. Such an assertion expresses that each rule in $m_i$,
$0\leq i\leq n$ is more preferred than any rule in $m_{i+1}$. 
Figure~\ref{light-olp} shows the
\textsc{olps} version of the the diagnostic problem from 
Example~\ref{ex:light}.
\par
\begin{figure}[htb]
\lstset{language=Prolog}
\lstset{frame=tb}
\begin{lstlisting}{}
FaultModel {
	-power.
	-bulb.
}
NormalOperation {
	power.
	bulb.
}
System {
	light :- power, bulb.
}
System < NormalOperation < FaultModel
Observations { -light :- light. }
\end{lstlisting}
\caption{The \textsc{olps} version of Example~\ref{ex:light}.}\label{light-olp}
\end{figure}
%
\par
The following results shed some light on the complexity of
the preferred answer set semantics.
\par
First we note that
checking whether $M$ is not a preferred answer set of an OLP $P$ is
in NP because
\begin{enumerate}
\item Checking that $M$ is an extended answer set of $P$, i.e.
verifying foundedness and verify that each non-satisfied rule is defeated,
can be done in deterministic polynomial time.
(E.g. foundedness can be verified using a marking algorithm that
repeatedly scans all rules in $P$, marking elements of $M$ that have a
``motivation'' based on already marked elements from $M$).
\item 
Guess a set $N\rleq M$, which can be done in polynomial time, and verify
that $N$ is an extended answer set.
\end{enumerate}
\par
Finding a preferred answer set $M$ can then be performed
by an NP algorithm that guesses $M$ and uses an NP oracle
to verify that it is not the case that $M$ is not a preferred
answer set. Hence the following theorem.
\begin{theorem}\label{inp2}
The problem of deciding, given an arbitrary ordered program $P$
and a literal $a$, whether $a$ occurs in any preferred answer set of
$P$ is in $\Sigma_2^P$.
\end{theorem}
\begin{theorem}\label{inpi2}
The problem of deciding, given an arbitrary ordered program $P$
and a literal $a$, whether $a$ occurs in every preferred answer set of
$P$ is in $\Pi_2^P$.
\end{theorem}
\begin{proof}
Finding a preferred answer set $M$ such that $a\not\in M$ is
in $\Sigma_2^P$ due to Theorem~\ref{inp2}. Thus, the complement is in
$\Pi_2^P$.
\end{proof}
\begin{theorem}\label{p2hard}
The problem of deciding, given an arbitrary ordered program $P$
and a literal $a$, whether $a$ occurs in any preferred answer set of
$P$ is $\Sigma_2^P$-hard.
\end{theorem}
\begin{proof}
The proof uses a reduction of the known $\Sigma_2^P$-hard problem of
deciding whether a quantified boolean formula 
$\phi = \Exists{x_1 ,\ldots, x_n}{\Forall{y_1, \ldots, y_m}{F}}$
is valid, where we may assume that $F = \lor_{c\in C}c$ with
each $c$ a conjunction of literals over $X\cup Y$ with 
$X = \set{x_1 ,\ldots, x_n}$ and $Y = \set{y_1 ,\ldots, y_m}$
($n,m>0$).
The construction is inspired by a similar result in
\cite{eitegottl93} for disjunctive logic programs.
\par
The program $P$ corresponding to $\phi$ is shown below
using a straightforward extension of the graphical
representation of Definition~\ref{def:olp}: the order in P is
defined by $P_4 < P_3 < P_2$ (note that the rules in $P_1$ are not
related to any other rules).
\[
\begin{array}{c|c}
P_1 = \set{\prule{x}{}\;\prule{\neg x}{}\mid x\in X} &
\begin{array}{c}
P_2 = \set{\prule{y}{}\;\prule{\neg y}{}\mid y\in Y}\\
\hline
P_3 = \prule{\neg sat}{sat}\\
\hline
P_4 = \set{\prule{sat}{c}\mid c\in C}
\end{array}
\end{array}
\]
Obviously, the construction of $P$ can be done in polynomial time.
Intuitively, the rules in $P_1$ and $P_2$ are used to guess a truth 
assignment for $X\cup Y$. 
\par
In the sequel, we will abuse notation by using $x_M$ and $y_M$
where $M$ is an answer set for $P$, to denote subsets of $M$,
e.g. $x_M = X\cap M$ and in expressions such as $F(x_M, y_M)$ 
which stands for $F(x_1 , \ldots,x_n,y_1,\ldots y_m)$
with $x_i = \mathbf{true}$ iff $x_i\in x_M$ and, similarly,
$y_j = \mathbf{true}$ iff $y_j\in y_M$. We will also sometimes
abbreviate the arguments of $F$, writing e.g. $F(x,y)$ rather than
$F(x_1 , \ldots,x_n,y_1,\ldots y_m)$. 
\par
The following properties of $P$ are straightforward to show:
\begin{enumerate}
\item\label{proofcomp_p1} If we have an extended answer set $M$ containing $sat\in M$,
then $F(x_M,y_M)$ must hold.
\item\label{proofcomp_p2} Any extended answer set satisfies all the
rules in $P_4$.
\item\label{proofcomp_p3} For extended answer sets $M_1$ and $M_2$, with
$M_1\cap X\not= M_2\cap X$, neither $M_1\rlt M_2$ nor $M_2\rlt M_1$ holds,
as the rules in $P_1$ are unrelated to any other rules.
\item\label{proofcomp_p4} If $M_1\rlt M_2$ for some extended answer
sets $M_1$ and $M_2$, then $M_1\cap X = M_2\cap X$ and, moreover,
$sat\in M_2\setminus M_1$, i.e. $(\prule{\neg sat}{sat})\in P_{M_1}\setminus P_{M_2}$.
\end{enumerate}
\par
We show that $\phi$ is valid iff $sat\in M$ for some preferred answer
set $M$ of $P$.
\par
To show the ''if'' part, assume that $M$ is a preferred answer set
with $sat\in M$. By (\ref{proofcomp_p1}) we have that $F(x_M,y_M)$ holds. To prove that
$\phi$ is valid it remains to show that $\Forall{y}{F(x_M,y)}$.
Suppose that, on the contrary, $\Exists{y}{\neg F(x_M,y)}$ and consider
the extended answer set $M' = (M\cap (X\cup\neg X))\cup y\cup \neg(Y\setminus y)$. 
It is easy to verify that $M'\rlt M$ since the rules in $P_1$ satisfied by $M$ are 
the same as the rules in $P_1$ satisfied by $M'$; and all the rules in 
$P_4$ are satisfied by both $M$ and $M'$ (due to (\ref{proofcomp_p2})); 
and the rule in $P_3$ is defeated by $M$ and satisfied by $M'$. 
But $M'\rlt M$ contradicts the fact that $M$ is a preferred answer set. 
Thus, $\phi$ is valid.
\par
To show the reverse, assume that $\phi$ is valid, i.e. there exists
some $x_M \subseteq X$ such that $\Forall{y}{F(x_M ,y)}$.
Consider $M = x_M \cup\neg(X\setminus x_M)\cup y\cup\neg(Y\setminus y)\cup \set{sat}$
where $y\subseteq Y$ is arbitrary. Clearly, $M$ is an extended answer
set. To show that $M$ is preferred, assume that, on the contrary,
$M'\rlt M$ for some extended answer set $M'$. By (\ref{proofcomp_p4}),
$M\cap X = M'\cap X$ and $sat\not\in M'$. These imply
that $\neg F(x_M , y_{M'})$, contradicting that 
$\Forall{y}{F(x_M ,y)}$.
\end{proof}
\begin{theorem}\label{pi2hard}
The problem of deciding, given an arbitrary ordered program $P$
and a literal $a$, whether $a$ occurs in every preferred answer set of
$P$ is $\Pi_2^P$-hard.
\end{theorem}
\begin{proof}
Reconsider the program $P$ in the proof of Theorem~\ref{p2hard}.
Let $a$ be a fresh atom not occurring in $P$ and define $P'$ as $P$
with two extra rules \prule{a}{} and \prule{\neg a}{} in the component $P_2$.
Clearly, showing that $a$ does not occur in every preferred answer set
is the same as showing that $\neg a$ occurs in any preferred answer
set of $P$. Deciding the latter is $\Sigma_2^P$-hard by Theorem~\ref{p2hard};
thus deciding the complement of the former is $\Pi_2^P$-hard.
\par
In the appendix an alternative proof is provided using quantified
boolean formulas.
\end{proof}
\par
The following is immediate from Theorem~\ref{inp2}, Theorem~\ref{inpi2},
and Theorem~\ref{p2hard}.
\begin{corollary}
The problem of deciding, given an arbitrary ordered program $P$
and a literal $a$, whether $a$ occurs in any proper preferred answer set of
$P$ is $\Sigma_2^P$-complete. 
The problem of deciding, given an arbitrary ordered program $P$
and a literal $a$, whether $a$ occurs in every proper preferred answer set of
$P$ is $\Pi_2^P$-complete. 
\end{corollary}
%
\subsection{Adding Negation as Failure}\label{olp-naf}
\par
In view of the results from Section~\ref{olp-basic} and
Section~\ref{compute}, it is natural to
wonder whether adding negation as failure to ordered programs
leads to a more expressive formalism. 
\par
To study this question, we first extend simple logic programs
to allow negation as failure in both the head and the body
of rules. 
The definition closely mirrors Definition~\ref{def:simple-program}: we only
generalize the notion of defeat to take into account the possible
presence of negation as failure in the head of a rule.
\begin{definition}\label{elp-def}
An \textbf{extended logic program} (ELP) is a countable set $P$ of 
\textbf{extended rules}
of the form \prule{\alpha}{\beta} where $\alpha\cup\beta$ is a finite set of
extended literals, and $|\alpha|\leq 1$, i.e. $\alpha$ is a
singleton or empty. 
\par
An extended rule $r = \prule{a}{\beta}$ is \textbf{defeated} w.r.t. $P$
and $I$ iff $P$ contains
an applied \textbf{competing rule} $r' = \prule{a'}{\beta'}$
such that \set{a,a'} is inconsistent.
\par
An interpretation $I$ is an \textbf{extended answer set} of $P$ iff
$I$ is an answer set (see Section~\ref{prelim}) of $P_I$ and
each unsatisfied rule from \setmin{P}{P_I} is defeated w.r.t. $I$.
\end{definition}
\par
\begin{example}\label{ex0b}
Consider the extended program $P$ containing the following rules.
\begin{program3c}
\srule{\neg a}{}   & \srule{\neg b}{}   & \srule{c}{} \\
\srule{a}{\Naf{b}} & \srule{b}{\Naf{a}} & \srule{\Naf{c}}{a} \\
\end{program3c}
For the interpretation $I = \set{a,\neg b}$,  $P_I$
contains all rules but \prule{\neg a}{} and \prule{c}{} which
are defeated (w.r.t. $I$) by the applied rules
\prule{a}{\Naf{b}} and \prule{\Naf{c}}{a}, respectively.
$I$ is then an extended answer set because \set{a, \neg b}
is an answer set of
$(P_I)^I = \set{\prule{\neg b}{},\; \prule{a}{}}$.
\par
$P$ has three more extended answer sets, namely
$J = \set{\neg a, b, c}$,  
$K = \set{\neg a, \neg b, c}$ and
$L = \set{a,\neg b, c}$.
Here, $P_J = \setmin{P}{\set{\prule{\neg b}{}}}$,
and $(P_J)^J$ contains
\prule{\neg a}{}, \prule{b}{}, \prule{c}{} and 
\prule{}{a}.
For $K$, we have that
$P_K = P\setminus\set{\prule{a}{\Naf{b}},\:\prule{b}{\Naf{a}}}$ and
$(P_K)^K$ contains \prule{\neg a}{}, \prule{\neg b}{}, \prule{c}{}
and \prule{}{a}.
Finally, $L$ yields that
$P_L = P\setminus\set{\prule{\neg a}{},\prule{\Naf{c}}{a}}$ and
$(P_L)^L$ contains \prule{a}{}, \prule{\neg b}{} and \prule{c}{}.
\end{example}
\par
Unlike for simple logic programs, extended answer sets for
extended logic programs are not necessary minimal w.r.t. subset
inclusion, as demonstrated by the previous example where $I\subset L$.
The same holds for extended disjunctive logic programs as shown in
\cite{inoue94}.
\par
Furthermore, the extended answer set semantics for extended logic 
programs is not universal, even for programs without constraints,
as witnessed by the following example.
\par
\begin{example}\label{ex9}
Consider the extended logic program $P$ containing the rules
\prule{a}{\Naf{b}} and
\prule{b}{a,\Naf{c}}.
Clearly, no extended answer set can contain $c$ or $\neg c$.
\par
For $I=\emptyset$ we obtain 
$(P_I)^I = \set{\prule{b}{a, \Naf{c}}}^I = \set{\prule{b}{a}}$,
which has a unique answer set $\emptyset = I$. 
However, \prule{a}{\Naf{b}} is neither satisfied nor defeated in $I$.
For $I=\set{a}$,
$(P_I)^I = \set{\prule{a}{\Naf{b}}}^I = \set{\prule{a}{}}$
which has a unique answer set $\set{a} = I$. However,
\prule{b}{a, \Naf{c}} is neither satisfied nor defeated in $I$.
For $I=\set{b}$,
$(P_I)^I = P^I = \set{\prule{b}{a}}$ which has
a unique answer set $\emptyset \neq I$.
Finally, for $I=\set{a}$ we obtain 
$(P_I)^I = P^I = \set{\prule{b}{a}}$
which has $\emptyset \neq I$ as a unique answer set.
\par
Thus, $P$ has no extended answer sets.
\end{example}
\par
Obviously, any traditional answer set of an ELP $P$ is
also an extended answer set. However, unlike for simple programs,
a consistent ELP, i.e. a program that has answer sets, may
also have additional extended answer sets, as in the
following example.
\begin{example}
Consider the following program $P$.
\begin{program2c}
\srule{\neg b}{a} & \srule{b}{\Naf{b}} \\
\srule{a}{\Naf{b}} & \srule{b}{\Naf{a}} \\
\end{program2c}
Clearly, $I = \set{b}$ is an answer set with $P^I$ containing
\prule{\neg b}{a} and \prule{b}{} which has \set{b} as a minimal
answer set. Since $P_I = P$, \set{b} is also an extended answer set.
\par
However, also $J = \set{a, \neg b}$ is an extended answer set because:
\begin{itemize}
\item
$P_J$ contains all rules but \prule{b}{\Naf{b}}. The
latter rule is defeated (w.r.t. $J$) by the
applied rule \prule{\neg b}{a}. 
\item
${(P_J)}^J$ contains just \prule{a}{} and \prule{\neg b}{a} and thus
$J$ is an answer set of $P_J$.
\end{itemize}
Clearly, though, $J$ is not an answer set of $P$.
\end{example}
\par
Note that the program in the above example does not contain
negation as failure in the head of a rule. 
In fact, negation as failure in the heads of rules can be
removed by a construction that is similar to the one
used in \cite{inoue98} for reducing DLP's with negation as failure in 
the head to DLP's without.
\begin{definition}\label{def:trans_eas_to_as}
For $P$ an ELP, define $E(P)$ as the
ELP, without negation as failure in the head, obtained from $P$ by
replacing each rule \prule{a}{\beta} by (for $a$ an ordinary literal,
$\anot{a}$ is a new atom) by
~\prule{a}{\beta,\,\naf{\neg a},\,\Naf{\anot{a}}}~~ when $a$ is an ordinary
literal; or by
~\prule{\anot{\lit{a}}}{\beta,\,\Naf{\lit{a}}}~~ when $a$ is a naf-literal.
\end{definition}
\par
Intuitively, one can ignore an applicable rule \prule{a}{\beta} if it is defeated
by evidence for either $\neg a$ or $\Naf{a}$, thus making either
$\Naf{\neg a}$ or $\Naf{\anot{a}}$ false and the rule 
\prule{a}{\beta,\Naf{\neg a},\Naf{\anot{a}}} not applicable.
\par
The extended answer sets of $P$ can then be retrieved from 
the traditional answer sets of $E(P)$.
\begin{theorem}\label{eas-sim-theorem}
Let $P$ be an ELP. Then, $S$ is an extended answer set of $P$
iff there is an answer set $S'$ of $E(P)$ such that
$S=S'\cap(\hbases{P})$.
\end{theorem}
\par
For example, let $P = \set{\prule{a}{},\;\prule{\naf{a}}{}}$, which
has two extended answer sets \set{a} and $\emptyset$.
Then $E(P) = \set{\prule{a}{\naf{\neg a},\naf{\anot{a}}},\;
\prule{\anot{a}}{\naf{a}}}$ which has two traditional answer sets
\set{a} and \set{\anot{a}} corresponding with \set{a} and $\emptyset$.
\par
The definitions from Section~\ref{olp-basic} can be reused
to define extended ordered logic programs and their preferred
answer set semantics.
However, unlike simple programs, an extended program $R$ can have 
extended answer sets $M_1\not= M_2$ while $R_{M_1} = R_{M_2}$. 
E.g., the program \set{\prule{a}{\naf{b}}\ ,\;\prule{b}{\naf{a}}}
has two (extended) answer sets \set{a} and \set{b} that both satisfy
all the rules. Intuitively, $M_1$ should be incomparable with $M_2$,
hence the extra condition $R_{M_1}\not= R_{M_2}$ 
in the definition of $\rleq$ between extended answer sets.
\begin{definition}\label{def:eolp}
An \textbf{extended ordered logic program} (EOLP) is a pair $\olp{R}{<}$ 
where $R$ is an extended program and 
$<$ is a well-founded strict
partial order on the rules in $R$\footnote{
  Strictly speaking, we should allow $R$ to be a multiset or,
  equivalently, have labeled rules, so that
  the same rule can appear in several positions in the order.
  For the sake of simplicity of notation, we will ignore this issue:
  all results also hold for the general multiset case.
  }.
\par
The partial order $\rleq$ between
subsets of $R$ is defined as in Definition~\ref{def:reduct-order}.
For $M_1 ,M_2$ extended answer sets of $R$, we define
$M_1 \rleq M_2$ iff $R_{M_1}\not= R_{M_2}$ and $R_{M_1}\rleq R_{M_2}$.
As usual, $M_1 \rlt M_2$ iff $M_1 \rleq M_2$ and $M_1 \neq M_2$.
\par
An \textbf{answer set} for an EOLP $P$ is any extended answer set of $R$.
An answer set for $P$ is called \textbf{preferred} if it
is minimal w.r.t. $\rleq$.
\end{definition}
\begin{example}\label{ex1b}
Reconsider the program from Example~\ref{ex0b} with the following
preference relation, yielding an extended ordered program \olp{P}{<}.
\begin{program3c}
\srule{\neg a}{} &
\srule{\neg b}{} &
\srule{\Naf{c}}{a} \\
\hline
\srule{a}{\Naf{b}} &
\srule{b}{\Naf{a}} &
\srule{c}{} \\
\end{program3c}
The reducts of the answer sets of $P$ are 
$P_I = P\setminus \set{\prule{c}{},\:\prule{\neg a}{}}$,
$P_J = P\setminus \set{\prule{\neg b}{}}$,
$P_K = P\setminus \set{\prule{a}{\Naf{b}},\:\prule{b}{\Naf{a}}}$ and
$P_J = P\setminus \set{\prule{\neg a}{},\prule{\Naf{c}}{a}}$,
which are ordered by ${P_J}\rleq{P_I}$, ${P_J}\rleq{P_K}$, 
${P_L}\rleq{P_I}$ and ${P_L}\rleq{P_K}$, making both $J = \set{\neg a, b, c}$
and $L = \set{a, \neg b, c}$ preferred over both $I = \set{a,\neg b}$ 
and $K = \set{\neg a, \neg b, c}$.
\end{example}
\par
An interesting interaction between defeat and negation as failure can
occur when default (minimally preferred) rules of the form ~~\prule{\naf{a}}{}~~ are used.
At first sight, such rules are useless because \naf{a} is true by
default. However, if present, such rules can also be used to defeat
others as in the following example.
\begin{example}\label{ex:nafhead}
Consider the following EOLP.
\begin{program}
\srule{\naf{a}}{} \\
\hline
\srule{a}{} \\
\hline
\srule{}{a}
\end{program}
This program has the empty set as its single preferred answer set, its
reduct containing the rules ~~\prule{\naf{a}}{}~~ and ~~\prule{}{a}.
Without ~~\prule{\naf{a}}{}~~, it would be impossible to defeat
~~\prule{a}{}~~, thus violating ~~\prule{}{a}~~ and thus the program
would not have any answer sets.
\end{example}
\par
Extended (unordered) programs can be regarded as EOLP's with
an empty order relation.
\begin{theorem}\label{ELPasEOLP}
For an ELP $P$, the extended answer sets of $P$ coincide with
the preferred answer sets of the EOLP \olp{P}{\emptyset}.
\end{theorem}
\begin{proof}
Trivial. If the order relation is empty, there are no rules to counter
defeated rules, so every extended answer set is also preferred.
\end{proof}
\par
Interestingly, negation as failure can be simulated using order alone. 
However, from Theorem~\ref{ELPasEOLP} and
Example~\ref{ex0b} (where $I\subset L$ are both preferred answer
sets), it follows that preferred answer sets for EOLP's are not necessarily
subset-minimal, which is not the case for the preferred answer sets
of the ordered programs from Section~\ref{olp-basic}.
Hence, simulating an EOLP with an OLP will necessarily involve
the introduction of fresh atoms.
\par
One might be tempted to employ a construction similar to the one
used in Definition \ref{naf-olp} for simulating negation
as failure using a two-level order. This would involve
replacing extended literals of the form ~\naf{a}~ by
fresh atoms \anot{a} and adding ``default'' rules to introduce
\anot{a}.
\par
E.g. the extended program $P =
\set{\prule{a}{\naf{b}},\;\prule{b}{\naf{a}}}$ would be
simulated by the ordered program $N(P)$
\begin{program2c}
\srule{\anot{a}}{} & \srule{\anot{b}}{} \\
\hline
\srule{a}{\anot{b}} & \srule{b}{\anot{a}} \\
\hline
\srule{\neg a}{\anot{a}} & \srule{\neg b}{\anot{b}} \\
\srule{\neg \anot{a}}{a} & \srule{\neg \anot{b}}{b}
\end{program2c}
where the rules on the lowest level act as constraints,
forcing one of the ``default rules'' in the top level to
be defeated in any proper answer set of $N(P)$.
The ``constraint rules'' also serve to indirectly introduce
competition between formally unrelated atoms: in the example,
we need e.g. a rule to defeat \prule{\anot{a}}, based on
the acceptance of $a$.
\par
This does not work, however, since it may introduce
unwanted answer sets as for the program \set{\prule{a}{\naf{a}}} which
would yield the OLP program
\begin{program}
\srule{\anot{a}}{} \\
\hline
\srule{a}{\anot{a}} \\
\hline
\srule{\neg a}{\anot{a}} \\
\srule{\neg \anot{a}}{a} \\
\end{program}
which has a (proper) preferred answer set \set{\anot{a}, \neg a} while
the original program has no extended answer sets.
\par
The above examples seem to point to contradictory requirements for
the corresponding OLP programs:
for the first example, rules implying \anot{a} should be (indirect)
competitors for $a$-rules  while for the second example, 
the \anot{a}-rule should \textit{not} compete with the $a$-rule,
in order not to introduce spurious answer sets.
\par
The solution is to add not only fresh atoms for extended literals of
the form \naf{a}, but also for ordinary literals. Thus each extended
literal $l$ will be mapped to an independent new atom $\phi(l)$.
A rule \prule{l}{\beta} will then be translated to
\prule{\phi(l)}{\phi(\beta)}, which does not compete with any
other such rule. Defeat between such rules is however supported indirectly
by adding extra rules
that encode the consequences of applying such a rule: for an original rule
of the form \prule{a}{\beta}, $a$ an ordinary literal,
we ensure that its replacement \prule{\phi(a)}{\phi(\beta)} can,
when applied, indirectly defeat $\phi(\neg a)$- and
$\phi(\naf{a})$-rules by adding both
\prule{\neg \phi(\neg a)}{\phi(\beta),\phi(a)} and 
\prule{\neg\phi(\Naf{a})}{\phi(\beta),\phi(a)}.
Similarly, for an original rule of the form \prule{\naf{a}}{\beta},
$a$ an ordinary literal, 
a rule \prule{\neg \phi(a)}{\phi(\beta),\phi(\naf{a})}
will be added along with its replacement
\prule{\phi(\naf{a})}{\phi(\beta)}.
Consistency is assured by introducing a new most preferred component
containing, besides translated constraints \prule{}{\phi(\beta)} for
the original ones, rules of the
form \prule{}{\phi(a),\phi(\naf{a})}~~and~~ 
\prule{}{\phi(a),\phi(\neg a)}.
In addition, this new component also contains translations
\prule{a}{\phi(a)} of the new atoms, that correspond to
ordinary literals, back to their original versions.
\par
Negation as failure can then be simulated by introducing
``default'' rules of the form \prule{\phi(\naf{a})}{}~~ in a new
least preferred component. 
\par
Spurious answer sets are prevented, as these new default rules
introducing $\phi(\naf{a})$, which do not have defeat-enabling accompanying
rules as described above, cannot be used to defeat transformed rules 
of the original program, but only to make them applicable. 
E.g. the program \set{\prule{a}{\naf{a}}} mentioned above would
be translated as
\begin{program2c}
\srule{\phi(\naf{a})}{}		&	\srule{\phi(\naf{\neg a})}{} \\
\hline
\srule{\phi(a)}{\phi(\naf{a})}	\\
\srule{\neg\phi(\naf{a})}{\phi(\naf{a}),\phi(a)} \\
\srule{\neg\phi(\neg{a})}{\phi(\naf{a}),\phi(a)} \\
\hline
\srule{}{\phi(a),\phi(\naf{a})}		&	\srule{a}{\phi(a)}\\
\srule{}{\phi(a),\phi(\neg{a})}		& 	\srule{\neg a}{\phi(\neg a)}\\
\srule{}{\phi(\neg a),\phi(\naf{\neg a})}		
\end{program2c}
which has no proper preferred answer sets.
\par
Formally, for an EOLP \olp{R}{<}, 
we define 
a mapping $\phi$ translating original extended literals by:
$\phi(a)  =  a'$, $\phi(\neg a)  =  a_\neg'$,
$\phi(\naf{a})  =  \anot{a}$ and 
$\phi(\naf{\neg a})  =  \anot{\neg a}$;
where for each atom $a\in\hbase{R}$, 
$a', {a_\neg}', \anot{a}$ and $\anot{\neg a}$ are fresh atoms.
We use $\phi(X)$, $X$ a set of extended literals, to denote \set{\phi(x)\mid x\in X}.
\begin{definition}
Let $P=\olp{R}{<}$ be an extended ordered logic program.
The OLP version of $P$, denoted $N_s(P)$, is defined by
$N_s (P) = \olp{R_n \cup R' \cup R_c}{R_c < R'_< < R_n}$,
where
\begin{itemize}
\item $R_n = \set{ \prule{\phi(\Naf{a})}{} \mid a\in\hbases{R} }$,
\item $R'$ is obtained from $R$ by replacing each rule 
	\begin{itemize}
	
	\item $\prule{a}{\beta}$, where $a$ is a literal, by the rules \prule{\phi(a)}{\phi(\beta)} and
	\prule{\neg \phi(\neg a)}{\phi(\beta),\phi(a)} and \prule{\neg\phi(\Naf{a})}{\phi(\beta),\phi(a)};
	
	\item $\prule{\Naf{a}}{\beta}$ by the rules \prule{\phi(\Naf{a})}{\phi(\beta)} 
	and \prule{\neg \phi(a)}{\phi(\beta),\phi(\Naf{a})};
	
	\end{itemize}
\item $R_c = \set{\prule{}{\phi(\beta)} \mid \prule{}{\beta} \in R} \cup
\set{\prule{}{\phi(a),\phi(\Naf{a})};\;
\prule{}{\phi(a),\phi(\neg a)};\;
\prule{a}{\phi(a)}\mid a\in\hbases{R}}$.
\end{itemize}
Furthermore, $R'_<$ stands for the original order on $R$ but defined
on the corresponding rules in $R'$.
\end{definition}
Note that $N_s(P)$ is free from negation as failure.
\par
\begin{example}\label{example5}
The OLP $N_s(P)$, corresponding to the EOLP of Example~\ref{ex1b} is
shown below.
\par
\begin{program3c}
\srule{\anot{a}}{}	&	\srule{\anot{b}}{}	&	\srule{\anot{c}}{} \\
\srule{\anot{\neg a}}{}	&	\srule{\anot{\neg b}}{}	&	\srule{\anot{\neg c}}{} \\
\hline
\srule{a_\neg'}{} 			&	\srule{b_\neg'}{} 		&	\srule{\anot{c}}{a'} \\
\srule{\neg a'}{a_\neg'}		&	\srule{\neg b'}{b_\neg'}	&	\srule{\neg c'}{a',\anot{c}} \\
\srule{\neg\anot{\neg a}}{a_\neg'} 	&	\srule{\neg\anot{\neg b}}{b_\neg'} \\
\hline
\srule{a'}{\anot{b}}			&	\srule{b'}{\anot{a}}			&	\srule{c'}{} \\
\srule{\neg a_\neg'}{\anot{b},a'}	&	\srule{\neg b_\neg'}{\anot{a},b'}	&	\srule{\neg c_\neg'}{c'} \\
\srule{\neg\anot{a}}{\anot{b},a'}	&	\srule{\neg\anot{b}}{\anot{a},b'}	&	\srule{\neg\anot{c}}{c'} \\
\hline
\srule{a}{a'}				&	\srule{b}{b'}				&	\srule{c}{c'} \\
\srule{\neg a}{a_\neg'}			&	\srule{\neg b}{b_\neg'}			&	\srule{\neg c}{c_\neg'} \\
\srule{}{a',\anot{a}}		&	\srule{}{b',\anot{b}}		&	\srule{}{c',\anot{c}} \\
\srule{}{a_\neg',\anot{\neg a}}	&	\srule{}{b_\neg',\anot{\neg b}}	&	\srule{}{c_\neg',\anot{\neg c}} \\
\srule{}{a',a_\neg'}		&	\srule{}{b',b_\neg'}		&	\srule{}{c',c_\neg'} \\
\end{program3c}
The OLP has two proper preferred answer sets 
$J'=\{\neg a, b, c, a_\neg', b',c',\neg a',$ $\anot{a},$ $\neg\anot{b},$
$\neg b_\neg',$ $\neg\anot{c}, \neg c_\neg', \neg\anot{\neg a},
\anot{\neg b}, \anot{\neg c}\}$ and
$L'=\{a,\neg b,c,a',b_\neg',c',\neg b',$ $\neg\anot{a},$ $\neg a_\neg',$
$\anot{b},$ $\neg\anot{c},$ $\neg c_\neg',$ $\anot{\neg a},$
$\neg\anot{\neg b},$ $\anot{\neg c}\}$, 
corresponding to the preferred answer sets $J$ and $L$ of $P$.
\end{example}
\par
In the above example, the preferred answer set of $P$ can be recovered
from the proper preferred answer set of $N_s(P)$ by selecting the literals
from \hbases{P}.
The following theorem shows that this is a general property of $N_s(P)$.
\begin{theorem}\label{eolp-to-olp-sim}
Let $P = \olp{R}{<}$ be an extended ordered logic program. Then,
$M$ is a preferred answer set of $P$ iff there exists a proper
preferred answer set $M'$ of $N_s(P)$, such that $M=M'\cap(\hbases{R})$.
\end{theorem}
\par
Since the construction of $N_s(P)$ is polynomial, the above result, 
together with Theorems~\ref{naf-olp} and~\ref{dlp-olp}, suggests that
order is at least as expressive as negation as failure,
even if the latter is used in combination with the former.
\section{Relationship to Other Approaches}\label{relationships}

\subsection{Brewka's Preferred Answer Sets}\label{b-pref}
Preferred answer sets have been introduced in the setting of
extended logic programs. In \cite{brewka99} a strict partial order on
the rules in a program is used to prefer certain traditional answer
sets above others. Intuitively, such preferred answer sets, which we
call B-preferred answer sets in what follows to avoid confusion, are
traditional answer sets that can be reconstructed by applying the rules
in order of their priorities, i.e. starting with a most specific
rule and ending with the least specific ones.
\par
First, we note that the semantics defined in \cite{brewka99} resides
at the first level of the polynomial hierarchy, i.e. $\Sigma^P_1$,
while the semantics from Section~\ref{olp-basic} is at the second
level, i.e. $\Sigma^P_2$.
The extra expressiveness of the latter semantics is useful
for diagnostic and abductive reasoning applications: e.g.
finding a subset minimal explanation is known to be
$\Sigma^P_2$-complete \cite{eiter97a}.
In addition, we illustrate with some simple examples the differences between 
both approaches. 
\par
Using the ordered programs from Section~\ref{slp} directly in the
setting of the B-preferred answer set semantics is not very useful 
since the latter applies only to consistent programs that do
have traditional answer sets.
However, if we first apply the translation from
Definition~\ref{slp-to-elp} to our programs, thus transforming the extended
answer set semantics into the traditional one, we have a means to
compare both approaches.
\par
We start with a brief, formal description of B-preferred answer sets.
The programs under consideration in \cite{brewka99} are prioritized
extended logic programs.
\begin{definition}
A \textbf{prioritized extended logic program} is a pair $P=(R,<)$, where $R$ is an
extended logic program (Definition~\ref{elp-def}) and $<$ is a strict partial 
order on the rules\footnote{
	Again we only consider grounded programs, thus avoiding the
	complex definitions in \cite{brewka99} dealing with the ground
	instantiations of prioritized rule bases.}
in $R$.
\par
The \textbf{prerequisites} of a rule
$r=\prule{a}{\beta}$ are the literals from 
$(\beta\setminus\naf{\beta^-})$, i.e. the ordinary literals in its body.
If $\beta$ contains no ordinary literals, $r$ is said to be
\textbf{prerequisite-free}.
\end{definition}
\par
The B-preferred answer set semantics is defined on
programs having a well-ordering\footnote{
	A (strict) partial order $<$ on $S$ is called a (strict) total order iff
	$\Forall{x,y\in S}{x\neq y \Rightarrow x<y~\lor~ y<x}$.
	A (strict) total order $S,<$ is called a well-ordering iff
	each nonempty subset of $S$ has a minimal element, i.e.
	$\Forall{X\subseteq S,X\neq\emptyset}{
	  \Exists{x\in X}{\Forall{y\in X}{(x=y \lor x < y)}}}$.
} relation on the rules. Therefore another definition is needed to go
from an ordinary prioritized program to one with a well-ordering.
\begin{definition}
A \textbf{full prioritization} of a prioritized program $P=(R,<)$ is any pair 
$P^\ast=(R,<^\ast)$ where $<^\ast$ is a well-ordering on $R$ compatible with $<$, 
i.e. $r_1 < r_2$ implies $r_1 <^\ast r_2$, for all $r_1,r_2\in R$.
By $\mathcal{FP}(P)$ we denote the collection of all full prioritizations of $P$.
We say that $P$ is fully prioritized, if $\mathcal{FP}(P)=\{P\}$, i.e. $P$ 
coincides with its unique full prioritization.
\end{definition}
\par
Like the traditional answer set semantics, B-preferred answer sets are
defined in two steps. In the first step, B-preferred answer sets are
defined for prerequisite-free programs, i.e. programs
containing only prerequisite-free rules.
A rule $r$ is \emph{blocked}\footnote{
	We use the term ``blocked'' instead of the original
	``defeat''~\cite{brewka99} to avoid
	confusion with the notion of defeat from Definition~\ref{def:simple-program}. 
} by a literal $l$ iff
$l\in{\BODY{r}^-}$. On the other hand, $r$ is blocked by a set of
literals $X$, iff $X$ contains a literal that blocks $r$.
\par
In the following construction, the rules in a full prioritization
are applied in the order of their priorities.
\begin{definition}\label{brewkadef}
Let $P=(R,<)$ be a full prioritization of a prerequisite-free prioritized program; 
let $S$ be a set of literals and let $(R,<)=\{r_\alpha\}_<$. We define the 
sequence $S_\alpha, 0\le \alpha < ord(<)$, of sets $S_\alpha\subseteq\hbases{R}$ 
as follows:
\[
S_\alpha = \left\{\begin{array}{ll}
\bigcup_{\beta < \alpha} S_\beta, & \textit{if $r_\alpha$ is blocked by $\bigcup_{\beta < \alpha} S_\beta$ or} \\
& \textit{$\HEAD{r_\alpha}\in S$ and $r_\alpha$ is blocked by $S$,} \\
\bigcup_{\beta < \alpha} S_\beta\cup\{\HEAD{r_\alpha}\} & \textit{otherwise.}
\end{array}\right.
\]
\par
The set $C_P(S)$ is the smallest set of ground literals, i.e. $C_P(S)\subseteq \hbases{R}$, such that
\begin{enumerate}
\item $\bigcup_{\alpha < ord(<)} S_\alpha \subseteq C_P(S)$, and
\item $C_P(S)$ is logically closed.
\end{enumerate}
\end{definition}
\par
The condition ``$\HEAD{r_\alpha}\in S$ and $r_\alpha$ 
is blocked by $S$'' in the above definition is necessary to avoid situations 
in which a literal in $S$ is derived by two rules $r_1$ and $r_2$ such that 
$r_1 < r_2$, but $r_2$ is applicable in $S$ and $r_1$ is not. This would result 
in the conclusion \HEAD{r_1} at a priority higher than effectively sanctioned by 
the rules, as in the following example.
\par
\begin{example}
Consider the following program, taken from \cite{brewka99}.
\[
\begin{array}{lll}
r_4 &  : & \prule{p}{\naf{\neg p}} \\
\hline
r_3 &  : & \prule{\neg p}{\naf{p}} \\
\hline
r_2 &  : & \prule{q}{\naf{\neg q}} \\
\hline
r_1 &  : & \prule{p}{\naf{q}} \\
\end{array}
\]
This program has two classical answer sets, i.e. $S_1=\set{p,q}$ and
$S_2=\set{\neg p, q}$. Without the condition ``$\HEAD{r_\alpha}\in S$ and $r_\alpha$ 
is blocked by $S$'' in Definition~\ref{brewkadef},
the sequence for $S_1$ would be
$\set{p}, \set{p,q}, \set{p, q}, \set{p, q}$ because
$r_1$ would be applied, even if it is blocked w.r.t. the final 
set $S_1$. This in turn blocks $r_3$.
On the other hand, using Definition~\ref{brewkadef} correctly, $S_1$ yields
the sequence $\set{}, \set{q}, \set{\neg p, q}, \set{\neg p, q}$ where
$r_1$ is not applied and $r_3$ becomes applicable. 
With Definition~\ref{brewkadef2}, this implies that $S_1$ is not
B-preferred.
\end{example}
\par
In general, $C_P$ does not necessarily return the consequences of $R$, 
i.e. an applied rule $r_\alpha$ may later be blocked by some less preferred rule
$r_\beta$ where $\alpha < \beta$.
However, if a normal answer set $A$ of $R$ is a fixpoint of $C_P$, 
then all preferences are taken into account, i.e. a rule whose head is not in $A$ is
blocked by a more preferred rule applied in $A$. Such answer sets will
be preferred.
\begin{definition}\label{brewkadef2}
Let $P=(R,<)$ be a full prioritization of a prerequisite-free prioritized program; 
and let $A$ be a normal answer set of $R$.
Then $A$ is a \textbf{B-preferred answer set} of $P$ iff $C_P(A)=A$.
\end{definition}
\par
The B-preferred answer sets for programs with prerequisites
are obtained using a reduction to prerequisite-free programs.
\begin{definition}
Let $P=(R,<)$ be a full prioritization of a prioritized program; 
and let $X\subseteq \hbases{R}$.
Then, $^X{\!P}=(^X{\!R},^X{\!<})$ is the fully prioritized program obtained
from $P$, where $^X{\!R}$ is the set of rules obtained from $R$ by
\begin{enumerate}
\item deleting every rule having a prerequisite $l$ such that $l\not\in X$, and
\item removing from each remaining rule all prerequisites,
\end{enumerate}
and where $^X{\!<}$ is inherited from $<$ by the mapping $f:\ ^X{\!R}\longrightarrow R$, 
i.e. $r_1'\ ^X{\!<} r_2'$ iff $f(r_1') < f(r_2')$, where $f(r')$ is the first rule 
in $R$ w.r.t. $<$ such that $r'$ results from $r$ by step 2.
\end{definition}
\par
In the above reduction, a rule \prule{a}{\beta} is removed from the
program w.r.t. a set of literals $X$ iff the prerequisite part of the
rule is not applicable w.r.t. $X$, i.e.
$(\beta\setminus\naf{\beta^-})\not\subseteq X$. 
The prerequisites of the remaining rules can then be safely removed.
\begin{definition}
A set of ground literals $A\subseteq\hbases{R}$ is a B-preferred answer set of 
a full prioritization $P=(R,<)$ of a prioritized program, if $A$ is a B-preferred 
answer set of $^A{\!P}$. $A$ is a B-preferred answer set of a prioritized program $Q$, 
if $A$ is a B-preferred answer set for some $P\in\mathcal{FP}(Q).$ We use 
$\mathcal{AS}_B(Q)$ to denote the set of all B-preferred answer sets of $Q$.
\end{definition}
The following examples suggest that there is no clear
relationship between the preferred (Section~\ref{olp}) and
B-preferred~\cite{brewka99} semantics.
\par
\begin{example}\label{ex_B_1}
Consider the OLP $P$ on the left side below.
\[
\begin{array}{ccc}
\begin{array}{rllrll}
\ssrule{\neg a}{} & \ssrule{\neg b}{} \\
\hline
\ssrule{a}{} & \ssrule{b}{} \\
\hline
\ssrule{\neg b}{\neg a} & \ssrule{\neg a}{\neg b} \\
\ssrule{a}{b} & \ssrule{b}{a} \\
\end{array} & \hspace{0.5cm} &
\begin{array}{rllrll}
\ssrule{\neg a}{\naf{a}} & \ssrule{\neg b}{\naf{b}} \\
\hline
\ssrule{a}{\naf{\neg a}} & \ssrule{b}{\naf{\neg b}} \\
\hline
\ssrule{\neg b}{\neg a,\naf{b}} & \ssrule{\neg a}{\neg b,\naf{a}} \\
\ssrule{a}{b,\naf{\neg a}} & \ssrule{b}{a,\naf{\neg b}} \\
\end{array} 
\end{array}
\]
\par
This program has only one preferred answer set, i.e. $I=\set{a,b}$. 
On the other hand, the transformed program $E(P)$ (see Definition
\ref{slp-to-elp}) with the same ordering among the rules as in $P$
(see the program on the right side above),
has two B-preferred answer sets, namely $I$ and $J=\set{\neg a, \neg b}$.
\end{example}
\par
While the above example illustrates that the B-preferred answer set
semantics may yield too many answers w.r.t. the preferred answer set 
semantics, the next example shows that both approaches can
yield different answers.
\par
\begin{example}\label{ex_B_2}
Consider the OLP $P$ on the left side below, and its consistent
version $E(P)$, with the same order among the rules, on the right.
\[
\begin{array}{ccc}
\begin{array}{rll}
\ssrule{\neg b}{} \\
\hline
\ssrule{b}{} \\
\hline
\ssrule{a}{b} \\
\hline
\ssrule{\neg a}{} \\
\end{array} & \hspace{1cm} &
\begin{array}{rll}
\ssrule{\neg b}{\naf{b}} \\
\hline
\ssrule{b}{\naf{\neg b}} \\
\hline
\ssrule{a}{b,\naf{\neg a}} \\
\hline
\ssrule{\neg a}{\naf{a}} \\
\end{array} 
\end{array}
\]
\par
Let us interpret $a$ as 
"we have light" and $b$ as "the power is on". The fact
that we do not have any light ($\neg a$) has
the highest priority, followed by the rule that, if there
is power, we should have light ($\prule{a}{b}$).
The weakest rules assert that probably the power is on ($b$),
but on the other hand, it may not be ($\neg b$).
\par
Clearly, the only preferred answer set $\set{\neg a, \neg b}$ of $P$ fits well
with our intuition: we have no light because
somehow, the power is cut (defeating the $\prule{b}{}$ rule).
On the other hand, the single B-preferred
answer set is \set{\neg a, b} which questions the rule
$\prule{a}{b}$, although this rule is more preferred than
\prule{b}{}. This example illustrates some form of contrapositive
reasoning: since it is preferable to satisfy \prule{a}{b}, and $\neg a$ holds,
any answer set that provides (a motivation for) $\neg b$ will
be preferred over one that does not. It is this capability
that is of interest in e.g. diagnostic applications
\cite{dvnv2003a,dvnv2003b}.
\end{example}
\par
In the literature two stricter version of the B-preferred
answer set semantics have been considered: the $D$-preferred answer set
semantics \cite{delgrande2000} and the $W$-preferred answer set
semantics \cite{wang2000}. A nice overview of and comparison between
the $B$-, $W$- and $D$- preferred answer sets can be found in \cite{schaub2001}.
It turns out (Theorem~8 in \cite{schaub2001}) that
$\mathcal{AS}_D(R,<) \subseteq \mathcal{AS}_W(R,<) \subseteq $ \mbox{$\mathcal{AS}_B(R,<)$} $ \subseteq \mathcal{AS}(R)$,
i.e. $B$-preferredness is more strict than traditional answer sets,
$W$-preferredness is more strict than $B$-preferredness
and $D$-preferredness is the most strict preference relation.
This is illustrated by Example~\ref{ex_B_1} where neither 
$I$ or $J$ is $W$- or $D$-preferred.
Nevertheless, in Example~\ref{ex_B_2} the $B$-preferred answer set $\set{\neg a,b}$
is also $W$- and $D$-preferred,
suggesting that there is no relationship between those approaches and
the preferred answer set semantics from Section~\ref{olp}.
\par
\cite{brewka99} also introduces two principles, which we rephrase
below using the present framework, that every
system based on prioritized defeasible rules should satisfy.
\par
\begin{principle}\label{P1}
Let $M_1$ and $M_2$ be two different extended answer sets
of an ordered logic program $P=\olp{R}{<}$, generated by the rules 
$R'\cup\set{d_1}$ and $R'\cup\set{d_2}$, where $d_1,d_2\not\in R'$,
respectively. If $d_1$ is preferred over $d_2$, then $M_2$ is not a
(maximally) preferred answer set of $P$.
\end{principle}
\begin{principle}\label{P2}
Let $M$ be a preferred answer set of an
ordered logic program $P=\olp{R}{<}$ and let $r$ be an inapplicable
rule w.r.t. $M$. Then $M$ is a preferred answer set of
\olp{R\cup\set{r}}{<'} whenever $<'$ agrees with $<$ on the preference
among the rules in $R$.
\end{principle}
\par
In this context, a rule $r$ is said to be generating w.r.t. an
extended answer set $M$, if it is applied w.r.t. $M$.
It turns out that the preferred answer set semantics from
Section~\ref{olp} violates Principle~\ref{P1}, as illustrated by the following example.
\par
\begin{example}
Let $P$ be the OLP
\begin{program}
\srule{a}{} \\
\srule{b}{} \\
\hline
\srule{\neg a}{} \\
\hline
\srule{\neg b}{\neg a} \\
\hline
\srule{b}{\neg  b} \\
\end{program}
and consider the interpretations $I=\set{a,b}$ and $J=\set{\neg a,b}$. 
The generating rules for $I$ are \set{\prule{a}{},\prule{b}{}}, while those for $J$ are
\set{\prule{\neg a}{},\prule{b}{}}. Furthermore, we have that 
$\prule{\neg a}{} < \prule{a}{}$. Since $I$ is a 
preferred answer set of $P$, this violates Principle~\ref{P1}.
\end{example}
The problem with Principle~\ref{P1} comes from the fact that it
only considers ``generating'', i.e. applied, rules while the
semantics of Section~\ref{olp} also takes into account unapplied but
satisfied rules, which have an equal influence on the shape of an
answer set. E.g. in the above example, the rule \prule{\neg b}{\neg a}
is violated by $J$ but not by $I$, which should lead one to prefer
$I$ over $J$.
\par
These considerations lead to a modified version
of Principle~\ref{P1} where ``generating rules'' is replaced
by ``reducts'' (Definition~\ref{def:reduct}) of the extended answer sets.
\begin{principlerecap}{1'}
Let $M_1$ and $M_2$ be two different extended answer sets
of an ordered logic program $P=\olp{R}{<}$, with reducts
$R_{M_1} = R'\cup\set{d_1}$ and $R_{M_2} = R'\cup\set{d_2}$, 
where $d_1,d_2\not\in R'$,
respectively. If $d_1$ is preferred over $d_2$, then $M_2$ is not a
(maximally) preferred answer set of $P$.
\end{principlerecap}
Clearly, the semantics of Section~\ref{olp} obeys this principle
(obviously, $R_{M_1} \rlt R_{M_2}$). The new principle is also vacuously
satisfied for approaches, such as the B-preferred semantics, that
start from consistent programs. However, if we apply
e.g. the B-preferred semantics ``indirectly'' on inconsistent programs using
the construction of Definition~\ref{slp-to-elp}, 
it appears from Example~\ref{ex_B_2} that this semantics
does satisfy Principle~1'.
\par
Obviously, the second principle holds for the preferred answer set
semantics.
\par
\begin{theorem}\label{sat_b_prin}
The preferred answer set semantics satisfies Principle~\ref{P2}. 
\end{theorem}
\begin{proof}
Trivial. Adding inapplicable rules w.r.t. a preferred answer set to an
ordered logic program does not change anything to its preferredness.
\end{proof}

%
%
\subsection{Logic Programming with Ordered Disjunction}\label{lpod}
Logic programming with ordered disjunction (LPOD)~\cite{brewka2002a,brewka2002b} 
combines qualitative choice logic \cite{brewka2002} with answer set programming by
adding a new connective, called ordered disjunction, to logic programs.  Using this connective,
conclusions in the head of a rule are ordered according to preference.
Intuitively, one tries to satisfy an applicable rule by using
its most preferred, i.e. best ranked, conclusion.
\begin{definition}
A \textbf{logic program with ordered disjunction} (LPOD) is a set of
rules of the form \prule{a_1 \times \dots \times a_n}{\beta}, $n\geq
1$, where the $a_i$'s are (ordinary) literals and $\beta$ is a finite set of
extended literals.
\end{definition}
\par
An ordered disjunctive rule 
\prule{a_1 \times \dots \times a_n}{\beta}
can intuitively be read as: if 
$\beta$ is true, then accept $a_1$, if possible; if not, then
accept $a_2$, if possible; ...; if none of $a_1,\dots,a_{n-1}$ are
possible, then $a_n$ must be accepted.
\par
Similarly to ordered programs, the semantics for LPOD's is defined in
two steps.  First, answer sets for LPOD's are defined, which are then
ordered by a relation that takes into account to what degree
rules are satisfied. The minimal elements in this ordering are also
called preferred answer sets. To avoid confusion, we will
use the term ``preferred LPOD answer sets'' for the latter.
\par
Answer sets for LPOD's are defined using \textit{split programs},
a mechanism first used in \cite{sakama94} to define the possible
model semantics for disjunctive logic programs.
\begin{definition}
Let $r = \prule{a_1\times\dots\times a_n}{\beta}$ be a LPOD rule. For
$k \le n$ we define the $k^{th}$ option of $r$ as 
$r^k = \prule{a_k}{\beta,\,\naf{\set{a_1,\dots,a_{k-1}}}}$.
\par
An extended logic program $P'$ is
called a \textbf{split program} of a LPOD $P$ if it is obtained by
replacing each rule in $P$ by one of its options.
An interpretation $S$ is then an (LPOD) \textbf{answer set} of $P$ if it is an answer
set of a split program $P'$ of $P$.
\end{definition}
\par
Note that the split programs defined above do not contain negation as
failure in the head of rules.
\begin{example}\label{lpodex1}
The LPOD
\begin{program}
\srule{b\times c\times d}{} \\
\srule{c\times a\times d}{} \\
\srule{\neg c}{b} \\
\end{program}
\par
has five answer sets, namely
$S_1 = \set{a,b,\neg c}$, $S_2=\set{b,\neg c,d}$, $S_3=\set{c}$,
$S_4 = \set{a,d}$ and $S_5=\set{d}$.
\par
Note that $S_5\subset S_4$, illustrating that answer sets for LPOD
programs need not be subset-minimal. Intuitively, only $S_1$ and $S_3$ are optimal
in that they correspond with a best combination of options for each
of the rules.
\end{example}
\par
The above intuition is formalized in the following definition of
a preference relation on LPOD answer sets.
\begin{definition}
Let $S$ be an answer set of a LPOD $P$. Then $S$ satisfies the rule
\prule{a_1 \times \dots \times a_n}{\beta}
\begin{itemize}
\item[-]to degree $1$ if $S\not\models\beta$. 
\item[-]to degree $j$ $(1 \le j \le n)$ if $S\models\beta$ and
$j = min\set{i\mid a_i\in S}$.
\end{itemize}
For a set of literals $S$, we define $S^i(P) = \set{ r\in P \mid deg_S(r) = i}$,
where $deg_S(r)$ is used to denote the degree to which $r$ is satisfied w.r.t. $S$.
\par
Let $S_1$ and $S_2$ be answer sets of $P$. Then $S_1$ is
preferred over $S_2$, denoted $S_1 \rlt_b S_2$, iff
there is a $k$ such that $S_2^k(P) \subset S_1^k(P)$, and for all $j < k$,
$S_1^j(P) = S_2^j(P)$. A minimal (according to $\rlt_b$) answer set
is called a (LPOD) \textbf{preferred answer set} of $P$.
\end{definition}
\begin{example}
In the LPOD from Example~\ref{lpodex1}, both $S_1$ and $S_3$
satisfy the third rule to degree~1. In addition, $S_1$ satisfies
the first rule to degree~1 and the second rule to degree~2, while
$S_3$ satisfies the second rule to degree~1 and the first rule
to degree~2. Thus $S_1$ and $S_3$ are incomparable w.r.t. $\rlt_b$.
All other answer sets are less preferred than either $S_1$ or $S_3$:
e.g. $S_5$ satisfies the first and second rule only to degree~3,
and the third rule to degree~1, from which $S_1\rlt_b S_5$ and
$S_3\rlt_b S_5$. It follows that $S_1$ and $S_3$ are both preferred.
\end{example}
\par
Next, we show that the preference relation which is implicit in
ordered disjunctive rules can be intuitively simulated using
preference between non-disjunctive rules in one single ordered
program. In \cite{brewka2002b} an algorithm is presented to compute
LPOD preferred answer sets using traditional answer set programming. 
The algorithm needs two different programs, i.e. a generator that 
computes answer sets and a tester for checking preferredness,
which are executed in an interleaved fashion. On the other hand, using
the following translation together with the \textsc{olps} solver (see Section~\ref{compute})
does the job in a single run.
\begin{definition}\label{lpod2olp}
The EOLP version of a LPOD $P$, denoted $L(P)$, is
defined by
$L(P) = \olp{P_r \cup P_1 \cup \dots \cup P_n \cup P_d}{P_r < P_1 < \dots < P_n < P_d}$,
where
\begin{itemize}
\item $n$ is the size of the greatest ordered disjunction in P;
\item $P_r$ contains every non-disjunctive rule $\prule{a}{\beta}\in P$.
In addition, for every ordered disjunctive rule 
$r=\prule{a_1 \times \dots \times a_n}{\beta}\in P$, $P_r$ contains a
rule \prule{a_i}{\beta,\naf{\set{a_1,\dots,a_n}\setminus\set{a_i}}}
for every $1\leq i\leq n$, a rule \prule{nap_r}{\naf{l}} for every
literal $l\in\beta$ and a rule \prule{nap_r}{l} for every
$\naf{l}\in\beta$.
\item $P_d$ contains for every ordered disjunctive rule 
$r=\prule{a_1 \times \dots \times a_n}{\beta}\in P$ a rule
\prule{\naf{nap_r}}{} and rules
\prule{\naf{a_i}}{\beta,\naf{\set{a_1,\dots,a_{i-1}}}} for every
$1\leq i\leq n$.
\item for $1\leq k\leq n$, $P_k$ is defined by
$P_k=\set{
  \prule{a_k}{\beta,\naf{\set{a_1,\dots,a_{k-1}}}}\mid
  \prule{a_1\times\dots\times a_m}{\beta} \in P\textit{ with } k\leq m\leq n}$.
\end{itemize}
\end{definition}
\par
Intuitively, the rules in $P_r$ ensure that all rules in $P$ are
satisfied, while the rules in $P_d$ allow to defeat a rule in one
of the $P_1,\dots,P_{n-1}$ in favor of a rule in a less preferred
component (see also Example~\ref{ex:nafhead}).
Finally, the rules in $P_1, \dots, P_n$ encode the intuition
behind LPOD, i.e. better ranked literals in an ordered disjunction are
preferred.
\par
The $nap_r$-literals (and their corresponding rules) serve
to prevent LPOD answer sets that block disjunctive rules to be
preferred over others in $L(P)$.
E.g. the program
\begin{program}
\srule{c\times d}{a}\\
\srule{a}{\naf{b}}\\
\srule{b}{\naf{a}}\\
\end{program}
has both \set{a,c} and \set{b} as preferred LPOD answer sets, with the
latter blocking the disjunctive rule. Applying
Definition~\ref{lpod2olp} yields the EOLP
\begin{program}
\srule{\naf{nap_{\prule{c\times d}{a}}}}{} \\
\srule{\naf{c}}{a} \\
\srule{\naf{d}}{a,\naf{c}} \\
\hline
\srule{d}{a,\naf{c}}\\
\hline
\srule{c}{a}\\
\hline
\srule{c}{a,\naf{d}} \\
\srule{d}{a,\naf{c}} \\
\srule{nap_{\prule{c\times d}{a}}}{\naf{a}} \\
\srule{a}{\naf{b}} \\
\srule{b}{\naf{a}} \\
\end{program}
which has two preferred answer sets \set{a,c} and \set{b,nap_{\prule{c\times d}{a}}}.
Without the $nap_{\prule{c\times d}{a}}$ construct, only \set{b} would be preferred
as \set{b} would satisfy all rules, while \set{a,c} would defeat \prule{\naf{c}}{a}.
\par
\begin{example}
The result of the transformation of the LPOD from
Example~\ref{lpodex1} is shown below, where $r=\prule{b\times c\times d}{}$
and $s=\prule{c\times a\times d}{}$.
\begin{program2c}
\srule{\Naf{b}}{}		&	\srule{\Naf{c}}{} \\
\srule{\Naf{c}}{\Naf{b}}	&	\srule{\Naf{a}}{\Naf{c}} \\
\srule{\Naf{d}}{\Naf{b},\Naf{c}}& 	\srule{\Naf{d}}{\Naf{c},\Naf{a}} \\
\srule{\Naf{nap_r}}{}		&	\srule{\Naf{nap_s}}{}\\
\hline
\srule{d}{\Naf{b},\Naf{c}}	&	\srule{d}{\Naf{a},\Naf{c}} \\
\hline
\srule{c}{\Naf{b}}		&	\srule{a}{\Naf{c}} \\
\hline
\srule{b}{}			&	\srule{c}{} \\
\hline
\srule{b}{\Naf{c},\Naf{d}}	&	\srule{c}{\Naf{a},\Naf{d}} \\
\srule{c}{\Naf{b},\Naf{d}}	&	\srule{a}{\Naf{c},\Naf{d}} \\
\srule{d}{\Naf{b},\Naf{c}}	&	\srule{d}{\Naf{a},\Naf{c}} \\
\srule{\neg c}{b} \\
\end{program2c}
This EOLP program has two proper preferred answer
sets, i.e. $S_1 = \set{a,b,\neg c}$ and $S_3 = \set{c}$.
\end{example}
\par
In general, the preferred LPOD answer sets for a LPOD program $P$ coincide
with the proper preferred answer sets of $L(P)$.
\begin{theorem}\label{lpodtheorem}
An interpretation $S$ is a preferred LPOD answer
set of a LPOD $P$ iff there exists a proper preferred answer set $S'$ of $L(P)$
such that $S=S'\cap(\hbases{P})$.
\end{theorem}
%
%
\subsection{Answer Set Programming with Consistency-Restoring Rules}\label{crrules}
In \cite{balduccini2003a,balduccini2003b} an extension of answer set programming with
consistency-restoring rules (cr-rules) and preferences\footnote{
	\cite{balduccini2003a,balduccini2003b} allow for both static
	and dynamic preferences; here we only consider the static case.
} is presented.
The approach allows problems to be described in a concise, and easy to
read, manner. While \cite{balduccini2003a} introduces cr-rules, \cite{balduccini2003b} combines them
with ordered disjunction (Section~\ref{lpod}).
A possible application is presented in
\cite{balduccini2004} where cr-rules are used to improve the quality of solutions in the
planning domain.
\par
Intuitively, cr-rules in a program are like normal rules, but they can 
only be used as a last resort to obtain solutions, i.e. only when the 
program without cr-rules is inconsistent we can apply some of the 
cr-rules to obtain a solution for the problem, taking into account 
the preferences among the cr-rules. Consider for example the following 
program where $\gets_{cr}$ is used to denote cr-rules:
\[
\begin{array}{rrcl}
r_1: & \crrule{p}{\naf{t}} \\
r_2: & \crrule{q}{\naf{t}} \\
r_3: & \ssrule{s}{}
\end{array}
\]
and $r_2$ is preferred over $r_1$. As the single rule $r_3$ is 
consistent, the above program has only one answer set, i.e. \set{s}. 
Adding the constraint
\[
r_4: \gets \naf{p}, \naf{q}
\]
causes the combination of $r_3$ and $r_4$ to be inconsistent.
In this case, one of the cr-rules $r_1$ or $r_2$ is selected
for application. Either of them yields, when combined with $r_3$ and $r_4$, a
consistent program, making both \set{p,s} and \set{q,s} candidate solutions 
for the program. However, since $r_2$'s application is preferred over $r_1$'s,
only the latter candidate solution should be sanctioned as an answer set
of the program.
\par
It turns out that the intuition behind consistency restoring rules can
be captured by the preferred answer set semantics using a translation
similar to the one used for diagnostic and abductive reasoning with
preferences \cite{dvnv2003a,dvnv2003b}. E.g., the above program can 
be captured by the following ordered program:
\[
\begin{array}{rcl}
\ssrule{p}{\naf{t},inconsistent} \\
\ssrule{q}{\naf{t},inconsistent} \\
\hline
\ssrule{\naf{q}}{inconsistent} \\
\hline
\ssrule{\naf{p}}{inconsistent} \\
\hline
\ssrule{s}{} \\
\ssrule{inconsistent}{\naf{s}}\\
\ssrule{}{\naf{p},\naf{q}}\\
\ssrule{inconsistent}{\naf{p},\naf{q}}
\end{array}
\]
\par
Intuitively, the non cr-rules, which must always be satisfied,
are placed in the most specific component, together with rules that
check when the normal program is inconsistent, so allowing
consistency-restoring rules to be applied.
The cr-rules are placed at the least preferred level because 
they will only be applied as
a last resort. In the middle levels we introduce rules that allow for
the defeat of the cr-rules, i.e. allowing an applicable cr-rule not to
be applied. The ordering relation on these rules is the opposite of 
the preference relation between cr-rules as preferring (the
application of) $r_2$ over $r_1$ corresponds
to preferring the satisfaction of \prule{\naf{p}}{} upon \prule{\naf{q}}{}.
It can be verified that the above ordered program has only one preferred answer set
\set{q,s,inconsistent} corresponding with the application of the most preferred
cr-rule, while removing $r_4$ from the ordered program will result in the single
preferred answer set \set{s} implying that no cr-rules are used.
\par
Example~\ref{examplebalduccini} (Section~\ref{olp-basic}) provides
another illustration of the above
translation\footnote{
	As the program, except the constraint, is positive we can use 
	classical negation ($\neg$) and dispense the $inconsistent$
	literals and rules.
}: the ordered program there
is obtained by applying the construction on the following program with
cr-rules:
\[
\begin{array}{rcl}
\crrule{lift\_weights}{}\\
\crrule{play\_ball}{}\\
\crrule{run}{}\\
\crrule{swim}{}\\
\ssrule{full\_body\_exercise}{lift\_weights,run}\\
\ssrule{full\_body\_exercise}{swim,play\_ball}\\
\ssrule{}{\naf{full\_body\_exercise}}\\
\end{array}
\]
with, additionally, $\acrrule{run}{} < \acrrule{swim}{}$
and $\acrrule{ball\_play}{} < \acrrule{lift\_weights}{}$\footnote{
  We use $r_1 < r_2$ to indicate that (the application of) the cr-rule
  $r_1$ is preferred over $r_2$.}.
\par
As with the translation of LPOD (Section~\ref{lpod}), this
illustrates how OLP can be used to encode other extensions of answer
set programming. Together with an OLP solver such
as \textsc{olps} (Section~\ref{compute}), it also provides a
convenient translation-based implementation of such higher level 
formalisms.
%
%
\subsection{Ordered Logic, $\mathcal{DOL}$ and $\mathcal{DLP}^<$}\label{doldlp}
Ordered logic programming has a long history: 
in \cite{laenens90b,gabbay91,laenens92} semantics are given for
ordered programs containing non-disjunctive rules, while
\cite{bucca98,bucca99} apply the same ideas to the disjunctive case.
\par
In all these approaches, the partial order on rules is used to decide conflicts
between contradictory rules, i.e. applicable rules that cannot
both be applied in a consistent interpretation.
Specifically, an applicable rule $r$ may be left unapplied, i.e.
defeated, iff there exists an applied competitor $r'$ that is more
preferred, i.e. $r < r'$\footnote{
	In some approaches one demands that the competitor $r'$ cannot
	be less preferred then $r$, i.e. $r'\not> r$, thus allowing
	rules on the same level or on unrelated levels to defeat each
	other. However, this does not change anything to the conclusions 
	made in this section.}.
Clearly, these classical semantics for ordered logic use the order
relation on a local basis, i.e. to resolve conflicts between
individual contradicting rules, while the semantics from Section~\ref{olp}
is based on a global comparison of the reducts corresponding to
the candidate answer sets.
\par
This difference in using the order relation has important consequences
on the complexity of the resulting formalisms:
while the semantics that use the order in a
''local'' way all stay in the same complexity class of the underlying
non-ordered language, i.e. $\Sigma^P_1$-complete for 
\cite{laenens90b,gabbay91,laenens92} and $\Sigma^P_2$-complete for
\cite{bucca98,bucca99}, the ``global order'' semantics from Section~\ref{olp}
increments the complexity in the polynomial hierarchy,
i.e. $\Sigma^P_2$-complete instead of $\Sigma^P_1$-complete (see
Section~\ref{compute}).
\par
The following example illustrates how the classical semantics
leads to different, and, in our opinion less intuitive, results
from the preferred answer set semantics.
\begin{example}
Consider the following ordered program.
\begin{program}
\srule{a}{}\\
\hline
\srule{\neg a}{}\\
\hline
\srule{b}{}\\
\hline
\srule{\neg b}{\neg a}
\end{program}
\par
Using the semantics from Section~\ref{olp} yields one preferred answer
set, i.e. $I=\set{a,b}$. However, in the setting of the classical
semantics $I$ cannot be a model as the rule \prule{\neg a}{} is
applicable, not applied and no better rule with opposite head exists.
On the contrary, those classical semantics will yield 
$J=\set{\neg a,\neg b}$ as a model, which is certainly not a preferred
answer set as the rule \prule{b}{} is defeated w.r.t. $J$, while it is
satisfied w.r.t. $I$.
\end{example}
%
%
\section{Application: Repairing Databases}\label{db-repair}
\par
The notion of database repair was first introduced in \cite{arenas99}
to address the problem of consistent query answering in inconsistent
databases, which was first mentioned in \cite{bry97}. 
\cite{arenas99} describes an algorithm to compute such
query answers using database repairs, while \cite{arenas2000} provides an
algorithm to compute such repairs themselves, using a complex
translation to
disjunctive logic programs with exceptions~\cite{kowalski90}.
\par
Here we show that database repairs can be obtained as the
preferred answer sets of a simple and intuitive
ordered program corresponding to the original database and constraints.
We review some definitions from \cite{arenas2000}, using
a simplified notation.
\begin{definition}\label{def:db}
A \textbf{database} is a consistent set of literals.
A \textbf{constraint} is a set $A$ of literals, to be interpreted
as a disjunction, $c = \lor_{a\in A}a$. The Herbrand base \hbase{C} of a set
of constraints $C$ is defined by $\hbase{C} = \cup_{c\in C}\hbase{c}$.
A database $D$ is \textbf{consistent} with a set of constraints $C$,
where $\hbase{C}\subseteq\hbase{D}$, just when
$D \models \land_{c\in C} c$, i.e. $D$ is a classical model of $C$.
A set of constraints $C$ is consistent iff there exists a database $D$
such that $D$ is consistent with $C$.
\end{definition}
\begin{definition}\label{def:db-order}
Let $D$ and $D'$ be databases with $\hbase{D}=\hbase{D'}$.
We use $\Delta_D (D')$ to
denote the difference $\setmin{D'}{D}$.
A database $D$ induces a partial order relation\footnote{
  The proof that $\leq_D$ is a partial order
  is straightforward.} $\leq_D$ defined by
\[
D_1 \leq_D D_2 \mbox{ iff } \Delta_D (D_1 ) \subseteq \Delta_D (D_2 )\enspace .
\]
\end{definition}
\par
Intuitively, $\Delta_D (D')$ contains the update operations that must
be performed on $D$ to obtain $D'$. A negative literal $\neg a$ in
$\Delta_D (D')$ means that the fact $a$ must be removed from $D$ while
$a\in\Delta_D (D')$ suggests adding $a$ to $D$.
\par
The $\leq_D$ relation represents the closeness to $D$:
$D_1 \leq_D D_2$ means that $D_1$ is a better approximation of $D$ than
$D_2$ (note that $D \leq_D D$).
\begin{definition}\label{def:repair}
Let $D$ be a database and let $C$ be a set of constraints
with $\hbase{C}\subseteq\hbase{D}$.
A database $D'$ is a \textbf{$C$-repair} of $D$ iff $D'\models C$ and
$D'$ is minimal in the $\leq_D$ partial order; i.e.
$D'' \leq_D D'$ implies that $D'' = D'$.
\end{definition}
\par
This definition differs from the one in \cite{arenas2000} where $\leq_D$
was defined based on the symmetric difference
$\Delta(D,D') = (\setmin{{D}^+}{{D'}^+} ) \cup (\setmin{{D'}^+}{{D}^+} )$
rather than our $\Delta_D (D')$.
\par
The following lemma shows that $\leq_D$ is the same using both relations.
\begin{lemma}\label{arenas-same-leq}
Let $D$ , $D_1$ and $D_2$ be databases over the same Herbrand base.
$\Delta_D (D_1 ) \subseteq \Delta_D (D_2 )$ iff $\Delta(D,D_1 ) \subseteq \Delta(D, D_2 )$.
\end{lemma}
\par
From Lemma~\ref{arenas-same-leq}, it follows that Definition~\ref{def:repair}
is equivalent to the one in \cite{arenas2000}.
\par
Next we provide a construction that maps a database $D$ and a set
of constraints $C$ to an ordered logic program $P(D,C)$ which, as shall be
shown further on, has the $C$-repairs of $D$ as preferred answer sets.
Using ordered logic instead of logic programs with exceptions
\cite{kowalski90} greatly simplifies (w.r.t. \cite{arenas2000})
constructing repairs: we can
dispense with the shadow versions of each predicate and we do
not need disjunction. Moreover, our approach handles constraints of
arbitrary size, while \cite{arenas2000} is limited to constraints
containing up to two literals.
\begin{definition}\label{def:pdc}
Let $D$ be a database and let $C$ be a consistent set of constraints
with $\hbase{C}\subseteq\hbase{D}$.
The \textbf{ordered version} of $D$ w.r.t. $C$, denoted $P(D,C)$,
is shown below.
{
\renewcommand{\arraystretch}{1.2}
\[
\begin{array}{cl}
\set{\prule{\neg a}{}\mid a\in D} & (n) \\
\hline
\set{\prule{a}{}\mid a\in D} & (d) \\
\hline
\set{\prule{a}{\neg(\setmin{A}{\set{a})}}\mid
	\lor_{a\in A} a \in C} & (c)

\end{array}
\]
}
\end{definition}
\par
Intuitively, the $c$-rules enforce the constraints (they are
also the strongest rules according to the partial order).
The $d$-rules simply input the database as ``default'' facts
while the $n$-rules will be used to provide a justification
for literals needed to satisfy certain $c$-rules, thus
defeating $d$-rules that would cause constraints to be violated.
\begin{theorem}\label{repair-is-aset}
Let $D$ be a database and let $C$ be a consistent set of constraints
with $\hbase{C}\subseteq\hbase{D}$. 
Each repair of $D$ w.r.t. $C$ is a preferred answer set of $P(D,C)$.
\end{theorem}
\par
The reverse of Theorem~\ref{repair-is-aset} also holds.
\begin{theorem}\label{aset-is-repair}
Let $D$ be a database and let $C$ be a consistent set of constraints
with $\hbase{C}\subseteq\hbase{D}$.
Each preferred answer set of $P(D,C)$ is a $C$-repair of $D$.
\end{theorem}

\begin{example}
Consider the propositional version of the example from \cite{arenas2000}
where the database $D = \set{ p, q, r}$ and the set of constraints
$C = \{
\neg p \lor q, 
\neg p \lor \neg q, 
\neg q \lor r, 
\neg q \lor \neg r, 
\neg r \lor p, 
\neg r \lor \neg p, 
\}$.
The program $P(D,C)$ is shown below.
\begin{program3c}
\srule{\neg p}{} &
\srule{\neg q}{} &
\srule{\neg r}{} \\
\hline
\srule{p}{} &
\srule{q}{} &
\srule{r}{} \\
\hline
\srule{\neg p}{\neg q} &
\srule{q}{p} &
\srule{\neg p}{q} \\

\srule{\neg q}{p} &
\srule{\neg q}{\neg r} &
\srule{r}{p} \\

\srule{\neg r}{q} &
\srule{\neg q}{r} &
\srule{p}{r} \\

\srule{\neg r}{\neg p} &
\srule{\neg p}{r} &
\srule{\neg r}{p} \\
\end{program3c}
It is easily verified 
that $R=\set{\neg p, \neg q, \neg r}$ is the only repair of $D$ w.r.t. $C$
and the only proper preferred answer set of $P(D,C)$.
\end{example}
\par
The example below illustrates that the simple
translation of Definition~\ref{def:pdc} for database repairs does not work with 
other ordered formalisms (see Section~\ref{relationships}), where a
rule may be left unsatisfied if there exists an applicable (or
applied) better rule with an opposite head.
\begin{example}
Consider the database $D = \set{\neg a, \neg b}$ and the following set
of constraints $C = \{a\lor \neg b, \neg a \lor b\}$. Obviously, $D$
itself is the only repair w.r.t. $C$. Now, consider $P(D,C)$ which is
shown below.
\begin{program2c}
\srule{a}{} & \srule{b}{} \\
\hline
\srule{\neg a}{} & \srule{\neg b}{} \\
\hline
\srule{a}{b} & \srule{b}{a} \\
\srule{\neg a}{\neg b} & \srule{\neg b}{\neg a} \\
\end{program2c}
\par
This program has only one proper preferred answer set, namely $D$ itself.
But, when we consider the same program in most other ordered
formalisms, e.g. \cite{laenens92}, we get two solutions, i.e. $D$ and 
$\set{a,b}$.
\end{example}
\section{Conclusions and Directions for Further Research}\label{conclusions}
The preferred answer set semantics for ordered programs is based on a few
simple intuitions: ignore rules that are defeated by other applied
rules and order answer sets according to the natural order between
the sets of rules that they satisfy. The resulting system is
surprisingly powerful and versatile and turns out to be useful
for several application areas such as database repair
(Section~\ref{db-repair}) or diagnostic systems \cite{dvnv2003a,dvnv2003b}.
It may also serve as a ``common base language'' for the
encoding and implementation (through an OLP solver) of other higher level extensions of
answer set programming such as ordered disjunctions
\cite{brewka2002a,brewka2002b} or programs with
consistency-restoring rules \cite{balduccini2003a,balduccini2003b}.
\par
A first implementation of an ordered logic program solver 
(\textsc{olps}) is available under the \textsc{gpl} at \url{http://tinf2.vub.ac.be/olp/}.
After grounding, \textsc{olps} computes (a selection of) the proper preferred answer sets of a
finite ordered program. 
\par
Besides research topics relating to the theory and
implementation of the preferred answer set semantics, there
are also some other application areas that still need to be explored.
One such area concerns updates of
logic programs, see e.g. \cite{eiter2000a,alferes2000}. 
Here, it is natural to consider a new update as an addition of a new
most specific level to the ordered program representing the previous
version. The preferred answer set semantics will then return solutions
that favor compatibility with more recent knowledge, possibly at the expense of
earlier rules.
\par
The formalism can also be extended to consider hierarchies of
preference orders. Such a system could be used to model
hierarchical decision making where each agent has her own preferences,
and agents participate in a hierarchy representing relative authority
or confidence. The semantics of such a system should 
reflect both individual preferences and the decision hierarchy.
\par
Finally, it would be interesting to apply the intuition of
Definition~\ref{def:reduct-order} to a system where preferences are
expressed on literals rather than on rules. Intuitively, in such a
formalism, preference would depend on the content of the candidate
answer sets, rather than how well they satisfy the underlying
rules\footnote{
  A system using preferences on literals has been proposed in \cite{sakama96}
  but the way these preferences are used to obtain a ranking of answer
  sets is different from our proposal.
  }.
In fact, a system that combines both rule- and literal-preferences,
perhaps in a complex hierarchy as mentioned above,
could be useful for certain applications.

\appendix

\section*{Appendix: Proofs}
\par
\begin{theoremrecap}{\ref{slp-vs-elp}}
Let $P$ be a SLP. The extended answer sets of $P$ coincide with the
answer sets of $E(P)$.
\end{theoremrecap}
\begin{proof}
\par
\fbox{$\Longrightarrow$} Let $M$ be an extended 
answer set of $P$. We must show that $M$ is an
answer set of $E(P)$, i.e. $M$ is an answer set
of the simple program $E(P)^M$
which is obtained from $E(P)$ by keeping (a) the rules
\prule{a}{\beta} from $P$ where $\neg a\not\in M$ and (b) the
constraints \prule{}{\beta} from $P$.
Thus it suffices
to show that $P_M^\star = {E(P)^M}^\star$. 
\par
Consider a rule $r = \prule{a}{\beta}$ in $E(P)^M$. Since
$\neg a\not\in M$, this rule cannot be defeated and hence $r\in P_M$.
Further, constraints $s=\prule{}{\beta}$ in $E(P)^M$ can never be
defeated and as $M$ is an extended answer set they are satisfied
w.r.t. $M$, so $s\in P_M$.
As a result $E(P)^M \subseteq P_M$ and, by the monotonicity of the
$\star$-operator, 
\begin{equation}\label{eq2}
{E(P)^M}^\star \subseteq P_M^\star\enspace .
\end{equation}
\par
Next, note that 
$P_M^\star = \set{r \mid \mbox{$r$ applied w.r.t. $M$}}^\star$, i.e.
only the applied rules in $P_M$ suffice to generate $M$.
Each applied rule $r=\prule{a}{\beta}$ in $P_M$ must
belong to $E(P)^M$ since $a\in M$, and thus $\neg a\not\in M$, because
$r$ is applied. Thus $P_M^\star \subseteq {E(P)^M}^\star$ and,
by (\ref{eq2}), $P_M^\star = {E(P)^M}^\star$. 
\par
\fbox{$\Longleftarrow$} Let $M$ be an answer set of $E(P)$, i.e. 
$M = {E(P)^M}^\star$ where $E(P)^M$ is as above.
We show that $M$ is an extended answer set of $P$.
Obviously, all rules
in $E(P)^M$ are satisfied by $M$ and thus $E(P)^M \subseteq P_M$ and,
by the monotonicity of the $\star$-operator, $M \subseteq P_M^\star$.
\par
The constraints in $E(P)$ do not contain negation as failure and,
obviously, they are satisfied by $M$. So, all rules $r$ not kept in
$E(P)^M$ have the form \prule{a}{\beta}.
\par
For a rule $r=\prule{a}{\beta}$ not in $E(P)^M$ we know that
$\neg a\in M$. If $\beta\not\subseteq M$, $r$ is
not applicable and thus $r\in P_M$. If, on the other hand,
$\beta\subseteq M$, $r$ is defeated because 
$\neg a\in M \subseteq P_M^\star$, which implies the existence
of an applied competitor in $P_M$. Thus each unsatisfied rule in $P$ is 
defeated w.r.t. $M$. It remains to be shown that $P_M^\star = M$.
This follows from $E(P)^M \subseteq P_M$ and
the fact that only rules that are not applicable w.r.t. $M$ are in
\setmin{P_M}{E(P)^M}.
\end{proof}

\begin{theoremrecap}{\ref{thm1}}
Let $<$ be a well-founded strict partial order
on a set $X$. The binary relation $\rleq$ on $2^X$  defined by
\mbox{$X_1 \rleq X_2$} iff
$\Forall{x_2\in \setmin{X_2}{X_1}}{
  \Exists{x_1\in \setmin{X_1}{X_2}}{x_1 < x_2}}$
is a partial order.
\end{theoremrecap}
\begin{proof}
The relation $\rleq$ is clearly reflexive. 
To show that it is antisymmetric, assume that both
$X_1 \rleq X_2$ and $X_2 \rleq X_1$. 
We show that $\setmin{X_1}{X_2} = \setmin{X_2}{X_1} = \emptyset$, from
which $X_1 = X_2$. Assume that, on the contrary,
$\setmin{X_2}{X_1}\neq\emptyset$ and let
$x_0$ be a minimal element in \setmin{X_2}{X_1} (such an element
exists because $<$ is well-founded).
Because $X_1 \rleq X_2$, there exists
$x_1\in\setmin{X_1}{X_2}$ for which $x_1 < x_0$. 
Since $X_2 \rleq X_1$, $x_1\in\setmin{X_1}{X_2}$
implies the existence of $x_2\in\setmin{X_2}{X_1}$ such
that $x_2 < x_1 < x_0$, contradicting the fact that
$x_0$ is minimal in \setmin{X_2}{X_1}.
\par
To show that $\rleq$ is transitive, let $X_1 \rleq X_2 \rleq X_3$ and
let $x_0 \in \setmin{X_3}{X_1}$.
We consider two possibilities.
\begin{itemize}
\item If $x_0 \in X_2$ then, because $X_1 \rleq X_2$, there exists
$x_1 \in\setmin{X_1}{X_2}$ with $x_1 < x_0$, where we choose
$x_1$ to be a minimal element satisfying these conditions.
Again there are two cases:
  \begin{itemize}
  \item If $x_1 \not\in X_3$ then $x_1\in\setmin{X_1}{X_3}$,
    showing that $X_1 \rleq X_3$.
  \item If $x_1 \in X_3$ then $x_1 \in\setmin{X_3}{X_2}$ and thus,
    since $X_2 \rleq X_3$, there exists $x_2 \in\setmin{X_2}{X_3}$
    such that $x_2 < x_1 < x_0$. If $x_2 \in X_1$ we are
    done. Otherwise $x_2 \in\setmin{X_2}{X_1}$ and thus,
    since $X_1 \rleq X_2$, there exists $x_3 \in \setmin{X_1}{X_2}$
    where $x_3 < x_2 < x_1 < x_0$, contradicting the minimality
    assumption on $x_1$.
  \end{itemize}
\item Otherwise, $x_0 \not\in X_2$ and thus, because $X_2 \rleq X_3$,
  there exists an element $x_1 \in\setmin{X_2}{X_3}$ such that
  $x_1 < x_0$ where we choose $x_1$ to be a minimal element satisfying
  these conditions. Again there are two cases:
  \begin{itemize}
  \item If $x_1 \in X_1$ then we are done as $x_1 < x_0$.
  \item Otherwise, since $X_1 \rleq X_2$, there exists an
    element $x_2 \in \setmin{X_1}{X_2}$ such that $x_2 < x_1$.
    If $x_2 \not\in X_3$ then we are done as $x_2 < x_0$.
    If, on the other hand, $x_2 \in X_3$, $X_2 \rleq X_3$ implies
    the existence of an element $x_3 \in\setmin{X_2}{X_3}$ with
    $x_3 < x_2$, contradicting our earlier assumption of the
    minimality of $x_1$.
  \end{itemize}
\end{itemize}
Hence, in call cases, there exists an element $x\in\setmin{X_1}{X_3}$
such that $x<x_0$ and thus $X_1 \rleq X_3$.
\end{proof}

\begin{theoremrecap}{\ref{naf-olp}}
Let $P$ be an (non-disjunctive) seminegative logic program
The ordered version of $P$, denoted $N(P)$ is defined by
$N(P) = \olp{P' \cup P_\neg}{<}$ with
$P_\neg = \{\prule{\neg a}{} \mid a\in\hbase{P}\}$
and $P'$ is obtained from $P$ by replacing each negated literal
$\naf{p}$ by $\neg p$. The order is defined by
$P' < P_\neg$, i.e. $\Forall{r\in P',r'\in P_\neg}{r<r'}$ (note that $P'\cap
P_\neg=\emptyset$). Then $M$ is a stable model of $P$ iff
$M\cup\neg(\setmin{\hbase{P}}{M})$ is a proper preferred answer set of
$N(P)$.
\end{theoremrecap}
\begin{proof}
\par
\fbox{$\Longrightarrow$} Let $M\subseteq\hbase{P}$ be 
a stable model of $P$. Thus, by definition,
${P^M}^* = M$ where $P^M$ is the result of the Gelfond-Lifschitz
transformation.
We show first that $M' = M\cup\neg(\setmin{\hbase{P}}{M})$
is an extended answer set of $R = P'\cup P_\neg$.
Besides \set{ \neg a \mid \neg a \in M'}
the reduct $R_{M'}$ also contains all rules from $P'$:
if not, there would be a rule \prule{a}{\beta} in $P^M$ 
where $\beta\subseteq M$ while $a\not\in M$, contradicting that
${P^M}^\star = M$. Also, it is is easy to see that applying
the $R_{M'}$-rules from \set{ \prule{\neg a}{} \mid \neg a \in M'}
blocks\footnote{
 A rule \prule{a}{\alpha} is blocked w.r.t. and interpretation
 $I$ if $\alpha\cap\neg I\neq\emptyset$.}
all rules corresponding to rules from \setmin{P}{P^M}. Consequently,
$R_{M'}^\star = (\set{ \prule{\neg a}{} \mid \neg a \in M'}\cup P^M)^\star = M'$.
\par
To show that $R_{M'}$ is minimal (and $M'$ is preferred), 
note for any $R$-extended answer set $X$ with $R_X\rlt R_{M'}$, it must be the case that
$R_X\supset R_{M'}$ and, moreover, $(\setmin{R_X}{R_{M'}})\subseteq P_\neg$
since the only rules defeated by $M'$ are in $P_\neg$.
Let $\prule{\neg a}{}\in\setmin{R_X}{R_{M'}}$.
Since $^\star$ is monotonic and $P'\subseteq R_X$, it follows
that $R_X^\star \supseteq\set{a,\neg a}$, contradicting that
$X$ is an extended answer set. From the fact that $P'\subseteq R_{M'}$, it 
follows that $M'$ is proper.
\par
\fbox{$\Longleftarrow$} Let $M'$ be a proper preferred answer set of
$N(P) = \olp{R=P'\cup P_\neg}{<}$. As $M'$ is proper, $P'\subseteq R_{M'}$.
Clearly $M'\cap\neg\hbase{P} = \set{\neg a\mid \prule{\neg a}{} \in
(P_\neg\cap R_{M'}}$ and $M'^+ = (R_{M'}^\star)^+ = Q^\star$, where
$Q$ is obtained from $P'\subseteq R_{M'}$ by removing all rules that
contain a literal $\neg a$ in the body such that $a\in M'$ and,
moreover, removing $\neg a\in M'$ from the bodies of the remaining
rules. Then $Q=P^{M^+}$, yielding that $M^+$ is a stable model of $P$.
\end{proof}

\begin{theoremrecap}{\ref{disj-olp}}
Let $P$ be a positive disjunctive logic program. $M$ is a minimal model
of $P$ iff $M'=M\cup\neg(\hbase{P}\setminus M)$ is a
proper preferred answer set of $D(P)$.
\end{theoremrecap}
\begin{proof}
\par
\fbox{$\Longrightarrow$} Let $M$ be a minimal model of $P$.
Clearly, $M'$ is total for $D(P)$ and, by construction, each
rule in $D(P)$ is either satisfied or defeated.  On the other hand,
the rules from $D(P)_{M'} \cap (P_+\cup P_-)$ suffice to
generate $M'$ which is thus founded. Hence, by
Definition~\ref{def:slp-aset}, $M'$ is an extended answer set of $D(P)$.
Since all rules in $P_p$ are satisfied by any model of $P$, $M'$ is
also proper.
\par
To show that $M'$ is preferred, assume it is not, i.e.
$N\rlt M'$ for some proper extended answer set $N$ of $D(P)$, for which,
obviously, $N^+$ is also a model of $P$. 
From the construction of $D(P)$ and
Definition~\ref{def:reduct-order}, we obtain that
$\setmin{D(P)_{M'}}{D(P)_N}\subseteq P_+$ from which
$N^+\subset M$, contradicting the fact that $M$ is a minimal model of
$P$.
\par
\fbox{$\Longleftarrow$} Let $M' = M \cup \neg(\setmin{\hbase{P}}{M})$
be a proper preferred answer set of $D(P)$.
Clearly, since $M'\models P_p$, $M$ must be a model of $P$ by
construction of the rules in $P_p$.
To show that $M$ is minimal, assume that it is not, i.e. $N\subset M$
for some model $N$ of $P$. A similar reasoning as above yields that
$N' = N\cup\neg(\setmin{\hbase{P}}{N})$ is a proper extended answer set of
$D(P)$. Moreover, from $N\subset M$, it is straightforward to show
that $N'\rlt M'$, contradicting that $M'$ is preferred.
\end{proof}

\begin{lemma}\label{dlp1}
Let $P$ be a seminegative DLP and let $M_1$ and $M_2$ be proper
extended answer sets of $D_n(P)$. Then $M_1^+\subset M_2^+$ iff
$M_1 \rlt M_2$.
\end{lemma}
\begin{proof}
\par
\fbox{$\Longrightarrow$} Assume that $M_1^+ \subset M_2^+$ and let
$r\in\setmin{D_n(P)_{M_2}}{D_n(P)_{M_1}}$ and let
$s\in\setmin{D_n(P)_{M_1}}{D_n(P)_{M_2}}$.
Since both $M_1$ and $M_2$ are proper, 
it must be the case that $r,s\in P_-\cup P_c$.
Obviously, 
$D_n(P)_{M_2}\cap P_- = \set{\prule{\neg a}{}\in P_-\mid \neg a\in \setmin{M_2}{M_2^+}}$.
As both $M_1$ and $M_2$ are total and $M_1^+\subset M_2^+$, we know
that
$\Forall{\neg a\in\setmin{M_2}{M_2^+}}{\neg a\in\setmin{M_1}{M_1^+}}$
and
$\Forall{a\in\setmin{M_2^+}{M_1^+}}{\neg a\in M_1^+}$.
Combining the above yields that
$D_n(P)_{M_1}\cap P_- = (D_n(P)_{M_2}\cap P_-)\cup
\set{\prule{\neg a}{}\in P_-\mid a\in\setmin{M_2^+}{M_1^+}}$.
Thus, $(D_n(P)_{M_2}\cap P_-) \subset (D_n(P)_{M_1}\cap P_-)$,
making $M_1$ preferred upon $M_2$, i.e. $M_1\rlt M_2$.
\par
\fbox{$\Longleftarrow$} Assume that $M_1\rlt M_2$
(from which, $M_1\neq M_2$) and let $a\in\setmin{M_1^+}{M_2^+}$.
Since $a\not\in M_2^+$, the $P_-$ rule $r_{\neg a} = \prule{\neg a}{}$
cannot be defeated and thus $\neg a\in M_2$ and
$r_{\neg a}\in {D_n(P)}_{M_2}$. On the other hand, 
$r_{\neg a}\not\in {D_n(P)}_{M_1}$ because the rule is defeated by
a rule introducing $a$ into $M_1$.
Thus
$r_{\neg a}\in\setmin{{D_n(P)}_{M_2}}{{D_n(P)}_{M_1}}$ but,
since $M_1$ and $M_2$ are proper, $r_{\neg a}$ is not countered by
${D_n(P)}_{M_1}$, contradicting that $M_1\rlt M_2$.
\end{proof}

\begin{theoremrecap}{\ref{dnp-pm}}
Let $P$ be a seminegative DLP. An interpretation $M$ is a
proper preferred answer set of $D_n(P)$ iff $M^+$ is a minimal
possible model of $P$.
\end{theoremrecap}
\begin{proof}
We show that the proper extended answer sets of $D_n(P)$ coincide with
the possible models of $P$, from which, by Lemma~\ref{dlp1},
the theorem readily follows.
\par
\fbox{$\Longrightarrow$} Let $M$ be a proper extended answer set of $D_n(P)$ and consider
$S_M(P) = ({D_n(P)}_M) \cap P_c$ where all occurrences of
negative literals $\neg a$ have been replaced by their
negation-as-failure counterparts \naf{a}. We claim that
$S_M(P)$ is a split program of $P$ because $S_M(P)$ contains
at least one split clause for each rule \prule{\alpha}{\beta}
from $P$. Indeed, if this were not the case, there would
exist a rule \prule{\alpha}{\beta} such that $\beta\subseteq M$
while $\alpha\cap M = \emptyset$, violating the corresponding
rules in $P_p$ and thus contradicting the assumption that $M$
is proper.
\par
To show that $M$ is a possible model, it then suffices to show
that $P'^\star = M^+$, 
where $P'= {{S_M(P)}^{M^+}}$ is the result of applying the
Gelfond-Lifschitz transformation to $S_M(P)$ and $M^+$.
However, we already know that
$P''^\star = M$ where
$P'' = ({D_n(P)}_M) \cap (P_c \cup P_-)$
because
rules from ${D_n(P)}_M \cap P_p$ are not needed to ``produce''
$M$ (indeed, for any applicable (and thus applied) rule
\prule{a}{\beta \cup\neg(\setmin{\alpha}{\set{a}})} in 
${D_n(P)}_M \cap P_p$,
there exists an ``equivalent'' applied rule
\prule{a}{\beta} in ${D_n(P)}_M \cap P_c$).
It is then not difficult to see that $P'$ can also be obtained
from $P''$ by
(a) removing all negative literals $\neg a$, where \prule{\neg a}{} is
in $P''$, from all rule bodies, and (b) removing any rules
that still contain negative literals, thus also the rules 
\prule{\neg a}{} which were not affected by (a).
Since this this operation can be viewed as part of the computation
of $(P''^\star)^+ = M^+$, it follows that $P'^\star = M^+$.
\par
\fbox{$\Longleftarrow$} Let $N\subseteq\hbase{P}$ be a possible model
corresponding to a split program $S_N(P)$ of
$P$ and let $M = N\cup\neg(\setmin{\hbase{P}}{N})$.
We use $P'$ to denote the program obtained from
$S_N(P)$ by replacing $\naf{\beta}$ by $\neg{\beta}$ in
all its rules.
It follows that
$P'' = {D_n(P)}_M = X \cup \set{\prule{\neg a}{}\mid a\in
(\setmin{\hbase{P}}{N}} \cup P_p$ where 
$P' \subseteq X \subseteq P_c$. To show that $M$
is a proper extended answer set we must show that (1) $M\models r$
for any $r\in P_p$, (2) $P''^\star = M$, and (3) each rule
in \setmin{D_n(P)}{P''} is defeated. Clearly, (1)
follows from the fact that $M$ is a model of $P$ (Proposition~3.1
in \cite{sakama94}).  To establish (2), it suffices to
note that $(P' \cup (P''\cap P_-))^\star = M$ because $M^+$
is a possible model, from which $P''^\star = M$ because
$(P' \cup (P''\cap P_-) \subseteq P''$ (and, by definition of $P''$,
$P''^\star \subseteq M$). (3) then follows immediately from
the fact that $M$ is total and (2).
\end{proof}

\begin{lemma}\label{lemma-witness}
Let \olp{P}{<} be an ordered program and $R\subseteq P$
and $T\subseteq P$ be sets of rules.
Then $T\not\rlt R$ iff $R$ has a witness against $T$,
which is equivalent to
$\Exists{X\in\witness{T}}{X\subseteq R}$.
\end{lemma}
\begin{proof}
From the definition of $\rlt$, it immediately
follows that $T\not\rlt R$ iff $R$ has a witness
against $T$.
\par
To show the second part of the lemma,
suppose $X\subseteq R$ for some $X\in\witness{T}$.
Then $X = \set{r}\cup (\down{\set{r}}\cap T)$ for some
$r\not\in T$. It is then straightforward to verify that
$r$ is a witness for $R$ against $T$ and thus $T\not\rlt R$.
\par
Conversely, if $T\not\rlt R$ then, according to the first part of
the lemma,
there exists a witness $r\in\setmin{R}{T}$ for which it
is easy to verify that $r\not\in T$ while $\down{\set{r}}\cap T\subseteq R$,
and thus $r\in\witness{T}$.
\end{proof}
\begin{lemma}\label{lemma-aux-2}
Let \olp{P}{<} be an ordered program, $\pair{R_i}{R_o}$ be a specification, $C$ a
constraint and $r$ a minimal (according to $<$) element from
\setmin{P}{(R_i\cup R_o)}.  Let $R$, with $r\in R$, be a set of rules. 
Then $R\in\min\expansion{\pair{R_i}{R_o}}{C}$ iff 
$R\in\min\expansion{\pair{R_i\cup\set{r}}{R_o}}{C}$.
\end{lemma}
\begin{proof}
\par
\fbox{$\Longrightarrow$} Let $R\in\min\expansion{\pair{R_i}{R_o}}{C}$. Then, 
by definition, $R\in\expansion{\pair{R_i\cup\set{r}}{R_o}}{C}$ because
$r\in R$.
Assume that, on the contrary, 
$R\not\in\min\expansion{\pair{R_i\cup\set{r}}{R_o}}{C}$.
Hence $\Exists{T\in\expansion{\pair{R_i\cup\set{r}}{R_o}}{C}}{T\rlt R}$.
But then also
$T\in\expansion{\pair{R_i}{R_o}}{C}$ because
$\expansion{\pair{R_i\cup\set{r}}{R_o}}{C}\subseteq\expansion{\pair{R_i}{R_o}}{C}$.
Consequently, $R$ would not be minimal in
$\expansion{\pair{R_i}{R_o}}{C}$, a contradiction.
\par
\fbox{$\Longleftarrow$} Let $R\in\min\expansion{\pair{R_i\cup\set{r}}{R_o}}{C}$
and assume that, on the contrary,\linebreak 
$\Exists{T\in \expansion{\pair{R_i}{R_o}}{C}}{T\rlt R}$.
Surely, $r\not\in T$, since otherwise $T\in\linebreak[0]\expansion{\pair{R_i\cup\set{r}}{R_o}}{C}$
and $R$ would not be minimal in 
$\expansion{\pair{R_i\cup\set{r}}{R_o}}{C}$.
Because $r$ is minimal in \setmin{P}{(R_i\cup R_o)}, $\down{\set{r}}\cap T = \down{\set{r}}\cap R$,
and thus $r\in R$ is a witness for $R$ against $T$.
It follows that  $T\not\rlt R$, contradicting our assumption.
\end{proof}
\begin{lemma}\label{lemma-aux-3}
Let \olp{P}{<} be an ordered program,
$\pair{R_i}{R_o}$ be a specification, $C$ a constraint and
$r$ a minimal (according to $<$) element from \setmin{P}{(R_i\cup R_o)}.
Let $R$, with $r\not\in R$ be a set of rules.\\ 
Then,
$R\in\min\expansion{\pair{R_i}{R_o}}{C}$ iff
$R\in\min\expansion{\pair{R_i}{R_o\cup\set{r}}}{C}$ and\linebreak
$\Forall{T\in\min\expansion{\pair{R_i\cup\set{r}}{R_o}}{C}}{T\not\rlt R}$.
\par
Moreover, if $C'$ is such that
\begin{eqnarray}\label{eq0}
\lefteqn{ \expansion{\pair{R_i}{R_o\cup\set{r}}}{C'} =} \nonumber\\
&& \set{T\mid T\in\expansion{\pair{R_i}{R_o\cup\set{r}}}{C} \land 
\Forall{m\in M}{\Exists{x\in\witness{m}}{x\subseteq T}}}
\end{eqnarray}
with $M=\min{\expansion{\pair{R_i\cup\set{r}}{R_o}}{C}}$, then
\begin{eqnarray}\label{eq1}
\lefteqn{\min{\expansion{\pair{R_i}{R_o\cup\set{r}}}{C'}} = } \nonumber\\
&& \set{T\mid T\in\min{\expansion{\pair{R_i}{R_o\cup\set{r}}}{C}} \land 
\Forall{m\in M}{\Exists{x\in\witness{m}}{x\subseteq T}}}
\end{eqnarray}
\end{lemma}
\begin{proof}
\par
\fbox{$\Longrightarrow$} Suppose that $R\in\min\expansion{\pair{R_i}{R_o}}{C}$.
Since $\expansion{\pair{R_i}{R_o\cup\set{r}}}{C} \subseteq
\expansion{\pair{R_i}{R_o}}{C}$ and
$R\in\expansion{\pair{R_i}{R_o\cup\set{r}}}{C}$,
$R\in\min\expansion{\pair{R_i}{R_o\cup\set{r}}}{C}$.
\par
Let $T\in\min{\expansion{\pair{R_i\cup\set{r}}{R_o}}{C}}$.
By Lemma~\ref{lemma-aux-2}, 
$T\in\min\expansion{\pair{R_i}{R_o}}{C}$.
Consequently, $R$ can
only be minimal in $\expansion{\pair{R_i}{R_o}}{C}$ if $T\not\rlt R$.
\par
\fbox{$\Longleftarrow$} Let $R\in\min{\expansion{\pair{R_i}{R_o\cup\set{r}}}{C}}$ such that
$\Forall{T\in\min\linebreak\expansion{\pair{R_i\cup\set{r}}{R_o}}{C}}{T\not\rlt R}$
and assume that, on the contrary, $S\rlt R$
for some $S\in\expansion{\pair{R_i}{R_o}}{C}$. Without loss of
generality, we can assume that $S$ is minimal in
$\expansion{\pair{R_i}{R_o}}{C}$.
Note that $r\not\in S$ is impossible since that would put
$S$ in $\expansion{\pair{R_i}{R_o\cup\set{r}}}{C}$, contradicting
that $R$ is minimal in $\expansion{\pair{R_i}{R_o\cup\set{r}}}{C}$.
It follows that $r\in S$ and thus, by our assumption that
$S$ is minimal in $\expansion{\pair{R_i}{R_o}}{C}$ and by Lemma~\ref{lemma-aux-2},
that $S$ is minimal in $\expansion{\pair{R_i\cup\set{r}}{R_o}}{C}$.
But then our assumption guarantees $S\not\rlt R$, a contradiction.
\par
\fbox{$\supseteq$ of (\ref{eq1})}
Let $T$ be any set in 
$\min{\expansion{\pair{R_i}{R_o\cup\set{r}}}{C}}$
that satisfies $\Forall{m\in M}{\linebreak\Exists{x\in\witness{m}}{T\cap\witness{m}\neq\emptyset}}$.
By (\ref{eq0}), $T\in\expansion{\pair{R_i}{R_o\cup\set{r}}}{C'}$.
If $T$ would not be minimal in $\expansion{\pair{R_i}{R_o\cup\set{r}}}{C'}$,
there would exist a $S\rlt T$ in $\expansion{\pair{R_i}{R_o\cup\set{r}}}{C'}$.
But, by (\ref{eq0}), $S\in\expansion{\pair{R_i}{R_o\cup\set{r}}}{C}$, contradicting
that $T$ is minimal in $\expansion{\pair{R_i}{R_o\cup\set{r}}}{C}$.
\par
\fbox{$\subseteq$ of (\ref{eq1})}
Assume, on the contrary, that
$S\in\min{\expansion{\pair{R_i}{R_o\cup\set{r}}}{C'}}$
while $S\not\in \set{T\mid T\in\min{\expansion{\pair{R_i}{R_o\cup\set{r}}}{C}} \land 
\Forall{m\in M}{\Exists{x\in\witness{m}}{x\subseteq T}}}$.
Obviously, $S\in\expansion{\pair{R_i}{R_o\cup\set{r}}}{C'}$ which,
combined with (\ref{eq0}), yields that 
$\Forall{m\in M}{\Exists{x\in\witness{m}}{x\subseteq S}}$ holds.
So, it must be the case that $S\not\in\min{\expansion{\pair{R_i}{R_o\cup\set{r}}}{C}}$,
but from (\ref{eq0}) we have that $S\in\expansion{\pair{R_i}{R_o\cup\set{r}}}{C}$,
thus $\Exists{U\in\expansion{\pair{R_i}{R_o\cup\set{r}}}{C}}{U\rlt S}$.
Since 
$S\in\min{\expansion{\pair{R_i}{R_o\cup\set{r}}}{C'}}$
we have that $U\not\in\min{\expansion{\pair{R_i}{R_o\cup\set{r}}}{C'}}$,
but also 
$U\not\in\expansion{\pair{R_i}{R_o\cup\set{r}}}{C'}$ and thus
$\Exists{m\in M}{\Forall{x\in\witness{m}}{x\not\subseteq U}}$.
As a result, $U$ has no witness against some $m\in M$, from which,
by Lemma~\ref{lemma-witness}, $m\rlt U$. Since
$\rlt$ is a partial order, transitivity with $U\rlt S$ yields $m\rlt S$,
contradicting that $S$ has a witness against any $m\in M$.
\end{proof}
\begin{lemma}\label{poas-in-aset}
Let \olp{P}{<} be an ordered program.
If $R$ is minimal w.r.t. $\rlt$ in 
$\expansion{\pair{R_i}{R_o}}{C}$ then 
$R \in \aset{\pair{R_i}{R_o}}{C}$.
\end{lemma}
\begin{proof}
Let $R\in\min{\expansion{\pair{R_i}{R_o}}{C}}$.  We show the result by induction on
the cardinality of \setmin{P}{(R_i\cup R_o)}.
\par
For the base case, we have that $R_i\cup R_o = P$ and, consequently,
$\expansion{\pair{R_i}{R_o}}{C}=\set{R}$ which is also returned by
$\aset{\pair{R_i}{R_o}}{C}$.
\par
For the induction step, take a minimal rule $r\in\setmin{P}{(R_i\cup R_o)}$,
such that\linebreak $\aset{\pair{R_i}{R_o}}{C}$ calls
$\aset{\pair{R_i\cup\set{r}}{R_o}}{C}$ and $\aset{\pair{R_i}{R_o\cup\set{r}}}{C'}$.
\par
We consider two cases.
\begin{itemize}
\item
If $r\in R$ then, by Lemma~\ref{lemma-aux-2},
$R\in\min{\expansion{\pair{R_i\cup\set{r}}{R_o}}{C}}$.
From the induction hypothesis, it follows
that $R\in\aset{\pair{R_i\cup\set{r}}{R_o}}{C}$.
It is clear that
$\aset{\pair{R_i\cup\set{r}}{R_o}}{C} \subseteq
\aset{\pair{R_i}{R_o}}{C}$ and thus
$R \in \aset{\pair{R_i}{R_o}}{C}$.
\item
If $r\not\in R$ then, by Lemma~\ref{lemma-aux-3},
$R\in\min{\expansion{\pair{R_i}{R_o\cup\set{r}}}{C}}$ and,
moreover, $S\not\rlt R$ for any $S\in\min{\expansion{\pair{R_i\cup\set{r}}{R_o}}{C}}$.
The latter condition is ensured by the definition of $C'$
in the code of Figure~\ref{basic-algo}: it is straightforward to show
that $R\models C'$ iff $R\models C$ and
$R$ has a witness against every $m\in M$. Hence
\begin{eqnarray*}
\lefteqn{ \expansion{\pair{R_i}{R_o\cup\set{r}}}{C'} =} \\
&& \set{T\mid T\in\expansion{\pair{R_i}{R_o\cup\set{r}}}{C} \land 
\Forall{m\in M}{\Exists{x\in\witness{m}}{x\subseteq T}}}
\end{eqnarray*}
\par
From (\ref{eq1}) in Lemma~\ref{lemma-aux-3} and the induction hypothesis, it
then follows that
$R\in\linebreak\aset{\pair{R_i}{R_o\cup\set{r}}}{C'}$, from which
$R\in\aset{\pair{R_i}{R_o}}{C}$ by the last line in Figure~\ref{basic-algo}.
\end{itemize}
\end{proof}

\begin{lemma}\label{aset-in-poas}
Let \olp{P}{<} be an ordered program, 
$\pair{R_i}{R_o}$ be a specification and $C$ a
constraint.
Then $\aset{\pair{R_i}{R_o}}{C} \subseteq \min\expansion{\pair{R_i}{R_o}}{C}$.
\end{lemma}
\begin{proof}
Clearly, the function terminates since, within
a call $\aset{\pair{R_i}{R_o}}{C}$,
any recursive call $\aset{\pair{R'_i}{R'_o}}{C'}$
satisfies $R_i\cup R_o \subset R'_i \cup R'_o$
while $R_i \cup R_o$ is bounded from above by the finite set $P$.
\par
Since the function \textit{aset} is recursive, we
can show the result by induction on the depth of the recursion.
\par
For the base case, we consider two possibilities:
\begin{itemize}
\item If $\pair{R_i}{R_o}$ is inconsistent with $C$
then 
$\expansion{\pair{R_i}{R_o}}{C} = \emptyset$ and
the lemma holds vacuously.
\item If $R_i \cup R_o = P$, $R_i^\star$ is an extended answer set
and $\pair{R_i}{R_o}$ is consistent with $C$ then
$\expansion{\pair{R_i}{R_o}}{C} = \set{R_i}$ and, again,
the lemma holds vacuously.
\end{itemize}
\par
For the induction step, let $R\in\aset{\pair{R_i}{R_o}}{C}$.
There are two possibilities.
\begin{itemize}
\item $R\in\aset{\pair{R_i\cup\set{r}}{R_o}}{C}$
and thus, by the induction hypothesis, 
$R\in\linebreak\min\expansion{\pair{R_i\cup\set{r}}{R_o}}{C}$.
Lemma~\ref{lemma-aux-2} then implies
$R\in\min\expansion{\pair{R_i}{R_o}}{C}$.
\item If $R\in\aset{\pair{R_i}{R_o\cup\set{r}}}{C'}$
where $C'$ satisfies 
\begin{eqnarray}\label{eq00}
\lefteqn{ \expansion{\pair{R_i}{R_o\cup\set{r}}}{C'} =} \nonumber\\
&& \set{T\mid T\in\expansion{\pair{R_i}{R_o\cup\set{r}}}{C} \land 
\Forall{m\in M}{\Exists{x\in\witness{m}}{x\subseteq T}}}
\end{eqnarray}
The induction hypothesis implies
$R\in\min\expansion{\pair{R_i}{R_o\cup\set{r}}}{C'}$. Together
with (\ref{eq00}) and Lemma~\ref{lemma-aux-3}, this implies
that
$R\in\min\expansion{\pair{R_i}{R_o}}{C}$.
\end{itemize}
\end{proof}

\begin{theoremrecap}{\ref{poas-eq-aset}}
Let \olp{P}{<} be an ordered program,
$\pair{R_i}{R_o}$ be a specification and $C$ a constraint.
Then $\aset{\pair{R_i}{R_o}}{C} = 
\min\expansion{\pair{R_i}{R_o}}{C}$.
\end{theoremrecap}
\begin{proof}
Immediate from Lemma~\ref{poas-in-aset} and
Lemma~\ref{aset-in-poas}
\end{proof}

\begin{theoremrecap}{\ref{pi2hard}}
The problem of deciding, given an arbitrary ordered program $P$
and a literal $a$, whether $a$ occurs in every preferred answer set of
$P$ is $\Pi_2^P$-hard.
\end{theoremrecap}
\begin{proof}
The proof uses a reduction of the known $\Pi_2^P$-hard problem of
deciding whether a quantified boolean formula 
$\phi = \Forall{x_1 ,\ldots, x_n}{\Exists{y_1, \ldots, y_m}{F}}$
is valid, where we may assume that $F = \land_{c\in C}c$ with
each $c$ a disjunction of literals over $X\cup Y$ with 
$X = \set{x_1 ,\ldots, x_n}$ and $Y = \set{y_1 ,\ldots, y_m}$
($n,m>0$).
\par
The program $P$ corresponding to $\phi$ is shown below.
\[
\begin{array}{c|c}
P_1 = \set{\prule{x}{}\;\prule{\neg x}{}\mid x\in X} &
\begin{array}{c}
P_2 = \prule{sat}{}\\
\hline
P_3 = \set{\prule{y}{}\;\prule{\neg y}{}\mid y\in Y}\\
\hline
P_4 = \prule{sat}{\neg sat}\\
\hline
P_5 = \set{\prule{\neg sat}{c'}\mid c\in C}
\end{array}
\end{array}
\]
where $c'$ is obtained from $c$ by taking the negation, i.e.
if $c=l_1\lor\dots\lor l_n$, then $c'$ denotes $\neg l_1\land\dots\land\neg l_n$.
\par
Obviously, the construction of $P$ can be done in polynomial time.
Intuitively, the rules in $P_1$ and $P_3$ are used to guess a truth 
assignment for $X\cup Y$. 
\par
In the sequel, we will abuse notation by using $x_M$ and $y_M$
where $M$ is an answer set for $P$, to denote subsets of $M$,
e.g. $x_M = X\cap M$ and in expressions such as $F(x_M, y_M)$ 
which stands for $F(x_1 , \ldots,x_n,y_1,\ldots y_m)$
with $x_i = \mathbf{true}$ iff $x_i\in x_M$ and, similarly,
$y_j = \mathbf{true}$ iff $y_j\in y_M$. We will also sometimes
abbreviate the arguments of $F$, writing e.g. $F(x,y)$ rather than
$F(x_1 , \ldots,x_n,y_1,\ldots y_m)$. 
\par
The following properties of $P$ are straightforward to show:
\begin{enumerate}
\item\label{proofcomppi_p1} A rule in $P_5$ is only applicable if $F$
does not hold. If we have an extended answer set $M$ containing $\neg sat \in M$,
then $F(x_M,y_M)$ does not hold.
\item\label{proofcomppi_p2} Any preferred answer set $M$ of $P$ always
satisfies all the rules in $P_5$, otherwise
$M'=(M\setminus\set{sat})\cup\set{\neg sat}$ would be better then $M$,
a contradiction.
\item\label{proofcomppi_p3} For each combination $X_i$ of $X$ we have
at least one preferred answer set containing $X_i$. For extended answer 
sets $M_1$ and $M_2$, with $M_1\cap X\not= M_2\cap X$, neither 
$M_1\rlt M_2$ nor $M_2\rlt M_1$ holds,
as the rules in $P_1$ are unrelated to any other rules.
\end{enumerate}
\par
We show that $\phi$ is valid iff $sat\in M$ for every preferred answer
set $M$ of $P$.
\par
To show the ''if'' part, assume that every preferred answer set $M$
contains $sat\in M$. Suppose $\phi$ is not valid, then there exists a
combination $X_i$ of $X$ such that $\Forall{y}{\neg F(X_i,y)}$. By
(\ref{proofcomppi_p3}) we know that there must exist at least one
preferred answer set $M'$ with $X_i\subseteq M'$. Combining
(\ref{proofcomppi_p1}) and (\ref{proofcomppi_p2}) with the fact that
$\Forall{y}{\neg F(X_i,y)}$ yields $\neg sat\in M'$, contradicting
that every preferred answer set contains $sat$. Thus, $\phi$ is valid.
\par
To show the reverse, assume that $\phi$ is valid.
Suppose there exists a preferred answer set $M$ of $P$ not containing
$sat$, i.e. $sat\not\in M$. Then, by construction of $P$, $\neg sat\in M$.
By (\ref{proofcomppi_p1}), this yields that $F(x_M,y_M)$ does not
hold. As $M$ is preferred any other preferred answer set $M'$
containing $x_M$ (we will have one for each combination of $Y$ due to
the rules in $P_3$) must also make $F$ not valid, otherwise $M$ could
not be preferred, as the one making $F$ valid would satisfy the rule
$P_4$ which is defeated w.r.t. $M$, making it obviously preferred upon $M$.
This yields that for the combination $x_M$ of $X$
no combination of $Y$ exists making $F$ true, i.e.
$\Exists{x}{\Forall{y}{\neg F(x,y)}}$, a contradiction. Thus, every
preferred answer set contains $sat$.
\end{proof}

\begin{theoremrecap}{\ref{eas-sim-theorem}}
Let $P$ be an $ELP$. Then, $S$ is an extended answer sets of $P$
iff there is an answer set $S'$ of $E(P)$ such that
$S=S'\cap(\hbases{P})$.
\end{theoremrecap}
\begin{proof}
\par
\fbox{$\Longrightarrow$} Let $S$ be an extended answer set of $P$ and
consider the interpretation 
$S' = S\cup\set{\anot{a}\mid\prule{\Naf{a}}{\beta}\in P\land S\models\beta\cup\set{\Naf{a}}}$. 
Then, $S'$ is an answer set of $E(P)$.
\par
First we show that every rule in $E(P)$ is satisfied w.r.t. $S'$. 
Constraint rules are clearly satisfied w.r.t. $S'$. Suppose there is a rule
$\prule{a}{\beta,\Naf{\neg a},\Naf{(\anot{a})}}\in E(P)$ such that
$S'\models\beta\cup\set{\Naf{\neg a},\Naf{(\anot{a})}}$ and
$S'\not\models a$. By construction of the rule, the corresponding rule
in $P$ is also applicable and not applied w.r.t. $S$. As $S$ is an
extended answer set, this means that there must be an applied rule
\prule{X}{\beta} with $X=\neg a$ or $X=\Naf{a}$ in $P$ defeating the
rule \prule{a}{\beta}. By construction of the rules in $E(P)$ and the
construction of $S'$, the corresponding rule in $E(P)$ is also applied
w.r.t. $S'$, i.e. either $S'\models\neg a$ or $S'\models\anot{a}$,
contradicting $S'\models\set{\Naf{\neg a},\Naf{(\anot{a})}}$. The same
reasoning can be done for the rules \prule{\anot{a}}{\beta,\Naf{a}}
in $E(P)$.
\par
From $S$ an extended answer set of $P$ we know that ${(P_S)^S}^\star=S$.
Consider the rules $\prule{a}{\beta}\in P$ with $a$ an ordinary
literal, such that $S\models\beta\cup\set{a}$ and thus
$\prule{a}{\beta\setminus\Naf{(\beta^-)}}\in (P_S)^S$.
This rule is represented in $E(P)$ as 
\prule{a}{\beta,\Naf{\neg a},\Naf{(\anot{a})}} and by construction of
$S'$ we have that $\prule{a}{\beta\setminus\Naf{(\beta^-)}}\in E(P)^{S'}$.
Thus, $S\subseteq {E(P)^{S'}}^\star$. Clearly, if
$S\models\beta\cup\set{\Naf{a}}$ for a rule
$\prule{\Naf{a}}{\beta}\in P$, then the corresponding rule
$\prule{\anot{a}}{\beta,\anot{a}}\in E(P)$ is also applicable w.r.t.
$S'$ by construction of $S'$, thus
$\prule{\anot{a}}{\setmin{\beta}{\Naf{(\beta^-)}}}\in E(P)^{S'}$,
yielding that 
$S'\setminus S\subseteq {E(P)^{S'}}^\star$.
Finally, $S'\subseteq {E(P)^{S'}}^\star$ and because all rules
are satisfied w.r.t. $S'$, ${E(P)^{S'}}^\star = S'$.
\par
\fbox{$\Longleftarrow$} Let $S'$ be an answer set of $E(P)$ and let 
$S = S'\cap(\hbases{P})$. We show that $S$ is an extended answer set
of $P$.
\par
First we show that every rule in $P$ is either satisfied or defeated
w.r.t. $S$. Again, constraints are clearly satisfied w.r.t. $S$.
Suppose there is a rule $\prule{a}{\beta}\in P$ such that
$S\models\beta$ and $S\not\models a$. The corresponding rule in
$E(P)$, i.e. either \prule{a}{\beta,\Naf{\neg a},\Naf{(\anot{a})}}
or \prule{\anot{a}}{\beta,\Naf{a}}, is satisfied w.r.t. $S'$,
yielding that either $S'\not\models\Naf{\neg a}$,
$S'\not\models\Naf{(\anot{a})}$ or $S'\not\models\Naf{a}$. Thus,
there is an applied rule $\prule{X}{Y}\in E(P)$ with $X=\neg a$,
$X=\anot{a}$ or $X=a$. By construction of the rules in $E(P)$ this
means that the corresponding rule in $P$ is also applied w.r.t. $S$,
defeating the rule \prule{a}{\beta}.
\par
Finally we have to show that ${(P_S)^S}^\star = S$, which is quite
obvious as the generating, i.e. applicable and applied, rules
$\prule{a}{\beta\setminus\Naf{(\beta^-)}}\in E(P)^{S'}$
for ${E(P)^{S'}}^\star$ are
also in $(P_S)^S$ and they clearly do not depend on any literal of the
form $\anot{a}$. Thus, ${(P_S)^S}^\star = S$.
\end{proof}


\begin{lemma}\label{form_of_nsp_as}
Let $P = \olp{R}{<}$ be an extended ordered logic program. 
Every proper extended answer set $S'$ of $N_s(P)$ is of
the form 
$S'= S \cup
\set{\phi(a),\neg\phi(\Naf{a}),\neg\phi(\neg a)\mid a\in S} \cup
\set{\phi(\Naf{a})\mid a\in(\hbases{R})\setminus S} \cup 
\set{\neg\phi(a)\mid\prule{\Naf{a}}{\beta}\in R \land S\models\beta\cup\set{\Naf{a}}}$,
where $S\subseteq\hbases{R}$.
\end{lemma}
\begin{proof}
Take $S'$ a proper extended answer set of $N_s(P)$.
Let $S= S'\cap(\hbases{P})$. 
\par
First consider $a\in S$. For $a$ to be in $S'$, we must have that
$\phi(a)\in S'$, by construction of the rules \prule{a}{\phi(a)} in
$R_c$. Combining $\phi(a)\in S'$, with the constraint
\prule{}{\phi(a),\phi(\Naf{a})} and the rule
\prule{\phi(\Naf{a})}{}, yields that $\neg\phi(\Naf{a})\in S'$
as otherwise $S'$ cannot be proper, i.e. \prule{\phi(\Naf{a})}{}
cannot be defeated.
\par
As $\phi(a)\in S'$, there must be an applied rule
$\prule{\phi(a)}{\phi(\beta)}\in N_s(P)$, i.e.
$\phi(\beta)\cup\set{\phi(a)}\subseteq S'$. Then, also
\prule{\neg\phi(\neg a)}{\phi(\beta),\phi(a)} is applicable and yields,
combined with the constraint \prule{}{\phi(a),\phi(\neg a)} and the
fact that $S'$ is proper, $\neg\phi(\neg a)\in S'$.
Thus, $X = \set{\phi(a),\neg\phi(\Naf{a}),\neg\phi(\neg a)\mid a\in S} \subseteq S'$.
\par
Let $Y = (\hbases{R})\setminus S$. By construction of the rules in
$N_s(P)$, we can never defeat the rules $\set{\prule{\phi(\Naf{a})}{}\mid a\in Y}$, 
yielding that $Z = \set{\phi(\Naf{a})\mid a\in Y} \subseteq S'$.
\par
Finally, one can easily see that the only rules left which can derive
something new into $S'$, i.e. something in 
$T = S'\setminus S\setminus X\setminus Z$, are of the form
\prule{\neg\phi(a)}{\phi(\beta),\phi(\Naf{a})}, where
$a\in(\hbases{R})\setminus S$. Thus, 
$T\subseteq\set{\neg\phi(a)\mid a\in Y}$.
\par
To end, we show that 
$T = \set{\neg\phi(a)\mid\prule{\Naf{a}}{\beta}\in R \land S\models\beta\cup\set{\Naf{a}}}$.
The $\supseteq$ direction is obvious, as the corresponding rules
\prule{\neg\phi(a)}{\phi(\beta),\phi(\Naf{a})} in $N_s(P)$ will be
applicable, thus applied by construction of the rules, by construction
of $S'$ and the fact that $S'$ is a proper extended answer set.
The $\subseteq$ direction is also quite obvious.
Take $\neg\phi(a)\in T$ with $a\in(\hbases{R})\setminus S$ and suppose
there is no rule \prule{\Naf{a}}{\beta} in $P$ that is applied w.r.t. $S$.
Then, by construction of the rules in $N_s(P)$, there is no applied rule 
in $N_s(P)$ with $\neg\phi(a)$ in the head, contradicting that $S'$ is
a proper extended answer set, as $T\subseteq S'$.
\end{proof}

\begin{lemma}\label{ext_as_coincide}
Let $P = \olp{R}{<}$ be an extended ordered logic program. 
Then, $S$ is an extended answer set of $P$ iff 
$S'= S \cup
\set{\phi(a),\neg\phi(\Naf{a}),\neg\phi(\neg a)\mid a\in S} \cup
\set{\phi(\Naf{a})\mid a\in(\hbases{R})\setminus S} \cup 
\set{\neg\phi(a)\mid\prule{\Naf{a}}{\beta}\in R \land S\models\beta\cup\set{\Naf{a}}}$
is a proper extended answer set of $N_s(P)$.
\end{lemma}
\begin{proof}
\par
\fbox{$\Longrightarrow$} Take $S$ an extended answer set of $P$ and
take $S'$ as defined. First we show that every rule $r$ of $N_s(P)$ is
either satisfied or defeated w.r.t. $S'$.
\par
\begin{itemize}
\item Every rule $\prule{\phi(\Naf{a})}{}\in R_n$ is either satisfied or
defeated w.r.t. $S'$. Suppose $\phi(\Naf{a})\not\in S'$. Then, by
construction of $S'$, $a\in S$. As $S$ is an extended answer set of
$P$, there is a rule $\prule{a}{\beta}\in P$ such that
$S\models \set{a}\cup\beta$. For this rule we have three corresponding
rules in $N_s(P)$, i.e. $\prule{\phi(a)}{\phi(\beta)},\; \prule{\neg\phi(\neg a)}{\phi(\beta),\phi(a)}$
and \prule{\neg\phi(\Naf{a})}{\phi(\beta),\phi(a)}.
By construction of $S'$ and
$\phi(\beta)$ we have that $\phi(\beta)\cup\set{\phi(a),\neg\phi(\Naf{a})}\subseteq S'$,
which yields that \prule{\neg\phi(\Naf{a})}{\phi(\beta),\phi(a)} defeats
\prule{\phi(\Naf{a})}{}.
\item The rules of type \prule{\phi(a)}{\phi(\beta)} are either satisfied or
defeated w.r.t. $S'$. Suppose that $\phi(\beta)\subseteq S'$ and
$\phi(a)\not\in S'$, implying that $a\not\in S'$, which yields that the corresponding rule
\prule{a}{\beta} in $P$ is applicable w.r.t. $S$ by construction of
the rules in $N_s(P)$ and the construction of $S'$. As $a\not\in S'$ we have $a\not\in S$,
implying that the rule \prule{a}{\beta} in $P$ must be defeated
w.r.t. $S$ by an applied rule \prule{\neg a}{\beta'} or \prule{\naf{a}}{\beta'}, 
as $S$ is an extended answer set of $P$. Again by
construction of $S'$ and the rules in $N_s(P)$, the rule
\prule{\phi(\neg a)}{\phi(\beta')} or \prule{\phi(\naf{a})}{\phi(\beta')}
in $N_s(P)$ is applied w.r.t. $S'$, making either the rule
\prule{\neg\phi(a)}{\phi(\beta'),\phi(\neg a)} or the rule
\prule{\neg\phi(a)}{\phi(\beta'),\phi(\naf{a})} applied,
defeating \prule{\phi(a)}{\phi(\beta)}.
\item The rules of type \prule{\phi(\Naf{a})}{\phi(\beta)} are either
satisfied or defeated w.r.t. $S'$. Suppose that $\phi(\beta)\subseteq S'$
and $\phi(\Naf{a})\not\in S'$. This implies that $a\in S$, by
construction of $S'$. As $S$ is an extended answer set, there must be
a rule $\prule{a}{\beta'}\in P$ such that $S\models\beta'\cup\set{a}$.
Then, by construction of $S'$, the rule 
$\prule{\phi(a)}{\phi(\beta')}\in R'$ is applied w.r.t. $S'$, but
also the rule $\prule{\neg\phi(\Naf{a})}{\phi(\beta'),\phi(a)}\in R'$ is,
defeating \prule{\phi(\Naf{a})}{\phi(\beta)}.
\item The rules left in $R'$, i.e. rules of type
$\prule{\neg\phi(\neg a)}{\phi(\beta),\phi(a)};\;
\prule{\neg\phi(\Naf{a})}{\phi(\beta),\phi(a)}$ and
\prule{\neg\phi(a)}{\phi(\beta),\phi(\Naf{a})}, are always 
satisfied w.r.t. $S'$ by construction of $S'$.
\item All the rules in $R_c$ are always satisfied w.r.t. $S'$.
By construction of $S'$, the constraints \prule{}{\phi(a),\phi(\Naf{a})}
and \prule{}{\phi(a),\phi(\neg a)} are always satisfied w.r.t. $S'$,
i.e. inapplicable.
Clearly, rules of the form \prule{a}{\phi(a)} will be either applied,
or inapplicable. Finally, the constraints in $P$ are satisfied w.r.t.
$S$, thus the constraints $\prule{}{\phi(\beta)}\in R_c$ are
satisfied w.r.t. $S'$, yielding that $S'$ is proper.
\end{itemize}
\par
The only thing left to proof is that $N_s(P)_{S'}^\star = S'$. First
of all, let $T = \set{\phi(\Naf{a})\mid \phi(\Naf{a})\in S'}$. Obviously,
$T\subseteq N_s(P)_{S'}^\star$ as $\prule{\phi(\Naf{a})}{}\in N_s(P)_{S'}$
for every $\phi(\Naf{a})\in S'$.
\par
We know that ${(R_S)^S}^\star = S$ as $S$ is an extended answer set of $P$.
A rule $\prule{a}{\beta}\in R$ with $a$ an ordinary literal that is kept 
in $(R_S)^S$ is of the form \prule{a}{\beta\setminus\naf{\beta^-}}\footnote{
	Constraints, i.e. \prule{}{\beta}, in $(R_S)^S$ are not
	important in here, as they do not play a role in ${(R_S)^S}^\star$ because
	$S$ is an extended answer set.
}. For all those rules in $(R_S)^S$
we have that $\set{\phi(\Naf{a})\mid a\in\beta^-}\subseteq T$, and by
construction of the rules in $N_s(P)$, the construction of $S'$,
the fact that $T$ can be derived immediately in
$N_s(P)_{S'}^\star$ and the fact that ${(R_S)^S}^\star = S$, 
this yields that $\set{\phi(a)\mid a\in S}\subseteq N_s(P)_{S'}^\star$,
implying that also $S\subseteq N_s(P)_{S'}^\star$ by the rules
\prule{a}{\phi(a)} in $R_c$.
\par
For every applied rule $\prule{\phi(a)}{\phi(\beta)}\in N_s(P)$ with $a\in S$,
we have that \prule{\neg\phi(\neg a)}{\phi(\beta),\phi(a)} and
\prule{\neg\phi(\Naf{a})}{\phi(\beta),\phi(a)} are satisfied w.r.t. $S'$
and thus are in $N_s(P)_{S'}$, yielding that
$\set{\neg\phi(\neg a),\neg\phi(\Naf{a})\mid a\in S}\subseteq
N_s(P)_{S'}^\star$.
\par
Finally, consider the applied rules $\prule{\Naf{a}}{\beta}\in P$. By
construction of $S'$ and the rules \prule{\phi(\Naf{a})}{\phi(\beta)} and
\prule{\neg\phi(a)}{\phi(\beta),\phi(\Naf{a})} in $N_s(P)$, both rules are
satisfied w.r.t. $S'$, thus in $N_s(P)_{S'}$.
From \prule{\Naf{a}}{\beta} applied w.r.t. $S$ we have that 
$S\models\beta\cup\set{\Naf{a}}$ and by construction of $S'$, both
rules in $N_s(P)_{S'}$ are applicable w.r.t. $S \cup
\set{\phi(a),\neg\phi(\Naf{a}),\neg\phi(\neg a)\mid a\in S} \cup
\set{\phi(\Naf{a})\mid a\in(\hbases{R})\setminus S}$, which is already
shown to be in $N_s(P)_{S'}^\star$. Thus,
$\set{\neg\phi(a)\mid\prule{\Naf{a}}{\beta}\in R \land S\models\beta\cup\set{\Naf{a}}}
\subseteq N_s(P)_{S'}^\star$. As a conclusion,
$S'\subseteq N_s(P)_{S'}^\star$ and $S' = N_s(P)_{S'}^\star$ by the
fact that each rule is satisfied or defeated w.r.t. $S'$.
\par
\fbox{$\Longleftarrow$} Take $S'$ a proper extended answer set of
$N_s(P)$. From Lemma~\ref{form_of_nsp_as} we know that $S'$ is
of the form $S'= S \cup
\set{\phi(a),\neg\phi(\Naf{a}),\neg\phi(\neg a)\mid a\in S} \cup
\set{\phi(\Naf{a})\mid a\in(\hbases{R})\setminus S} \cup 
\set{\neg\phi(a)\mid\prule{\Naf{a}}{\beta}\in R \land S\models\beta\cup\set{\Naf{a}}}$,
where $S\subseteq\hbases{R}$.
The only thing we have to show is that $S$ is an extended answer set
of $P$.
\par
By construction, all constraints in $\prule{}{\beta}\in P$ are satisfied w.r.t. $S$ as the
constraints $\prule{}{\phi(\beta)}\in R_c$ are satisfied w.r.t. $S'$.
Every rule \prule{a}{\beta} in $P$, with $a$ an extended literal,
is either satisfied or defeated w.r.t. $S$. Suppose not, i.e.
$S\models\beta\land S\not\models a$ and there is no defeater w.r.t. $S$
in $P$ for \prule{a}{\beta}. By construction of the rules in
$N_s(P)$, the corresponding rule $\prule{\phi(a)}{\phi(\beta)}\in N_s(P)$
is applicable w.r.t. $S'$, but not applied by construction of $S'$.
As $S'$ is a proper extended answer set, the rule
\prule{\phi(a)}{\phi(\beta)} must be defeated w.r.t. $S'$, i.e.
there must be an applied rule \prule{\neg\phi(a)}{\beta',\phi(X)} in $N_s(P)$
with $X=\neg a$ or $X=\Naf{a}$ when $a$ is an ordinary literal
and $X=\lit{a}$ when $a$ is an extended literal. This yields that
the corresponding rule in $P$ is also applied w.r.t. $S$, defeating
$\prule{a}{\beta}\in P$, a contradiction.
\par
Finally, we show that ${(R_S)^S}^\star = S$. For every applied rule
$\prule{\phi(a)}{\phi(\beta)}\in N_s(P)$ w.r.t. $S'$ and $a$ an ordinary
literal, we have that $\set{a\mid\phi(\Naf{a})\in\phi(\beta)}\cap S = \emptyset$ 
by construction of $S'$ and thus of $S$. This implies that
$\prule{a}{\beta\setminus\naf{\beta^-}}\in (R_S)^S$ and 
$\beta\setminus\naf{\beta^-}\subseteq S$. 
We know that $N_s(P)_{S'}^\star = S'$ and that
$\set{\phi(a)\mid a\in S}$ are only produced by the applied rules of
the form \prule{\phi(a)}{\phi(\beta)} with $a$ an ordinary literal,
starting from \set{\phi(\Naf{a})\mid a\in(\hbases{R})\setminus S}.
\par
Now, combining the above with the construction of the rules in
$N_s(P)$ yields that $S\subseteq {(P_S)^S}^\star$.
As every rule in $P$ is either applied or defeated w.r.t. $S$ we have
$S = {(P_S)^S}^\star$.
\end{proof}

\begin{theoremrecap}{\ref{eolp-to-olp-sim}}
Let $P = \olp{R}{<}$ be an extended ordered logic program. Then,
$M$ is a preferred answer set of $P$ iff there exists a proper
preferred answer set $M'$ of $N_s(P)$, such that $M=M'\cap(\hbases{R})$.
\end{theoremrecap}
\begin{proof}
\par
\fbox{$\Longrightarrow$} Take $S$ a preferred answer set of $P$ and 
take $S'$ as defined in Lemma~\ref{ext_as_coincide}. By that same
lemma we get that $S'$ is a proper extended answer set of $N_s(P)$. 
Suppose $S'$ is not preferred, then there exists a proper extended 
answer set $S''\not=S'$ of $N_s(P)$ such that $S''\rleq S'$. 
Let $T = S''\cap(\hbases{P})$.
From Lemma~\ref{ext_as_coincide} we know that $T$ is an extended
answer set of $P$.
As both $S'$ and $S''$ are proper, we have that 
$N_s(P)_{S'}\setminus N_s(P)_{S''}$ and $N_s(P)_{S''}\setminus N_s(P)_{S'}$
only contain rules from $R'$ and $R_n$.
\par
Suppose $N_s(P)_{S'}\setminus N_s(P)_{S''}$ contains a rule $r$ from $R'$.
Then, $S''\rleq S'$ implies that there exists a rule 
$r'\in N_s(P)_{S''}\setminus N_s(P)_{s'}$ such that $r' < r$.
From the proof of Lemma~\ref{ext_as_coincide} we know that
only rules of the type $\prule{\phi(a)}{\phi(\beta)}\in R'$,
with $a$ an extended literal, can be defeated and
we also know from the same proof that the rules in $P$
corresponding with the defeated rules in $R'$ w.r.t. $S'$ ($S''$) 
are also defeated w.r.t. $S$ ($T$), yielding that $T\rleq S$, a contradiction.
\par
Suppose $N_s(P)_{S'}\setminus N_s(P)_{S''}$ contains only rules from $R_n$,
which implies that $P_S\setminus P_T=\emptyset$. Then $S''\rleq S'$ yields
that $N_s(P)_{S''}\setminus N_s(P)_{S'}$ must
contain at least a rule from $R'$ to counter the $R_n$ rules in
$N_s(P)_{S'}\setminus N_s(P)_{S''}$. As a result, $P_S \not= P_T$,
implying that, combined with $P_S\setminus P_T=\emptyset$,
we have $T\rleq S$, a contradiction.
\par
Thus, $S'$ is a proper preferred answer set of $N_s(P)$ and $S = S'\cap(\hbases{P})$.
\par
\fbox{$\Longleftarrow$} Take $S'$ a proper preferred answer set of $N_s(P)$.
Let $S = S'\cap(\hbases{P})$. Again from Lemma~\ref{ext_as_coincide} we know that 
$S$ is an extended answer set of $P$.
Suppose $S$ is not preferred. Then, there exists an extended answer
set $T$ of $P$ such that $T \rleq S$, which implies $P_S\not= P_T$.
Take $T'$ as described by Lemma \ref{ext_as_coincide} and we get 
that $T'$ is a proper extended answer of
$N_s(P)$ with the same defeated rules in $R'$ as the corresponding
defeated rules in $P$. The same holds for defeated rules w.r.t. $S$
and $S'$. Then $T\rleq S$ implies $T'\rleq S'$, a contradiction.
\end{proof}


\begin{lemma}\label{LPOD_as_propextas_LP}
Let $P$ be an LPOD. $S$ is answer set of $P$ iff 
$S'=S\cup\set{nap_{\prule{a_1\times\dots\times a_n}{\beta}}\mid\prule{a_1\times\dots\times a_n}{\beta}\in P\land S\not\models\beta}$ 
is a proper extended answer set of $L(P)$.
\end{lemma}
\begin{proof}
\par
\fbox{$\Longrightarrow$} Take $S$ an answer set of $P$, yielding that
there exists a split program $P'$ of $P$ such that $S$ is an answer
set of $P'$. Take $S'$ as defined. We first show that every rule in 
$L(P)$ is either satisfied or defeated w.r.t. $S'$.
\par
Obviously, all rules in $P_r$ corresponding with normal rules in $P$ 
are satisfied as these rules are contained in every split program. 
The rules \prule{a_i}{\beta,\naf{\setmin{\set{a_1,\dots,a_n}}{a_i}}}
for every $1\leq i\leq n$ for each ordered disjunctive rule are
clearly satisfied w.r.t. $S'$ as every split program contains an option
that is satisfied w.r.t. $S$, thus also w.r.t. $S'$.
Furthermore, the rules \prule{nap_r}{\naf{l}} and \prule{nap_r}{l} in $P_r$ 
are also satisfied w.r.t. $S'$ by construction of $S'$, i.e. when an
ordered disjunctive rule $r$ is not applicable w.r.t. $S$, we have
$nap_r\in S'$ making the rules clearly satisfied, while in the case
$r$ is applicable, none of the rules are applicable.
\par
The rules $\prule{\naf{nap_r}}{}\in P_d$ for which $nap_r\not\in S'$ 
are clearly satisfied, while the ones for which $nap_r\in S'$ are
clearly defeated by an applied rule \prule{nap_r}{l} or
\prule{nap_r}{\naf{l}}, this by construction of $S'$.
Further, the rules $\prule{\NAF\ a_i}{\beta,\naf{\set{a_1,\dots,a_{i-1}}}}\in P_d$ 
for which $S'\models\beta\cup\naf{\set{a_1,\dots,a_{i-1}}}$ and $a_i\in S'$ are 
defeated by the corresponding applied rule
$\prule{a_i}{\beta,\naf{\set{a_1,\dots,a_{i-1}}}}\in P_i$.
\par
Finally, if a rule $\prule{a}{\beta}\in P_1\cup\dots\cup P_n$ is applicable but not
applied, it is defeated by a rule $\prule{\NAF\ a}{\beta}\in P_d$, by
construction of $P_d$.
\par
The only thing left to show is ${(L(P)_{S'})^{S'}}^\star = S'$. By construction
of $L(P)$ we get that $P'\subseteq L(P)$. This yields, as all rules in
$P'$ are satisfied w.r.t. $S$, that $P'^S\subseteq (L(P)_{S'})^{S'}$.
As a result, $S\subseteq {(L(P)_{S'})^{S'}}^\star$. Clearly, for each
unapplicable ordered disjunctive rule $r=\prule{a_1\times\dots\times a_n}{\beta}$,
i.e. $\Exists{l\in\beta}{S\not\models l}$, we have that
$\prule{nap_r}{\naf{l}}$ (when $l$ is a literal) or 
$\prule{nap_r}{l}$ (when $l$ is a naf-literal) is in $L(P)_{S'}$, yielding that
$S'\subseteq {(L(P)_{S'})^{S'}}^\star$. As all rules are satisfied or
defeated w.r.t. $S'$, this results in $S' = {(L(P)_{S'})^{S'}}^\star$.
\par
\fbox{$\Longleftarrow$} Take $S'$ a proper extended answer set of $L(P)$
and let $S=S'\cap(\hbases{P})$. Consider $L(P)_{S'}$, i.e. all satisfied 
rules in $L(P)$ w.r.t. $S'$. Take $P'\subseteq L(P)_{S'}$ in the following way:
\begin{enumerate}
\item\label{rules1}Take all rules in $P_r$ corresponding with normal rules in $P$.
As $S'$ is proper and the satisfaction of the selected rules does not
change w.r.t. $S$, every normal rule in $P$ is also in $P'$.
\item\label{rules2}Take every applicable and applied rule $r$ from the $P_i$'s.
By construction of $L(P)$ at most one option for every ordered
disjunctive rule is taken, as options in $P_j$ with $j > i$ are not
applicable, because they contain $\NAF\ H(r)$ in the body; and options $r'$
in $P_j$ with $j < i$ must be defeated, otherwise $r$ would not be applicable because
it contains $\NAF\ H(r')$ in the body.
\item\label{rules3}For every ordered disjunctive rule that is not
applicable w.r.t. $S'$ (or $S$) take an arbitrary option to be in $P'$.
\end{enumerate}
Clearly, $P'$ as constructed above, is a split program of $P$. 
\par
The rules $\prule{c_i}{\beta,\NAF\ (\set{c_1,\dots,c_n}\setminus\set{c_i})}\in P_r$
that are applicable and applied are not needed to produce the $S$ part in ${(L(P)_{S'})^{S'}}^\star$,
as also the rule $\prule{c_i}{\beta,\NAF\ \set{c_1,\dots,c_{i-1}}}\in P_i$ is applicable
and applied, thus in $(L(P)_{S'})^{S'}$. Furthermore, the rules with
$nap_r$ in the head are also not needed to derive $S$.
Thus, all rules needed to produce $S$ in $L(P)$ are also in $P'$, 
yielding that ${P'^S}^\star = S$, making $S$ an answer set of $P$.
\end{proof}
\par
\begin{lemma}\label{alternative_pref}
Let $P$ be a LPOD and let $S \not= T$ be proper extended answer sets for $L(P)$.
Then, the following is equivalent:
\begin{enumerate}
\item\label{alternative_pref1} $S \rleq T\enspace ,$
\item\label{alternative_pref2} $L(P)_S \not= L(P)_T$ and $\Forall{r\in L(P)_T\setminus L(P)_S,\exists r'\in L(P)_S\setminus L(P)_T}{r' < r}\enspace ,$
\item\label{alternative_pref3} $L(P)_S \not= L(P)_T$ and $\Exists{k}{Sat_{L(P)}^k(T)\subset Sat_{L(P)}^k(S) \land \Forall{j < k}{Sat_{L(P)}^j(T) = Sat_{L(P)}^j(S)}}\enspace .$
\end{enumerate}
where $Sat_{L(P)}^k(S)$ denotes the set of rules in $P_k\subseteq L(P)$ that are satisfied w.r.t. $S$.
\end{lemma}
\begin{proof}
By definition $(\ref{alternative_pref1}) \Leftrightarrow (\ref{alternative_pref2})$ holds.
\par
\fbox{$(\ref{alternative_pref2}) \Rightarrow (\ref{alternative_pref3})$} 
First, it can never happen that $\setmin{L(P)_T}{L(P)_S}=\emptyset$.
Suppose it is, than together with $L(P)_T\not= L(P)_S$, it implies $L(P)_T\subset L(P)_S$. 
We have three cases:
\begin{itemize}
\item $r=\prule{a_i}{\beta,\naf{\set{a_1,\dots,a_{i-1}}}}\in P_i$ with
$r\in \setmin{L(P)_S}{L(P)_T}$ yields $T\models\beta\cup\naf{\set{a_1,\dots,a_{i-1}}}$
and $T\not\models a_i$. By virtue of Lemma \ref{LPOD_as_propextas_LP}
we have that $T\cap(\hbases{P})$ is an LPOD answer set, thus we must
have a satisfied, w.r.t. $T$, option 
$r'=\prule{a_j}{\beta,\naf{\set{a_1,\dots,a_{j-1}}}}\in P_j$ with
$j>i$. For this rule, the corresponding rule
$r''=\prule{\naf{a_j}}{\beta,\naf{\set{a_1,\dots,a_{j-1}}}}\in P_d$ is
defeated w.r.t. $T$. Further, all rules
$\prule{\naf{a_k}}{\beta,\naf{\set{a_1,\dots,a_{k-1}}}}\in P_d$ with
$k\leq j$ are satisfied w.r.t. $T$, thus also w.r.t. $S$.
Now consider $S$ which satisfies the rule $r$.
We get two cases:
	\begin{itemize}
	\item $\Exists{k\leq i}{\prule{a_k}{\beta,\naf{\set{a_1,\dots,a_{k-1}}}}}\in P_k$ is applied
	w.r.t. $S$. However, this implies the rule $\prule{\naf{a_k}}{\beta,\naf{\set{a_1,\dots,a_{k-1}}}}\in P_d$
	is defeated w.r.t. $S$, a contradiction.
	\item $S\not\models\beta$, i.e. the ordered disjunctive rule $s$ corresponding with
	the rules $r$, $r'$ and $r''$ is not applicable w.r.t. $S$.
	Then, Lemma \ref{LPOD_as_propextas_LP} yields that
	$nap_s\not\in T$, while $nap_s\in S$. The former implies that
	the rule $\prule{\naf{nap_s}}{}\in P_d$ is satisfied w.r.t.
	$T$ and while it should also be satisfied by $S$, the latter
	clearly shows its defeated, a contradiction.
	\end{itemize}
\item $r=\prule{\naf{a_i}}{\beta,\naf{\set{a_1,\dots,a_{i-1}}}}\in P_d$ with
$r\in \setmin{L(P)_S}{L(P)_T}$ yields that there also must be a rule
$r'=\prule{a_j}{\beta,\naf{\set{a_1,\dots,a_{j-1}}}}\in P_j$ with
$r'\in \setmin{L(P)_S}{L(P)_T}$, which is handled in the previous case.
\item $r=\prule{\naf{nap_s}}{}\in P_d$ with $r\in \setmin{L(P)_S}{L(P)_T}$ yields 
that $s$ is applicable w.r.t. $S$ and not applicable w.r.t. $T$.
However, $s$ applicable w.r.t. $S$ implies at least one option 
$\prule{a_i}{\beta,\naf{\set{a_1,\dots,a_{i-1}}}}\in P_i$ of $s$
will be applied w.r.t. $S$, defeating the rule
$t=\prule{\naf{a_i}}{\beta,\naf{\set{a_1,\dots,a_{i-1}}}}\in P_d$.
However, $s$ not applicable w.r.t. $T$ implies $t$ is satisfied w.r.t.
$T$, yielding $t\in\setmin{L(P)_T}{L(P)_S}$, a contradiction.
\end{itemize}
\par
As $\setmin{L(P)_T}{L(P)_S}\not=\emptyset$, $S\rleq T$ implies that
$\setmin{L(P)_S}{L(P)_T}\not=\emptyset$.
Take the most specific rule $r'\in L(P)_S\setminus L(P)_T$. 
Clearly, $r' < r$ for every $r\in L(P)_T\setminus L(P)_S$.
As $S$ and $T$ are proper we must have $r' \in P_k$ for a certain $k\in[1\dots n]$.
As $r'$ is the most specific rule in $L(P)_S\setminus L(P)_T$, 
we have $\Forall{j < k}{(L(P)_S\setminus L(P)_T)\cap P_j = (L(P)_T\setminus L(P)_S)\cap P_j = \emptyset}$, 
which is equivalent with $\Forall{j < k}{Sat_{L(P)}^j(T) = Sat_{L(P)}^j(S)}$. As $r'$ is the most specific
rule in $L(P)_S\setminus L(P)_T$, every rule $r\in L(P)_T\setminus L(P)_S$ must
belong to a $P_i$ with $i > k$ (or to $P_d$), which yields that 
$(L(P)_T\setminus L(P)_S)\cap P_k \subset (L(P)_S\setminus L(P)_T)\cap P_k$;
and as a result $Sat_{L(P)}^k(T)\subset Sat_{L(P)}^k(S)$. 
\par
\fbox{$(\ref{alternative_pref3}) \Rightarrow (\ref{alternative_pref2})$}
Take a $k$ so that (\ref{alternative_pref3}) holds. From $\Forall{j < k}{Sat_{L(P)}^j(T) = Sat_{L(P)}^j(S)}$ it follows that
$\Forall{j < k}{(L(P)_S\setminus L(P)_T)\cap P_j = (L(P)_T\setminus L(P)_S)\cap P_j = \emptyset}$.
From $Sat_{L(P)}^k(T)\subset Sat_{L(P)}^k(S)$, we get $(Sat_{L(P)}^k(S)\setminus Sat_{L(P)}^k(T)) \subset ({L(P)}_S\setminus {L(P)}_T)$
and $(L(P)_T\setminus L(P)_S)\cap P_k = \emptyset$, which directly yields
that $r' < r$ for every $r'\in Sat_{L(P)}^k(S)\setminus Sat_{L(P)}^k(T)$ and every $r\in {L(P)}_T\setminus {L(P)}_S$.
\end{proof}

\begin{theoremrecap}{\ref{lpodtheorem}}
An interpretation $S$ is a preferred LPOD answer
set of a LPOD $P$ iff there exists a proper preferred answer set $S'$ of $L(P)$
such that $S=S'\cap(\hbases{P})$.
\end{theoremrecap}
\begin{proof}
\par
\fbox{$\Longrightarrow$}
Take $S$ a preferred LPOD answer set of $P$. From Lemma~\ref{LPOD_as_propextas_LP} we have that
there exists a proper extended answer set $S'$ of $L(P)$ such that $S=S'\cap(\hbases{P})$. 
Suppose $S'$ is not preferred, i.e. there exists a proper extended answer set $T' \not= S'$
of $L(P)$ such that $T' \rleq S'$. Again, by virtue of Lemma~\ref{LPOD_as_propextas_LP},
we have that $T=T'\cap(\hbases{P})$ is an LPOD answer set of $P$.
\par
Applying Lemma \ref{alternative_pref} to $T'\rleq S'$ w.r.t. $L(P)$
yields
\begin{equation}
\Exists{k}{Sat_{L(P)}^k(S')\subset Sat_{L(P)}^k(T') \land \Forall{j < k}{Sat_{L(P)}^j(S') = Sat_{L(P)}^j(T')}}\enspace .\label{lpodproof1}
\end{equation}
By construction of $L(P)$ we have that 
$Sat_{L(P)}^1(S') = S^1(P)\setminus\set{\prule{a}{\beta}\in P}$ and
$Sat_{L(P)}^j(S')\setminus Sat_{L(P)}^{j-1}(S') = S^j(P)$\footnote{
	This is not completely correct as rules in $Sat_{L(P)}^j(S')\setminus Sat_{L(P)}^{j-1}(S')$
	are options from ordered disjunctive rules, while $S^j(P)$
	contains ordered disjunctive rules. However, as only one
	option for such a rule can be in the former, there is a
	one-to-one mapping between the elements in the former and the
	elements in the latter.
	} (and the same for $T$ and $T'$).
Combined with (\ref{lpodproof1}) this yields 
\begin{equation}
\Exists{k}{S^k(P)\subset T^k(P) \land \Forall{j < k}{S^j(P) = T^j(P)}}\enspace .\label{lpodproof2}
\end{equation}
But, (\ref{lpodproof2}) implies that $T$ is LPOD-preferred upon $S$ for $P$, a contradiction.
\par
\fbox{$\Longleftarrow$}
Take $S'$ a proper preferred answer set of $L(P)$. By Lemma
\ref{LPOD_as_propextas_LP}, $S=S'\cap(\hbases{P})$ is an LPOD answer set of $P$. Suppose
$S$ is not LPOD preferred, i.e. there exists an LPOD answer set $T$
such that $T \rlt_b S$. This yields
\begin{equation}
\Exists{k}{S^k(P)\subset T^k(P) \land \Forall{j < k}{S^j(P) = T^j(P)}}\enspace .\label{lpodproof3}
\end{equation}
\par
Again from Lemma~\ref{LPOD_as_propextas_LP} we have that there must
exist a proper extended answer set $T'$ of $L(P)$ such that
$T=T'\cap(\hbases{P})$.
By construction of $L(P)$, for every ordered disjunctive rule $r$ that is satisfied to degree $k$,
the corresponding options in $P_1,\dots,P_{k-1}$ are defeated, while
the corresponding options in $P_k,\dots,P_n$ are satisfied, yielding
that (it also hold for $T$ and $T'$)
\begin{eqnarray}
Sat_{L(P)}^1(S') = S^1(P)\setminus\set{\prule{a}{\beta}\in P}\label{lpodproof4} \enspace ,\\
Sat_{L(P)}^k(S') = Sat_{L(P)}^{k-1}(S') + S^k(P)
\label{lpodproof5}\enspace .
\end{eqnarray}
Combining (\ref{lpodproof4}), (\ref{lpodproof5}) and
(\ref{lpodproof3}) results in
\begin{equation}
\Exists{k}{Sat_{L(P)}^k(S')\subset Sat_{L(P)}^k(T') \land \Forall{j < k}{Sat_{L(P)}^j(S') = Sat_{L(P)}^j(T')}}\enspace .\label{lpodproof6}
\end{equation}
Using (\ref{lpodproof6}) with Lemma~\ref{alternative_pref} yields $T' \rleq S'$, a contradiction.
\end{proof}


\begin{lemmarecap}{\ref{arenas-same-leq}}
Let $D$ , $D_1$ and $D_2$ be databases over the same Herbrand base.
$\Delta_D (D_1 ) \subseteq \Delta_D (D_2 )$ iff $\Delta(D,D_1 ) \subseteq \Delta(D, D_2 )$.
\end{lemmarecap}
\begin{proof}
To show the ``only if'' part, assume that
$p\in\Delta (D, D_1 ) = (\setmin{D^+}{{D_1}^+}) \cup (\setmin{{D_1}^+}{D^+})$.
We consider two possibilities:
\begin{enumerate}
\item
	If $p\in\setmin{D^+}{{D_1}^+}$ then $p\in D$ and 
	$\neg p\in D_1$.
	Hence, by definition, $\neg p\in\Delta_D (D_1 )$
	Since $\Delta_D (D_1 ) \subseteq \Delta_D (D_2 )$,
	$\neg p\in\setmin{D_2}{D}$ and thus
	\mbox{$p\in\setmin{D^+}{{D_2}^+} \subseteq \Delta(D,D_2 )$}.
\item
	If $p\in\setmin{{D_1}^+}{D^+}$ then 
	$p\in\Delta_D (D_1 ) \subseteq \Delta_D (D_2 )$.
	Since $p$ is an atom, this implies that 
	$p\in\setmin{{D_2}^+}{D^+} \subseteq \Delta(D,D_2 )$.

\end{enumerate}
To show the ``if'' part, assume that
$p\in\Delta_D (D_1 ) = \setmin{D_1}{D}$.
We consider two cases.
\begin{enumerate}
\item	
  If $p$ is an atom, then 
  $p\in\setmin{{D_1}^+}{D^+}\subseteq \Delta(D, D_1 )\subseteq\Delta(D, D_2 )$.
  Thus \mbox{$p\in\setmin{{D}^+}{{D_2}^+}$} or
  \mbox{$p \in\setmin{{D_2}^+}{{D}^+} $}. 
  The former is impossible because $p\not\in D^+$.
  The latter implies that $p\in(\setmin{D_2}{D}) = \Delta_D (D_2 )$.
\item 	
  If $p$ is a negative literal, then 
  $\neg p\in(\setmin{D^+}{{D_1}^+})\subseteq\Delta(D,D_1 )$.
  Thus $\neg p\in\Delta(D, D_2 )$ and, because 
  $\neg p\in D$, $\neg p \in\setmin{D^+}{{D_2}^+}$.
  Consequently, $p\in(\setmin{{D_2}^+}{D^+})\subseteq\Delta_D (D_2 )$.
\end{enumerate}
\end{proof}

\begin{lemma}\label{lemma2}
Let $D$ be a database and let $C$ be a consistent set of constraints
with $L_C$ the set of literals occurring in $C$.
For any $C$-repair $R$ of $D$, we have that 
$\setmin{R}{L_C} = \setmin{D}{L_C}$, i.e. $D$ and $R$ agree on
\setmin{\hbase{D}}{\hbase{L_C}}.
\end{lemma}
\begin{proof}
Straightforward. Suppose e.g. that $D$ and $R$ do not agree on
$l\not\in L_C$, i.e. $l\in R$ and $\neg l\in D$.
Since $l$ does not occur in any constraint,
$R' = (\setmin{R}{\set{l}})\cup{\neg l}\models C$ and, moreover
$R'\leq_D R$, contradicting that $R$ is a repair.
\end{proof}

\begin{theoremrecap}{\ref{repair-is-aset}}
Let $D$ be a database and let $C$ be a consistent set of constraints
with $\hbase{C}\subseteq\hbase{D}$. 
Each repair of $D$ w.r.t. $C$ is a preferred answer set of $P(D,C)$.
\end{theoremrecap}
\begin{proof}
Let $R$ be a repair of $D$ w.r.t. $C$. 
By definition, $\hbase{R} = \hbase{D}$ and $R$ is consistent. 
On the other hand,
$P(D,C)_R$ obviously contains a rule \lrule{d}{a}{} for each
$a\in R\cap D$ and a rule \lrule{n}{\neg a}{} for each 
$a\in\setmin{R}{D}$. Thus $P(D,C)_R^\star = R$.
\par
We next show that $R$ satisfies or defeats each rule in 
$P(D,C)$. By definition, and the construction of $P(D,C)$,
all $c$-rules are satisfied. On the other
hand, any $n$-rule \lrule{n}{a}{} which is not satisfied is
defeated by an applied $d$-rule \lrule{d}{\neg a}{}. As to
the $d$-rules, Lemma~\ref{lemma2} implies that
each rule \lrule{d}{a}{} where $a$ does not occur in $C$
is applied.
\par
The remaining case concerns $d$-rules \lrule{d}{a}{} where
$a$ occurs in $C$ and $a\not\in R$, i.e. the rule is not
satisfied. We consider two possibilities: either
$\neg a$ occurs in $C$ and thus, by the construction of
$P(D,C)$, there exists a satisfied $c$-rule
\lrule{c}{\neg a}{\alpha}, defeating \lrule{d}{a}{}, or
$\neg a$ does not occur in $C$. The latter case is impossible since
then, by Lemma~\ref{lemma2},
$R$ should agree with $D$ on $\neg a$, contradicting
our assumption that $\neg a\in\setmin{R}{D}$.
\par
Hence, $R$ is an extended answer set.
\par
To show that $R$ is minimal w.r.t \rleq, assume that, on the contrary,
there exists an extended answer set $M\rleq R$ of $P(D,C)$ such that
$M\neq R$.
Thus, by Definition~\ref{def:reduct-order},
\begin{equation}\label{eq3}
\Forall{r\in\setmin{P(D,C)_R}{P(D,C)_M}}{
   \Exists{r'\in\setmin{P(D,C)_M}{P(D,C)_R}}{r'<r}}\enspace .
\end{equation}
Note that $\setmin{P(D,C)_R}{P(D,C)_M}\neq\emptyset$ (otherwise,
$M=R$ would follow).
Since $c$-rules cannot be defeated, any $r$ as in (\ref{eq3}) must
have a label $d$ or $n$. 
By definition, $R$ satisfies all $c$-rules and thus any $r'$
satisfying (\ref{eq3}) must also be a $d$ or $n$ rule.
Combining these observations yields that any $r$ and $r'$ satisfying
(\ref{eq3}) must be an $n$-rule and a $d$-rule, respectively.
But this implies that $\Delta_D (M) \subset \Delta_D (R)$,
contradicting the fact that $R$ is a $C$-repair of $D$.
\end{proof}

\begin{lemma}\label{lemma1}
Let $D$ be a database and let $C$ be a consistent set of constraints
with $\hbase{C}\subseteq\hbase{D}$.
Each preferred answer set $M$ of $P(D,C)$ satisfies $C$, i.e. $M\models C$.
\end{lemma}
\begin{proof}
Clearly, as $C$ is consistent, there must exist a proper extended answer
set $I$, i.e. $I\models C$. By Lemma \ref{lemma-proper}, each preferred
answer set must be proper, from which this lemma follows.
\end{proof}

\begin{theoremrecap}{\ref{aset-is-repair}}
Let $D$ be a database and let $C$ be a consistent set of constraints
with $\hbase{C}\subseteq\hbase{D}$.
Each preferred answer set of $P(D,C)$ is a $C$-repair of $D$.
\end{theoremrecap}
\begin{proof}
Let $M$ be a preferred answer set of $P(D,C)$. From Lemma \ref{lemma1}
it follows that $M\models C$. Assume that, on the contrary, 
$M$ is not a $C$-repair of $D$, i.e.
there exists a $C$-repair $R$ such that $\Delta_D (R)\subset \Delta_D (M)$
(Definition~\ref{def:repair}).
Let $l$ be a literal in \setmin{\Delta_D (M)}{\Delta_D (R)}.
Thus $l\in M$ while $\neg l\in D\cap R$. By construction,
$P(D,C)$ contains an $d$-rule $r=\lrule{d}{\neg l}{}$ in
\setmin{P_R}{P_M}. Since $M$ is preferred and, by
Theorem~\ref{repair-is-aset}, $R$ is an extended answer set,
$M\rleq R$ implies the existence of a rule $r'<r$ with
$r'\in\setmin{P_M}{P_R}$. But any such rule $r'$ must
be a $c$-rule which, by definition, is satisfied by $R$,
a contradiction.
\end{proof}

\end{document}